# Education Paper

# The Astrobiology Primer: An Outline of General Knowledge—Version 1, 2006

**EDITOR-IN-CHIEF**

Lucas J. Mix, Berkeley, CA (LM; Introduction, 3A, 4A–D, 6B, 7, and 7B)

**EDITORS**

John C. Armstrong, Weber State University (JA; 5 and 5C)
Avi M. Mandell, Pennsylvania State University (AvM; 5B)
Annika C. Mosier, Stanford University (AnM; 6 and 6C)
Jason Raymond, Arizona State University (JR; 4 and 4B–E)
Sean N. Raymond, University of Colorado, Boulder (SnR; 1, 2, and 2A)
Frank J. Stewart, Harvard University (FS; 3, 3A–C, and 6B)
Kaspar von Braun, California Institute of Technology (KvB; 1 and 1A–C)
Olga Zhaxybayeva, Dalhousie University (OZ; 4 and 4B–D)

**AUTHORS**

Linda Billings, SETI Institute (LB; 7C)
Vyllinniskii Cameron, Pennsylvania State University (VC; 2B)
Mark Claire, University of Washington (MC; 2C and 2E)
Greg J. Dick, University of California, San Diego (GD; 6B)
Shawn D. Domagal-Goldman, Pennsylvania State University (SG; 2D)
Emmanuelle J. Javaux, University of Liege (EJ; 4F)
Orion J. Johnson, University of Southern California (OJ; 6A)
Chris Laws, University of Washington (CL; 5A)
Margaret S. Race, SETI Institute (MR; 7C)
Jon Rask, Ames Research Center (JoR; 7A)
John D. Rummel, NASA Headquarters (JRm; 7C)
Rachel T. Schelble, University of Southern California (RS; 2C)
Steve Vance, University of Washington (SV; 5D)

## CONTRIBUTORS

Zach Adam, University of Washington
Peter Backus, SETI Institute
Luther Beegle, Jet Propulsion Laboratory
Janice Bishop, SETI Institute
Kristie Boering, University of California, Berkeley
Michael Briley, University of Wisconsin, Oshkosh
Wendy Calvin, University of Nevada
David Catling, University of Washington
Carol Cleland, University of Colorado, Boulder
K. Estelle Dodson, Ames Research Center
Julie Fletcher, Ames Research Center
Eduardo de la Fuente Acosta, University of Guadalajara, Mexico
Ico de Zwart, University of Colorado, Boulder
Jennifer Eigenbrode, Carnegie Institute
Jack Farmer, Arizona State University
Siegfried Frank, Potsdam Institute of Climate Impact Research, Germany
Peter Gogarten, University of Connecticut
Edward Goolish, Ames Research Center
Rosalind Grymes, Ames Research Center
Nader Haghighipour, University of Hawaii
Troy Hudson, Jet Propulsion Laboratory
Vladimir Ivkovic, Institute of Anthropological Research, Croatia
Muffarah Jahangeer, George Mason University
Bruce Jakosky, University of Colorado, Boulder
Scott Kenyon, Smithsonian Astrophysical Observatory
Steven Kilston, Ball Aerospace
Andrew Knoll, Harvard University
Eric Korpela, University of California, Berkeley
David Lamb, University of California, Santa Barbara
Jospeh Lazio, Naval Research Laboratory
Richard Lenski, Michigan State University
Lindsey Link, University of Colorado, Boulder
Karen Lloyd, University of North Carolina, Chapel Hill
Jonathan Lunine, University of Arizona
Michael Manga, University of California, Berkeley
Tim McCoy, Smithsonian Institution
Karen Meech, University of Hawaii
Gustavo Mello, Observatorio do Valongo, Brazil
Steven Mojzsis, University of Colorado, Boulder
David Morrison, Ames Research Center
Oliver Morton
Duane Moser, Desert Research Institute
Kenneth Nealson, University of Southern California
Francis Nimno, University of California, Los Angeles
Ray Norris, CSIRO Australia Telescope National Facility
Erika Offerdahl, Arizona State University
Tom Olien, Humber Institute, Canada
Norman Pace, University of Colorado, Boulder
Susan Pfiffner, University of Tennessee
Cynthia Philips, SETI Institute
Shanti Rao, Jet Propulsion Laboratory
Debora Rodriquez, Michigan State University
John Rummel, NASA Headquarters
Bill Schopf, University of California, Los Angeles
Sara Seager, Carnegie Institute
Norman Sleep, Stanford University
Mitchell Sogin, Marine Biological Laboratory, Woods Hole
Nina Solovaya, Slovak Academy of Science, Slovak Republic
Woodruff Sullivan, University of Washington
Brian Thomas, University of Kansas
Thorsteinn Thorsteinsson, Orkustofnun, Iceland
Carmen Tornow, German Aerospace Center
Michael Wevrick
Nick Woolf, University of Arizona
Kosei Yamaguchi, Institute for Frontier Research on Earth Evolution, Japan
Michael Zerella, University of Colorado, Boulder

## FOREWORD TO VERSION 1 (2006)

**The Astrobiology Primer has been created as a reference tool for those who are interested in the interdisciplinary field of astrobiology. The field incorporates many diverse research endeavors, but it is our hope that this slim volume will present the reader with all he or she needs to know to become involved and to understand, at least at a fundamental level, the state of the art. Each section includes a brief overview of a topic and a short list of readable and important literature for those interested in deeper knowledge. Because of the great diversity of material, each section was written by a different author with a different expertise. Contributors, authors, and editors are listed at the beginning, along with a list of those chapters and sections for which they were responsible. We are deeply indebted to the NASA Astrobiology Institute (NAI), in particular to Estelle Dodson, David Morrison, Ed Goolish, Krisstina Wilmoth, and Rose Grymes for their continued enthusiasm and support. The Primer came about in large part because of NAI support for graduate student research, collaboration, and inclusion as well as direct funding. We have entitled the Primer version 1 in hope that it will be only the first in a series, whose future volumes will be produced every 3–5 years. This way we can insure that the Primer keeps up with the current state of research. We hope that it will be a great resource for anyone trying to stay abreast of an ever-changing field. If you have noticed any errors of fact or wish to be involved in future incarnations of the project, please contact Lucas Mix (e-mail: lucas@flirble.org).**

## TABLE OF CONTENTS

## Introduction (LM)

Astrobiology, the study of life as a planetary phenomenon, aims to understand the fundamental nature of life on Earth and the possibility of life elsewhere. To achieve this goal, astrobiologists have initiated unprecedented communication among the disciplines of astronomy, biology, chemistry, and geology. Astrobiologists also use insights from information and systems theory to evaluate how those disciplines function and interact. The fundamental questions of what "life" means and how it arose have brought in broad philosophical concerns, while the practical limits of space exploration have meant that engineering plays an important role as well.

The benefits of this interdisciplinary collaboration have been, and continue to be, immense. The input of scientists from multiple areas has forced researchers to become aware of their basic assumptions and why they do science the way they do. Cooperation has led to insights about the many connections between life and the atmosphere, oceans, and crust of Earth. Comparisons of geologic and atmospheric features on Earth, Mars, and Venus have provided insight into these planets' varying histories and what part life may or may not have played. Likewise, the growing body of data about planets orbiting distant stars (as well as smaller bodies orbiting our own star) has revealed much about the formation of our own planetary system and how stars and planets interact.

There is, however, a steep learning curve in astrobiology due to the variety of backgrounds brought to the field. Conferences provide unique insights but can be tedious and unproductive when presenters and audience fail to connect on terminology and basic assumptions. The purpose of the Astrobiology Primer is to bridge some of these gaps by creating a common foundation in knowledge.

### What the Primer Is

Our intention was to create a short, accessible, and very limited reference for astrobiologists. We hope that many of you will, at one time or another, come to the Primer with a specific question and find the answer or the location of an answer in the literature. We also hope that you will come to see the Primer as a kind of reference guide that will help you to brush up on the basics before attending a talk or visiting a colleague in a different field.

The Primer originated at the 2001 general meeting of the NASA Astrobiology Institute (NAI). NAI worked hard on the education and incorporation of graduate students, and this project stemmed from a discussion as to the range of background knowledge an astrobiologist may be expected to have. Several of us decided that graduate students entering into the field would benefit from a short volume that would not only act as an introduction to astrobiology but also provide background in areas where an astrobiologist may wish to be competent. As the project evolved, we came to believe that such a reference would be invaluable to the entire community and very likely contribute to the developing discussion between disciplines.

Above all, the Primer is intended to be an ongoing project. This volume includes only the most recent and limited version of a larger endeavor. If the project is well received, a new version of the Primer will be released every 3–5 years. Each version will incorporate new research and correct the mistakes of earlier versions No one volume, of course, can contain the vast amount of information brought to play in astrobiology, but we believe that the Primer will provide a forum and a language around which the community will have the opportunity to develop a consensus about central issues.

### What the Primer Is Not

The Astrobiology Primer is not a textbook. Many good textbooks exist, and we would like to recommend the following:

#### *Astronomy*

#### *Biochemistry*

#### *Biology*

A publication this size cannot provide a complete picture of any one topic. It is our wish, however, that the Primer will be viewed as a table of contents for the vast literature available. Sections of the Primer are written to represent a synthesis and point readers to resources and terms that will allow them to find the current state of the art. The articles listed below each section are valuable places to look for more information, but they are not, in general, citations. Each author brings a great deal of additional information to his or her section.

Unfortunately, the Primer cannot be comprehensive. Each chapter will demonstrate the biases and limited knowledge of the editors. Where we have erred, we hope that you will contact us and help us create a reliable central index of information for astrobiologists.

Neither is the Primer an official representation of NASA/NAI research, positions, or goals. While NAI has been greatly helpful in facilitating communication and resources, the contents of the Primer are a product of independent work. We are deeply indebted to NAI and the International Astronomical Union Commission 51: Bioastronomy for their efforts.

## How the Primer Was Arranged

The Primer was constructed collaboratively. Ninety researchers from around the world contributed information with regard to what they expected from other astrobiologists and what they would like to know themselves but still had difficulty understanding (see Contributors). Those submissions were read and considered by the editors who produced a list of seven general categories of knowledge, represented by the seven chapters in the Primer.

Each chapter has been divided into small, digestible segments with references for further information. The brevity of the Primer means that much important research was left out and some material has been oversimplified. The editors would like to apologize for the undoubted errors and omissions that resulted from our attempts to condense the material.

The initials appearing at the beginning of each chapter reflect the editor responsible for gathering and organizing the information in that chapter. Likewise, the initials at the beginning of each section indicate the author for that section. In both cases, however, input from contributors and editors has seriously impacted the content and tone of the material.

## A Note on Terminology

One of the challenges inherent to any multidisciplinary work is that a number of key words hold different meanings within the various disciplines. We have tried, in this work, to be consistent throughout, though some ambiguities remain.

The word "terrestrial" has been particularly problematic with regard to the Primer, and therefore "terrestrial" or earthlike planets are referred to as rocky planets and "terrestrial" or Earth-located organisms and environments are referred to as terrean. Also, when referring to organisms living on land or in soil (as opposed to aquatic or oceanic), "terrestrial" has been replaced by land-dwelling or similar words. The word "terrestrial" was not used except in references and proper names (*i.e.*, Terrestrial Planet Finder).

## Abbreviation

NAI, NASA Astrobiology Institute.

# Chapter 1. Stellar Formation and Evolution (KvB, SnR)

## Introduction: Stars and Astrobiology

Stars like the Sun appear to the eye as pinpoints of light in the sky. They are, in fact, enormous balls of gas, each emitting vast amounts of energy every second in a losing battle against the relentless inward gravitational force generated by their own titanic masses. The radiation emitted by the Sun constitutes the primary source of energy for most life forms on earth (see Fig. 1.1). There are between 100 and 400 billion stars with vastly different astrophysical properties in the Milky Way Galaxy, which itself is one of hundreds of billions of very different galaxies in the observable universe.

In the Milky Way, stars are continually being born and dying—some quiet and some extremely violent deaths. The material of which stars are made is physically and chemically altered within the stars throughout the course of their lifetimes and at their deaths. Practically, all elements heavier than helium—and thus all components of the organic molecules necessary for life to form—are produced by stars. (For a specific definition of "organic," which is not equivalent to "biological," see Sec. 3A.) At the end of a star's life, this material is partially returned to the place from whence it came: the interstellar medium (ISM). The ISM contains a mixture of gas (single atoms and small molecules) and dust (larger chunks of matter, near 1 $\mu$m in size) that penetrates the entire galaxy. It is in the ISM where the formation of stars, planets, and ultimately life begins.

[Editor's note: Astronomers call all elements except hydrogen (H) and helium (He) metals.

**FIG. 1.1. Local numbers.**

| | |
|---|---|
| Mass of Earth = 5.976 x $10^{24}$ kg | Mass of Sun = 1.989 x $10^{30}$ kg |
| Mean radius of Earth = 6.371 x $10^{6}$ m | Diameter of Sun = 1.392 x $10^{6}$ km |
| Escape speed = 11.2 km/s | Luminosity of Sun = 3.827 x $10^{26}$ W |
| Mass of Moon = 7.348 x $10^{22}$ kg | The speed of light (c) = 2.998 x $10^{8}$ m/s |
| Mean radius of Moon = 1.738 x $10^{6}$ m | Light Year = 9.460 x $10^{12}$ km |
| Escape speed = 2.4 km/s | (1 Parsec = 3.262 light years) |
| Avg. Distance from Sun to Earth = 1 Astronomical Unit (AU) = 1.496 x $10^{8}$ km | |
| Gravitational Constant (G) = 6.67 x $10^{-11}$ $Nm^2/kg^2$ | |
| Acceleration due to gravity at Earth surface (g) = 9.8 $m/s^2$ | |
| Acceleration due to gravity at Moon surface = 1.6 $m/s^2$ | |
| Acceleration due to gravity at Mars surface = 3.7 $m/s^2$ | |

Chemists call metal all elements except hydrogen, halogens (B, C, N, O, Si, P, S, As, Se, Te, F, Cl, Br, I, At), and noble gasses (He, Ne, Ar, Kr, Xe, Rn). Biologists will occasionally refer to elements other than those basic to life (C, H, N, O, P, S) as metals if they are associated with a biomolecule; however, usage is not consistent.]

## 1A. The Formation of Stars (KvB)

### *The ISM and Molecular Clouds (MCs)*

Stars are born within MCs, dark interstellar clouds within the ISM made of gas and dust. The density of these MCs is $10^2$–$10^4$ molecules/cm$^3$, about a factor of 1,000 higher than typical interstellar densities. Molecular gas makes up about 99% of the mass of the clouds; dust accounts for a mere 1%. Typical masses of MCs are about $10^3$–$10^4$ solar masses extended across 30 light years. The chemical composition of MCs varies, but a good approximation is that they are made up mostly of hydrogen (74% by mass), helium (25% by mass), and heavier elements (1% by mass).

Shock waves within MCs (caused by supersonic stellar winds or massive supernovae, for example) are chiefly responsible for the compression and subsequent clumping of gas and dust inside the cloud. Gravity condenses local clumps to form protostars and, eventually, stars. This is only possible if the cloud is sufficiently cold and no other mechanisms (such as strong magnetic fields or turbulence) are present to halt collapse. Typical temperatures within MCs are 10–30 K.

MCs of many thousands of solar masses typically produce a large number of protostars and are commonly referred to as stellar nurseries. The protostellar cloud loses mass, and star formation efficiency—the proportion of the original mass retained by the final star—is on the order of a few percent. Some consequences of this fact are the following:

- Stars preferentially form in clusters of most likely hundreds of stars since MCs will not just form one star. Over time (10–100 million years), these clusters may disperse since the gravitational pull between stars is not strong enough to keep them together once the gas dissipates.
- A substantial fraction of stars (half or more of all stars in the Milky Way) are members of binary or multiple star systems.
- High-mass stars (4 solar masses or higher) form much less frequently than their lower-mass counterparts. High-mass stars require the existence of much more massive prestellar cores (described below) than do low-mass stars. Furthermore, high-mass stars have significantly shorter lifetimes.

### *Protostars and the Birth of Stars*

A blob of gas several times the size of our Solar System, called a prestellar core, contracts under its own gravity to form a protostar. Half of the system's initial gravitational energy disperses through radiation, while the other half is converted into heat. The temperature of the core begins to rise. After a few thousand years of collapse, low-mass protostars ($<$4 solar masses) typically reach temperatures of 2,000–3,000 K and begin to emit light. As they are still enshrouded in a cloud of gas and dust, however, these protostars are not observable in visible wavelengths but only the infrared (IR), where dust is relatively transparent.

Over the course of 1–10 million years, the temperature in the core of a low-mass protostar reaches a few million K by converting the potential energy lost due to gravitational contraction into heat. At that point, a new source of energy is tapped: thermonuclear fusion sets in and starts converting H to He, thereby releasing enormous amounts of energy. Increased temperature halts gravitational collapse, and when hydrostatic equilibrium is reached (gravitational forces = forces created by thermal pressure), the protostar becomes a main sequence star (see Sec. 1B).

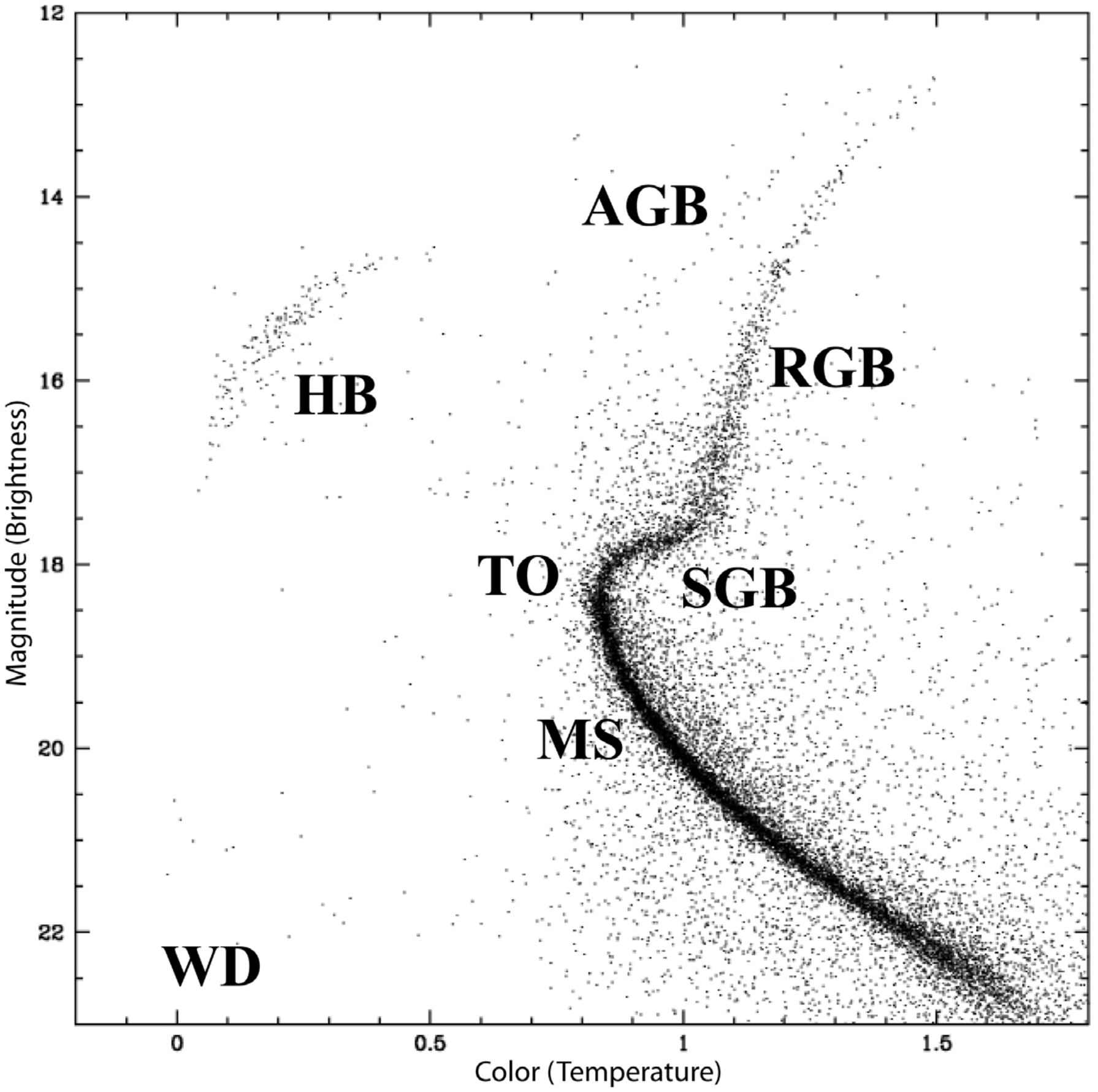

**FIG. 1.2. Color-magnitude diagram (CMD).** The observational CMD of the globular cluster M12 is shown. The $x$-axis indicates color as a proxy for temperature, with blue/hot to the left. The $y$-axis indicates magnitude/brightness, with bright toward the top. MS, main sequence; TO, turnoff point; SGB, subgiant branch; RGB, red giant branch; HB, horizontal branch; AGB, asymptotic giant branch; WD, theoretical white dwarfs region (not observed in the data used for this figure). It is important to point out that the data represent a snapshot of a stellar population's distribution of stars, and not the evolutionary track of any one star. The $x$-axis indicates color as the difference between magnitudes measured in green and near-infrared light. The smaller the value of the color, the bluer and hotter the star is. The $y$-axis indicates apparent brightness in units of magnitudes where lower brightness corresponds to numerically higher magnitudes. A difference of 5 magnitudes corresponds to a factor of 100 in flux (energy received per second for a given detector size), so a star of magnitude 5 (limit of visibility to the naked eye) appears 100 times fainter than a star of magnitude 0. Brightness/magnitude for a given star can be measured in different filters, such as blue, green, red, infrared, etc. The difference between two magnitudes in different filters for a given star is indicative of its color and thus surface temperature. For instance, a hot, blue star would appear brighter in the blue than in the red.

Less frequent, more massive (4–15 or more solar masses) protostars evolve differently. The collapse and subsequent heating occurs much more rapidly, and the onset of H-burning in the core is reached after only 20,000 to 1 million years.

A word of caution: the image of dust and gas simply falling straight onto the forming protostar is incorrect. All material in the universe has some angular momentum, and the gravitational collapse of gas and dust onto the prestellar core is a complex phenomenon. Conservation of angular momentum causes the infalling material to flatten and form a "circumstellar disk" of material rotating around the protostar. Particles in this disk will interact with each other, which causes them to lose energy and migrate inward, and as a result matter is accreted onto the protostar. It is believed that such circumstellar disks provide the environment necessary for planet formation.

## 1B. The Evolution of Stars (KvB)

### *The Main Sequence*

A star reaches the "main sequence" when its primary source of energy is thermonuclear fusion in the stellar core. The main sequence is a region in a diagram that shows the luminosities (energy emitted by the star per second) of stars as a function of their surface temperature—a Hertzsprung-Russell Diagram (HRD) (not shown here). Stars spend about 90% of their lives on the main sequence. As they evolve, their luminosities and surface temperature change, and they leave the main sequence to occupy different regions of the HRD. The observational equivalent of the HRD is the color-magnitude diagram (CMD), which lists magnitude (apparent brightness) as a function of color (indicative of effective surface temperature: blue = hotter, red = cooler). Figure 1.2 shows a CMD of the globular cluster (GC) M12 with the various evo-lutionary stages (discussed below) of a star's life labeled. All stars in a GC are located at about the same distance from Earth, have nearly identical chemical compositions, and were born around the same time. Since all stars in the GC are located at the same distance from Earth, their observed brightness reflects their intrinsic brightness (or luminosity). Stars of different masses evolve at different rates, however, and the CMD of a GC includes stars at different evolutionary stages (despite their same age).

**FIG. 1.3. Stellar properties.** Various astrophysical parameters for main sequence stars of various spectral types, having a chemical composition not too different than that of the Sun. MS lifetime reflects average time [in billions of years (Ga)] on the main sequence. The last two columns list peak wavelength emitted and the dominant color (observed near the star). IR, infrared; UV, ultraviolet. The calculations of the approximate main sequence lifetimes are based on the assumption that the Sun will be on the main sequence for 10 Ga. The luminosities are calculated using the Stephan-Boltzmann Law, *i.e.*, integrated over all wavelengths. Values for stars that have left the main sequence (Sec. 1C), brown dwarfs (Sec. 1B), Jupiter, and Earth are listed for comparison. (*At specific wavelengths only.)

| Spectral type | mass [$M_{Sun}$] | radius [$R_{Sun}$] | luminosity [$L_{Sun}$] | surface temp. [K] | MS lifetime [Ga] | $\lambda_{peak}$ [nm] | peak color (appearance) |
|---|---|---|---|---|---|---|---|
| O5 | 60 | 12 | 400,000 | 42,000 | 0.0004 | 70 | UV (blue) |
| B0 | 17.5 | 7.4 | 40,000 | 30,000 | 0.008 | 100 | UV (blue) |
| B5 | 5.9 | 3.9 | 700 | 15,200 | 0.1 | 190 | UV (blue) |
| A0 | 2.9 | 2.4 | 50 | 9,790 | 0.7 | 300 | UV (blue) |
| A5 | 2.0 | 1.7 | 10 | 8,180 | 1.8 | 350 | UV (blue) |
| F0 | 1.5 | 1.5 | 6 | 7,300 | 3.6 | 400 | violet (white) |
| F5 | 1.4 | 1.3 | 3 | 6,650 | 4.3 | 440 | blue (white) |
| G0 | 1.05 | 1.1 | 1.3 | 5,940 | 8.9 | 490 | blue (white) |
| G5 | 0.92 | 0.92 | 0.7 | 5,560 | 12 | 520 | green (white) |
| K0 | 0.79 | 0.85 | 0.5 | 5,150 | 18 | 560 | yellow (white) |
| K5 | 0.67 | 0.72 | 0.2 | 4,410 | 27 | 660 | red (white) |
| M0 | 0.51 | 0.60 | 0.07 | 3,840 | 54 | 750 | red (yellow) |
| M5 | 0.21 | 0.27 | 0.007 | 3,170 | 490 | 910 | IR (orange) |
| Pulsar | 1.5 | $1.5 \times 10^{-5}$ | 1,000* | several M | - | - | |
| Red Giant | 1-3 | 20 | 100 | 4,000 | - | 720 | red (yellow) |
| White Dwarf | 0.6 | 0.004 | 0.001 | 10,000 | - | 290 | UV (blue) |
| Brown Dwarf | 0.04 | 0.1 | $1 \times 10^{-5}$ | 1,000 | 300 | 3,000 | IR (red) |
| Jupiter | $1 \times 10^{-3}$ | 0.1 | $1 \times 10^{-6}$ | 180 | - | 16,000 | IR (invisible) |
| Earth | $3 \times 10^{-6}$ | 0.004 | $1 \times 10^{-9}$ | 280 | - | 10,000 | IR (invisible) |

The single most important stellar parameter, mass, defines where a star is located on the main sequence and what its evolution will be. The more massive a star is, the greater are its surface temperature and luminosity. The relationship is approximately $L \sim M^{3.5}$, where $L$ is luminosity and $M$ is stellar mass. A higher stellar mass furthermore implies a shorter lifetime and more violent final stages. Masses of main sequence stars range between 0.08 solar masses and 100 solar masses. Below 0.08 solar masses, the conditions for H-burning are not reached. However, for masses between 0.013 ($\sim$13 Jupiter masses) and 0.08 solar masses, less energetic deuterium burning may take place. These substellar objects, called brown dwarfs, separate giant planets from stars. Above 100 solar masses, internal pressures overcome gravity, and the prestellar core will not collapse sufficiently. Such high-mass stars are extremely rare.

Stellar mass can only rarely be observed directly, so stars are typically classified by "spectral type." The Sun's spectral type, for instance, is G2 V. Stars are sorted into the following spectral types in order of decreasing mass and surface temperature: O, B, A, F, G, K, M, L, T. Within these spectral types, smaller distinctions are made by adding a number between 0 and 9 to the spectral type. Thus, the surface temperature of a B star is hotter than that of an F star, and the surface temperature of a G2 star is slightly hotter than that of a G7 star. Some astrophysical parameters associated with spectral types of main sequence stars are given in the stellar properties table (Fig. 1.3). Furthermore, stars are categorized into luminosity classes, which are essentially a measure of evolutionary stage. Suffice it to say here that all main sequence stars are considered "dwarfs" and have a luminosity class of V (the Roman numeral for 5). The Sun—a G2 V star—is thus a main sequence star with a surface temperature of around 5,800 K.

By definition, a star is said to leave the main sequence when its H supply in the core has run out (Sec. 1C).

### *Astrophysical Properties of Stars on the Main Sequence*

Every star on the main sequence burns H to He in its core by one of two processes: the proton-proton chain or the carbon-nitrogen-oxygen cycle. Stars of roughly 1 solar mass or less reach up to about 16 million Kelvin in their cores and burn H via the proton-proton chain reaction. H atoms react directly to form He. More massive stars have higher core temperature ($T_{core}$) values and favor the carbon-nitrogen-oxygen cycle to produce energy. These stars contain a significant amount of carbon, nitrogen, and oxygen (about 1/1,000 atoms) produced by previous generations of stars. Atoms transitioning from one element to another catalyze the fusion of H.

The structure of stars also differs as a function of mass. Energy is produced in the stellar core and needs to be transported outward, by either convection (circulation of a fluid) or radiative diffusion (electromagnetic radiation). For stars with less than 0.8 solar masses, energy is transported via convection all the way to the surface and radiated into space. Stars with masses between 0.8 and 4 solar masses have a radiative zone around the core and a convective envelope around the radiative zone. Finally, high-mass stars (>4 solar masses) transport the energy from their cores first convectively and then radiatively to their surfaces.

Other astrophysical parameters vary tremendously between main sequence stars of different masses. Some are shown in the stellar properties table (Fig. 1.3).

### *Post-Main Sequence Evolution*

Over the course of its time on the main sequence, a star gradually uses up the H fuel in the core and consequently undergoes changes in luminosity, radius, and surface temperature. The decrease of atomic nuclei in the core (4 H nuclei form 1 He nucleus) lowers the pressure, which counteracts the gravitational force and results in a shrinking and heating of the core. Increased $T_{core}$ accelerates the rate of H-burning and causes the star to increase in size, surface temperature, and luminosity. Over the course of the Sun's 4.6 billion years on the main sequence, it has increased its radius by 6%, its surface temperature by 300 K, and its luminosity by 40%. The relative faintness of the young Sun has raised concerns about the temperature and habitability of the early Earth (the "faint young Sun" problem; see Sec. 2E).

Eventually, a star's H supply in the core runs out, and it leaves the main sequence at the turnoff point (TO in Fig. 1.2). Its time on the main sequence covers approximately 90% of a star's lifetime. The post-main sequence evolution takes up a relatively short time and, as on the main sequence, progresses much more rapidly for massive stars than low-mass stars. When the fuel in the core is exhausted, a star starts burning H to He in a shell of H surrounding the core. During this time, the He core contracts and heats up, while the layers above the H-burning shell expand and cool. Energy output of the shell and gravitational energy released from the shrinking core drive the luminosity of the star significantly higher. Thus, the star moves from left to right along the subgiant branch (SGB in Fig. 1.2) and up the red giant branch (RGB in Fig. 1.2). During this phase, a star's radius increases to many times its main sequence size, which results in increased luminosity in spite of decreased temperature. The Sun, for instance, will become so large that it will eventually swallow Mercury and Venus and reach Earth's orbit (which corresponds to about 200 times the Sun's current radius). Furthermore, the Sun will lose mass from its outer layers in the form of stellar wind at a much higher rate than during its time on the main sequence (around $10^{-7}$ $M_{Sun}$/year vs. $10^{-14}$ $M_{Sun}$/year on the main sequence).

As the He produced by the H-burning shell rains down onto the He core, the core grows in mass and heats up until, at around 100 million K, He starts thermonuclear fusion to form C and O (which happens toward the tip of the RGB; Fig. 1.2). This re-ignition of core fusion causes the cessation of H-shell burning. The star shrinks in radius, its surface temperature rises, and its luminosity decreases. Consequently, it moves across to the middle of the CMD (Fig. 1.2), from right to left along the horizontal branch, until the He supply in the core is depleted. During this phase, a star can become unstable to pulsation in its outer layers (these stars are known as RR Lyrae or Cepheid variables). At this point, the final stage of evolution of a star begins, which differs significantly between a low-mass star ($<4$ $M_{Sun}$) and a high-mass star ($>4$ $M_{Sun}$).

## 1C. The Death of Stars (KvB)

### *The Death of Low-Mass Stars*

When the He in the core of a low-mass star is depleted (it has all been fused into heavier elements), the He shell that surrounds the core starts thermonuclear fusion. History repeats itself, and the star enters a second red giant phase, this time crossing the CMD back along the asymptotic giant branch (AGB in the CMD). The star's mass loss rate is around $10^{-4}$ $M_{Sun}$/year, its surface temperature ~3,000 K, and the luminosity around 10,000 $L_{Sun}$. The very final stage of the low-mass star is the gradual shedding of all of the outer H and He layers, so that only the core remains. This core is roughly Earth-sized, with a surface temperature of ~100,000 K, and is comprised entirely of carbon and oxygen. The expanding outer layers that were shed are referred to as a planetary nebula, and the remaining core is known as a white dwarf (WD in the CMD). There are an estimated 20,000–50,000 planetary nebulae in the galaxy that return a total of around 5 $M_{Sun}$/year of material to the ISM (~15% of all matter expelled by the various sorts of stars).

### *The Death of High-Mass Stars and Replenishment of Metals in the ISM*

The $T_{core}$ of high-mass stars can reach sufficiently high values (600 million K) to initiate the thermonuclear fusion of C, which produces O, Ne, Na, and Mg as by-products. If $T_{core}$ reaches 1 billion K or more, even more thermonuclear reactions can take place, such as Ne-burning. If $T_{core}$ reaches 1.5 billion K, then O-burning will take place (creating S), and at 2.7 billion K, Si-burning will create a number of nuclei all the way up to Fe (this occurs for $M_{star} > 8$ $M_{Sun}$). New shell-burning and red giant phases occur before every new stage of core-burning. One may think of an iron core surrounded by a number of nuclei-burning shells (the "onion-model"). The result of these ever-expanding outer layers is a supergiant (luminosity class I) whose radius can reach 1,000 $R_{Sun}$ (one example is Betelgeuse in Orion).

During the burning of increasingly heavier nuclei, thermonuclear reactions produce a number of different chemical elements and isotopes by a process called slow neutron capture, during which neutrons collide and combine with charged protons. This sequence, however, ends at iron (for $M_{star} > 8$ $M_{Sun}$) since the fusion of Fe nuclei will not give off any net energy. Stars with $M_{star} < 8$ $M_{Sun}$ undergo a similar evolution to the low-mass stars and divest themselves of their outer layers (H, He, C, Ne, O, etc.) in a planetary nebula.

For stars with $M_{star} > 8$ $M_{Sun}$, the inert Fe core signals the star's impending doom. The core, no longer able to sustain thermonuclear fusion, contracts rapidly to produce the heat and energy required to stabilize the system. Once the Fe core reaches about 1.4 $M_{Sun}$, it is no longer capable of supporting itself against gravity, and the core rapidly contracts, going from Earth size to the size of a city in less than a second. The collapse is then halted by quantum effects. The outer layers of the star crash onto the core and bounce off, creating a shock wave as the outward moving material runs into additional infalling layers. The shock wave expands in a gigantic explosion in which the outer layers of the star are ejected in a supernova. Supernovae pack astonishing amounts of energy and typically outshine entire galaxies (100 billion stars) for a few days or weeks. Depending on how much mass remains in the stellar core, the center of a supernova forms either a neutron star (if the mass of the stellar core <3 $M_{Sun}$) or a black hole (if >3 $M_{Sun}$).

During this cataclysmic explosion, the shock wave that travels outward through the burning layers initiates a new wave of thermonuclear reactions by which many new elements are created, including many elements heavier than Fe. These "new" elements are thus returned to the ISM, the source of the star's original constituents during its formation. Finally, the shockwave of the supernova compresses regions of the ISM and fuels the formation of new stars. Thus, the process has completed one full circle and is ready to begin again with the formation of a new star.

## Abbreviations

AGB, asymptotic giant branch; CMD, color-magnitude diagram; GC, globular cluster; HRD, Hertzsprung-Russell Diagram; IR, infrared; ISM, interstellar medium; MC, molecular cloud; RGB, red giant branch; SGB, subgiant branch; $T_{core}$, core temperature; TO, turnoff point; WD, white dwarf.

# Chapter 2. Planet Formation and Evolution (SnR)

This chapter describes the formation and evolution of planets. In Sec. 2A, we present the current understanding of the formation and the dynamical (orbital) evolution of rocky and gas giant planets. Asteroids and comets, the "leftovers" of planet formation, are vitally important to astrobiology in terms of impacts on Earth, and are discussed in Sec. 2B. Section 2C introduces fundamental concepts in geology. The geophysical environments of the early Earth, which span a huge range of conditions, are examined in Sec. 2D. Section 2E reviews the formation and evolution of Earth's atmosphere, highlighting the rise of oxygen.

## 2A. Planet Formation and Dynamical Evolution (SnR)

Prior to the discovery of extrasolar planets, theories of planet formation needed only to explain our Solar System. In this chapter, we give an overview of the current paradigm for the formation of both rocky and gas/ice giant planets. We discuss the contents of our Solar System and the nomenclature of orbital dynamics, current theories of giant planet formation and migration, and potentially habitable, rocky planets. Known extrasolar planets are discussed in detail in another chapter (Sec. 5A).

### *Overview of the Solar System*

Our Solar System contains four rocky planets (Mercury, Venus, Earth, and Mars), two gas giant planets (Jupiter and Saturn), two ice giant planets (Uranus and Neptune), and three reservoirs of small bodies (the asteroid belt, the Kuiper Belt including Pluto, and the Oort Cloud) (see Fig. 2.1). The orbits of all the planets lie in roughly the same plane, and they all orbit the Sun in the same direction.

The rocky planets range from 0.4 to 1.5 astronomical units (AU), or the distance from Earth to the Sun, and consist largely of nickel-iron cores with silicate rock mantles and crusts. They vary in density (mainly because of the size of their cores), surface age, atmospheric composition, orbital eccentricity (the amount their orbits vary from perfect circles), rotation rate, magnetic field strength, and number of satellites (moons).

The giant planets range from 5 to 30 AU and consist mainly of hydrogen and helium in roughly the same proportions as the Sun. Saturn possesses a rocky core of about 10 Earth masses, and there is debate as to whether Jupiter also has a core. The ice giants consist largely of water ($H_2O$), ammonia ($NH_3$), methane ($CH_4$), and rock, with thin H-He atmospheres. Volatiles like water and ammonia are present in liquid state.

The small body reservoirs are thought to be the remnants of planet formation. They are populated by small, rocky/icy bodies, which are the source of impacts on Earth. The asteroid belt, Kuiper Belt, and Oort Cloud are discussed in detail in Sec. 2B.

### *Orbits and Resonances*

Kepler's laws explain that all bodies orbit the Sun in ellipses, with the Sun at one focus. An orbit is characterized by six orbital elements, three of which are important to know:

1. The semimajor axis $a$ is the mean distance between the body and the Sun.
2. The eccentricity $e$ measures the deviation of an orbit from a perfect circle. A circular orbit has $e = 0$; an elliptical orbit has $0 < e < 1$. Objects with eccentricity greater than 1 are said to be in "escape orbit" and will not remain in the system.

## A) Orbital Parameters

| *Planet/Obj* | *Parent Object* | *Semimajor Axis (AU)* | *Eccentricity* | *Inclination (deg)* | *Obliquity (deg)* |
|---|---|---|---|---|---|
| Mercury | Sun | 0.39 | 0.21 | 7 | 7 |
| Venus | Sun | 0.72 | 0.01 | 3.4 | 177.4 |
| Earth | Sun | 1 | 0.02 | 0 | 23.45 |
| Mars | Sun | 1.52 | 0.09 | 1.9 | 23.98 |
| | | | | | |
| Jupiter | Sun | 5.2 | 0.05 | 1.3 | 3.08 |
| Saturn | Sun | 9.54 | 0.06 | 2.5 | 26.73 |
| Uranus | Sun | 19.19 | 0.05 | 0.8 | 97.92 |
| Neptune | Sun | 30.06 | 0.01 | 1.8 | 28.8 |
| | | | | | |
| Asteroid Belt | Sun | 2.1-3.2 | 0.01-0.3 | 0-30 | - |
| Kuiper Belt (Pluto) | Sun | 35-50 | 0.03-0.3 | 1-30 | - |
| Oort Cloud | Sun | $2 \times 10^3$-$1 \times 10^5$ | >0.999, <1 | 0-360 | - |
| | | | | | |
| Moon | Earth | $3.8 \times 10^5$ km | 0.06 | 5.12 | - |
| Europa | Jupiter | $6.7 \times 10^5$ km | 0.01 | 0.47 | - |
| Titan | Saturn | $1.2 \times 10^5$ km | 0.03 | 0.33 | - |

## B) Planetary Properties

| *Planet/Obj* | *Mass ($M_{Earth}$)* | *Radius ($R_{Earth}$)* | *Comp. Density (g/cm$^3$)* | *Surface Temp. (K)* | *Main Atm. Components* | *Main Solid Components* |
|---|---|---|---|---|---|---|
| Mercury | 0.06 | 0.38 | 5.43 | 100 – 700 | - | Fe, $SiO_4$ |
| Venus | 0.82 | 0.95 | 5.25 | 730 | $CO_2$, $N_2$ | $SiO_4$, Fe |
| Earth | 1 | 1 | 5.52 | 260 – 310 | $N_2$, $O_2$, $H_2O$ | $SiO_4$, Fe, $H_2O$ |
| Mars | 0.11 | 0.53 | 3.93 | 150 – 310 | $CO_2$, $N_2$, A | $SiO_4$, Fe |
| | | | | | | |
| Jupiter | 317.8<br>9 | 11.19 | 1.33 | 120 | H, He | H, Fe?, $SiO_4$? |
| Saturn | 95.18 | 9.46 | 0.71 | 88 | H, He | H, Fe?, $SiO_4$? |
| Uranus | 14.54 | 4.01 | 1.24 | 59 | H, He, $CH_4$, $NH_3$ | $H_2O$, Fe, $SiO_4$ |
| Neptune | 17.13 | 3.81 | 1.67 | 48 | H, He, $CH_4$, $NH_3$ | $H_2O$, Fe, $SiO_4$ |
| | | | | | | |
| Asteroid Belt | - | <0.14 | 1.5 – 5 | - | - | Fe, Ni, $SiO_4$, $CO_3$ |
| Kuiper Belt (Pluto) | - | 0.02? – 0.36 | 1 – 2? | - | - | $H_2O$, Fe, $SiO_4$ |
| Oort Cloud | - | ? | 1 – 2? | - | - | $H_2O$, Fe, $SiO_4$ |
| | | | | | | |
| Moon | 0.012 | 0.27 | 3.34 | 120 – 380 | - | $SiO_4$, Fe |
| Europa | 0.008 | 0.25 | 3.04 | 100 – 120 | $O_2$ | $H_2O$, Fe, $SiO_4$ |
| Titan | 0.023 | 0.4 | 1.88 | 94 | $N_2$, $CH_4$ | $H_2O$, $CH_4$, Fe, $SiO_4$ |

**FIG. 2.1. Solar System bodies: (A) orbital parameters and (B) planetary properties.** AU, astronomical unit.

3. The inclination $i$ measures the angle between a body's orbit and the plane of Earth's orbit (the *ecliptic*).

Although the Sun is the primary gravitational influence in the Solar System, the dynamical importance of the planets and other massive bodies cannot be ignored. Orbital resonances (also called mean-motion resonances) occur when the orbital periods of two bodies form an integer ratio (*e.g.*, 2:1, 5:3, etc.). Resonances are found throughout the Solar System and may be stable or unstable.

### *Rocky Planet Formation and Volatile Delivery*

The planets formed from a disk of gas and dust known as the solar nebula. The lifetime of the gas component of analogous disks around other stars has been measured to be roughly 1–10 million years (Briceño *et al.*, 2001). The giant planets are, therefore, restricted to have formed within the lifetime of the gaseous disk, but the time scale for rocky planet formation is considerably longer, roughly 100 million years. Rocky planets form in several distinct stages (for a detailed review, see Chambers, 2004, and references therein):

1. Solid grains condense from the nebula and grow by sticky collisions with other grains. They settle to the denser mid-plane of the disk. Volatiles (water, methane, etc.) form as ices in the outer regions, while materials such as iron and silicates that can condense at high temperatures are more common in the inner disk.
2. Grains grow to kilometer-sized *planetesimals*, via either pairwise accretion (piece-by-piece buildup of small grains) or possibly gravitational collapse (the rapid coagulation of a large number of grains).
3. Planetesimals grow to Moon- to Mars-sized *planetary embryos* in a process called "oligarchic growth." Larger bodies have increasing mass (resulting in greater gravity) and surface area (resulting in more collisions), all of which leads to very rapid, runaway growth. At the end of this stage, roughly 100 planetary embryos exist in the inner Solar System.
4. Rocky planets form from collisions among planetary embryos. For two planetary embryos to collide, one or both must increase its eccentricity via gravitational interactions with other embryos or with the already-formed giant planets (resonances with Jupiter are particularly important). Radial mixing and giant impacts continue until only a few survivors remain in well-separated orbits.

There is a Catch-22 in the formation of habitable rocky planets with respect to water. A habitable planet resides in the habitable zone (HZ) of its star: the region in which liquid water may exist on the planet's surface given certain assumptions (Sec. 5B). The HZ for the Solar System lies between roughly 0.9 and 1.4 AU. Rocky planets acquire water by the accretion of smaller, water-rich bodies. The temperature in the HZ, however, is too hot for ice to form. How, then, does a planet acquire water and become truly habitable?

One scenario is that Earth and other habitable planets acquire water relatively late in the formation process. Water was delivered to Earth in the form of planetesimals and planetary embryos from past the snow line, where the temperature was below 170 K and water could condense into ice (see Morbidelli *et al.*, 2000). Isotopic evidence (hydrogen ratios) suggests that most of Earth's water originated in the outer asteroid region past roughly 2.5 AU, rather than in the more distant, cometary region. These water-rich, rocky bodies were subsequently transported to the inner Solar System by orbital dynamics.

The final assembly of the rocky planets is strongly affected by the gravitational influence of the giant planets. These typically remove (via collisions and ejections) about half of the total mass in the inner Solar System and contribute to the orbital eccentricities of the rocky planets. Current surveys have shown us a glimpse of the diversity of systems of giant planets in the galaxy (see Sec. 5A). We expect similar diversity in rocky planets.

### *Giant Planet Formation and Migration*

Giant planets are thought to form much more quickly than rocky planets. Two mechanisms for the formation of gas giant planets have been proposed: a bottom-up model ("core-accretion") and a top-down one ("fragmentation" or "disk instability"). Both are reviewed in Boss (2002).

The orbit of a giant planet may change significantly in its lifetime, usually moving inward toward the host star, because of gravitational torques (see *Orbital migration of giant planets*, below) within the gas disk. Additional orbital migration may occur after the dissipation of the gas disk because of interactions with remnant planetesimals. Indeed, the dynamical structure of the Kuiper Belt suggests that this may have occurred in the Solar System.

#### *Giant planet formation (bottom-up model)*

In one model, a roughly 10 Earth-mass core forms in the outer planetary region (past roughly 5 AU) via accretion of planetary embryos. The gravity of the core is sufficient to trap the local cold gas directly from the nebula, but this process is slow because the core must radiate away the heat of accretion. Gas becomes trapped slowly over a long period (often several Ma) until the gas envelope mass equals the core mass, and the process becomes runaway. The planet may subsequently accrete hundreds of Earth masses of gas in only a few thousand years and develop into a gas giant. If the buildup phase is too long, however, the gas disk will have dissipated before the onset of runaway accretion, and the planet will resemble Uranus and Neptune.

#### *Giant planet formation (top-down model)*

In the fragmentation scenario, giant planets form directly from the protoplanetary disk via a gravitational instability (Boss, 1997), which causes a patch of gas and dust to collapse under its own gravity and form a massive planet. The conditions for disk instability involve a balance of higher local density and higher local temperature. Formation times may be very short [1,000 years or less (see Mayer *et al.*, 2002)]. The model requires a relatively massive circumstellar disk and has trouble forming planets smaller than Jupiter (Boss, 2002).

#### *Orbital migration of giant planets*

The first extrasolar planet discovered around a main sequence star was a "hot Jupiter," a giant (half Jupiter-mass) planet orbiting its host star at a distance of 0.05 AU (Mayor and Queloz, 1995). The formation of giant planets is thought to be impossible so close to a star, where the temperature is very high and few solids can condense. Lin *et al.* (1996) proposed that the planet may have formed farther out in the disk and migrated inward via gravitational torques with the disk. A planet embedded in a massive gaseous disk feels torques from the gas orbiting both inside and outside its orbit. The outer gas orbits the star more slowly than the planet and, therefore, tends to slow the planet's orbit and cause its orbit to spiral inward. Conversely, the inner gas orbits more quickly and causes the planet's orbit to spiral outward. It has been shown that the outer torque is usually (but not always) the stronger of the two, which means that this effect causes an inward migration of the planet.

## 2B. Asteroids, Comets, and Impacts on Earth (VC)

Asteroids and comets are leftovers from the formation of the Solar System and, as such, can provide clues toward understanding the formation of planetary systems. They may also have been a source of organic molecules on the early Earth. This section begins with an overview of asteroids, comets, and meteorites and ends with a discussion of their specific importance for astrobiology.

### *Asteroids*

The asteroid belt lies between the orbits of Mars and Jupiter, with most asteroids residing in the main asteroid belt, between 2.1 and 3.3 AU from the Sun. Asteroids can be thought of as irregular-shaped, planetesimal-sized (~1 km) bodies composed of accreted material not incorporated into the planets.

Asteroid classes are based on composition, a property strongly correlated with location in the Solar System. The most common asteroids can be separated into four classes: M, S, C, and P. Inner main belt asteroids (between 2.1 and 2.8 AU) are typically metallic (M type)—consistent with Ni-Fe composition and moderate albedo (reflectivity)—or silicaceous (S type)—consistent with olivine, pyroxene, and metals composition and moderate albedo. Outer main belt asteroids (between 2.5 and 3.3 AU) are mostly carbonaceous (C type)—hydrated minerals with up to 20% water by mass—and thought to be related to carbonaceous chondrite meteorites. P-type asteroids—apparently lacking hydrated minerals—are also found in the outer belt. Both C and P types have low albedo.

In addition to the main belt asteroids, three other asteroid populations are known: near-Earth asteroids, Trojans, and Centaurs. Near-Earth asteroids approach Earth's orbit and pose the primary impact threat to Earth. Two groups of asteroids called Trojans orbit the Sun at the same distance as Jupiter, roughly 60° ahead of and behind the planet. Trojans commonly include asteroids of types P and D—the latter being similar to P, but redder. Centaurs are outer Solar System asteroids with elliptical orbits between Saturn and Uranus. Based on composition and comparisons to comets, it is thought that Centaurs may have originated in the Kuiper Belt and traveled toward the inner Solar System.

### *Comets*

Like asteroids, comets are leftover planetesimals from the primordial solar nebula. Comets, however, occupy a different region in the Solar System and are physically and compositionally diverse enough to be classified as a distinct group. Comets, upon entering the inner Solar System, display a coma and tail. They possess a core or nucleus composed of volatiles (materials easily converted to gas), metals, and dust; the release of gas and dust caused by heating as the comet approaches the Sun forms a coma, or mini-atmosphere, around the nucleus, as well as a tail, material driven off the comet by solar radiation or magnetic fields. Volatile components primarily include ices (water, carbon dioxide, ammonia, etc.), while nonvolatile components include micron-size aggregates of silicates, carbonates, and other minerals. These materials suggest that comets may be the most primitive bodies in the Solar System.

There are two main reservoirs of comets: the Oort Cloud and the Kuiper Belt. Two cometary groups are associated with these regions: ecliptic or short-period comets and nearly isotropic or long-period comets. Short-period comets have orbital periods of less than 200 years and relatively small orbital inclinations (they lie mainly in the plane of the planets, see Sec. 2A) and are associated with the Kuiper Belt. Long-period comets have orbital periods greater than 200 years, can come from any direction in the sky (they have random inclinations), and are sourced from the Oort Cloud.

The Kuiper Belt (including the so-called "scattered disk") extends for several hundred AU be-yond Neptune's orbit and may contain $10^9$–$10^{10}$ comets [called Kuiper Belt Objects (KBOs) or trans-neptunian objects]. Most KBOs lie inside 50 AU and have inclinations between 0 and 30°. Until recently, Pluto, at 2,300 km in diameter, was the largest known KBO. Its orbit crosses that of Neptune but avoids close encounters because it is in a 2:3 orbital resonance with Neptune (see Sec. 2A for a definition of resonance). Six other objects larger than 1,000 km have been observed. Sizing of the recently discovered 2003 UB313 has proved problematic, but estimates have it at least as large as Pluto.

The Oort Cloud occupies a spherical domain roughly $10^3$–$10^5$ AU from the Sun and may contain $10^{12}$ comets that, in total, equal ~5 Earth masses. Small icy bodies formed in the outer Solar System, comets were gravitationally scattered by the giant planets (but see Sec. 7B on comets for details of the Stardust mission). The orbits of these comets, only loosely bound to the Sun, were perturbed by encounters with nearby stars, molecular clouds, and the galactic tidal field. Their orbital orientations were thus randomized into the Oort "Cloud."

### Meteorites

A meteorite is defined as an extraterrean (extraterrestrial) body that strikes Earth's surface. It is generally accepted that the vast majority of meteorites are fragments of asteroids. The compositional and physical characteristics of meteorites provide clues to their origin as well as the composition of other objects in the Solar System.

The most general scheme classifies meteorites into three categories: stones, irons, and stony-irons. "Stones" are composed primarily of silicate minerals. Their composition suggests that they have undergone partial or complete melting within a large differentiated body such as a planet or a moon. "Irons" are composed of iron-nickel compounds and probably originate from M-type asteroids. "Stony-irons" have approximately equal amounts of both silicates and metal, and may originate in S-type asteroids (or the similar A-type with high olivine content).

Stones are subdivided into chondrites and achondrites. Chondrites contain "chondrules," tiny silicate or metal droplets formed during the period of planetary accretion; achondrites do not contain silicate or metal droplets. Chondrites can be further subdivided into three classes: ordinary, enstatite, and carbonaceous. Enstatite chondrites contain enstatite—a magnesium silicate mineral—and constitute only about 1% of all meteorites. Carbonaceous chondrites contain unusually high amounts of hydrocarbons and are, therefore, the most interesting astrobiologically. Isotopic fractionation indicates that the hydrocarbons were formed in the ISM and subsequently underwent aqueous alteration in the solar nebula and planetesimals. Roughly 80% of meteorites can be divided into high- and low-iron ordinary chondrites (in terms of falls). Because chondrites have not been subjected to melting or igneous differentiation, they preserve information about early Solar System processes such as condensation, evaporation, fractionation, and mixing.

### Earth Impacts and Delivery of Organics

Geologic and astronomic evidence suggests that meteorite impacts were significant to the evolution of life on Earth. Early impacts may have delivered the large quantities of organic molecules necessary for life. Large impacts have caused major extinction events (see Sec. 4D) and may have delayed the origin of life.

#### Earth impacts

Earth formed via accretion, the collisional buildup of many small bodies into a larger one (see Sec. 2A for details), which can include mas-sive events such as the Mars-sized impactor that is thought to have formed the Moon. Evidence remains for post-accretionary impacts on the Moon, primarily in the form of geochemical data from moon rocks brought back by the Apollo and Russian Luna missions. The Late Heavy Bombardment (LHB) period between 4.1 and 3.8 billions of years ago (Ga) was the most intense bombardment Earth suffered after its formation. Evidence for this comes from lunar samples and meteorites. Ages obtained from lunar impact melts and highland samples all cluster around the same age, ~3.9 Ga, which suggests that the rate of impacts was much higher at this time. What caused the LHB is uncertain, but a recent theory ties the LHB to perturbation of KBOs with the outward migration of Neptune (Gomes *et al.*, 2005).

The impact rate on the Earth has been, to some extent, in a steady state since the end of the LHB: roughly 20,000 tons of extraterrean material impact Earth each year. Impact frequency is a function of impactor size: extinction-level collisions with large bodies are rare, while small body collisions occur constantly. An impactor larger than a certain size (thought to be roughly 5 km, depending on composition, impact speed, and impact location) would cause worldwide devastation. For this reason, there is an endeavor underway (the "Spaceguard Goal") to find 90% of all near-Earth objects larger than 1 km. Extinction-level impacts with >1-km objects are thought to occur roughly once every 50–100 million years. The most recent extinction-level impact was the Cretaceous-Tertiary (K-T) impact 65 millions of years ago (Ma). The K-T impact involved a ~10-km object that, when it struck the Yucatan Peninsula in Mexico, released ~$10^8$ megatons of energy and created a crater (Chicxulub) 180–300 km in size. This impact resulted in a global catastrophe that led to the extinction of the dinosaurs as well as a host of environmental problems (wildfires, changes in atmospheric and ocean chemistry, short-term climatic changes as a result of large quantities of dust ejected into the atmosphere, generation of gigantic tsunamis). Other mass extinction events that may have been caused by large impacts occurred around ~365 Ma (Frasnian/Famennian Boundary) and ~250 Ma (Permian-Triassic boundary); both of these events resulted in the extinction of >70% of ocean and land species. There seems to be a correlation in occurrence among geologic events, mass extinctions, and impact events in that they all show a regular periodicity of ~30 million years. Reasons for this remain unclear but may involve the Sun's orbital period about the galactic center (Steel, 1997).

### *Comets and meteorites as sources of organic compounds*

A variety of organic species have been identified in meteorites and comets. The origin of organic materials found in meteorites is attributed to processes that involve interstellar, nebular, and planetary abiotic syntheses. They are not thought to have links to biology, in spite of the fact that some species display identical properties to terrean biomolecules (Pizzarello, 2004). Similar organics are found in comets. Perhaps most interesting are the volatile prebiotic molecules detected in cometary comas, including water ($H_2O$), carbon monoxide/dioxide ($CO/CO_2$), formaldehyde ($H_2CO$), nitrogen ($N_2$), hydrogen cyanide (HCN), hydrogen sulfide ($H_2S$), and methane ($CH_4$). The formation of organic volatiles is thought to require long periods (~$10^9$ years) and involve both radiation-induced processes and hot atom chemistry (Kissel *et al.*, 1997). Models estimate that the early Earth accreted $10^6$–$10^7$ kg/year of organics from comets (Chyba *et al.*, 1990) and ~$8 \times 10^4$ kg/year from carbonaceous chondrites (Pizzarello, 2004). These estimates do not take into account inputs from other sources of extraterrean organics such as interplanetary dust particles, interstellar dust, and other meteor types.

## 2C. Introduction to Geology (RS, MC)

### *Physical Geology and Earth*

Geology is the study of Earth, the processes that shape it, and the composition, structure, and physical properties that characterize it. When the Solar System formed, Earth, along with three other rocky planets, accreted solid rock, which later differentiated into layers. Earth is composed of three primary layers: core, mantle, and crust. The core consists of Fe and Ni, with a solid center and a liquid outer shell. The mantle is composed primarily of Fe, Mg, Si, and O.

Heat is transferred from the core to the crust through a mantle convection system. Hotter rock moves toward the outer edges of Earth, where it cools and sinks back down toward the core. The crust is brittle and broken up into plates that move over the surface of Earth, driven by mantle convection. Two types of crust exist—oceanic and continental—each of which exhibits different properties and behave differently at plate boundaries.

Three different types of interactions can occur: the plates can move in divergent directions (spreading center), they can converge (subduction zones or orogenic belts), or they can move parallel to one another (strike-slip). At spreading centers, hot mantle wells up and forms new crust at midocean ridges. When two oceanic plates or an oceanic plate and a continental plate collide, the denser oceanic plate is subducted (pulled under), which results in a trench (*e.g.*, the Marianas Trench). When two continental plates collide, mountains form (*e.g.*, the Himalayas). If two plates are sliding past one another, faults form and cause earthquakes (*e.g.*, the San Andreas Fault system).

### *Rocks and Minerals*

Minerals and rocks are different. A mineral occurs naturally. It is inorganic, homogeneous, and solid with a definite chemical composition and an ordered atomic arrangement. Rocks are composed of minerals and/or organic material. Three main rock types exist on Earth: sedimentary, igneous, and metamorphic. Sedimentary rocks form either by chemical precipitation or the gradual settling of materials (fragments of older rocks and marine shells) over time. The resulting sediments become lithified (made into solid rock instead of particles) through the process of compaction as new material forms on top and gravity presses the particles together. Organisms can become fossilized when the sediment turns to stone. Common sedimentary rocks include limestone, shale, conglomerate, and sandstone.

Igneous rocks form when molten (liquid) rock cools to the point of solidification. This can happen above ground (from lava flows, as in the formation of basalt) or below (in magma chambers, as in the formation of granite). Metamorphic rocks begin as sedimentary or igneous rocks, but their properties change as a result of high temperatures and pressures deep below the surface of Earth. Common metamorphic rocks include quartzite, marble, and gneiss.

### *Redox Geochemistry*

There are two broad classes of chemical reactions—reactions that transfer protons between reactants (defined as "acid-base" reactions) and reactions that transfer electrons between reactants (defined as "redox" reactions). In a redox reaction, the reactant that gains an electron is "reduced," while the reactant that loses an electron has been "oxidized." Redox reactions are coupled by definition; the only way for one molecule to be oxidized is for another molecule to be reduced.

Another important definition is the "oxidation state" of an element. The oxidation states possi-ble for an element are determined by fundamental physics, but the oxidation state in which that element exists at a given time is determined by environment. For example, quantum mechanics dictates that iron (Fe) can exist as $Fe^0$, $Fe^{2+}$, or $Fe^{3+}$, which means that iron will readily donate none, two, or three of its electrons, depending on what it encounters in a chemical reaction. For instance, iron rusts [$4Fe(OH)_2 + O_2 \rightarrow 2Fe_2O_3 + 4H_2O$] when it is oxidized from the $Fe^{2+}$ oxidation state to the $Fe^{3+}$ oxidation state. A major confusion is that the terms "oxidized/oxidizing/oxidant" need not have anything to do with the element oxygen; $FeCl_2 + CeCl_4 \rightarrow FeCl_3 + CeCl_3$ is a redox reaction in which $FeCl_2$ (the "reducing agent" or "reductant") is oxidized by $CeCl_4$ (the "oxidizing agent" or "oxidant"). In exchange for oxidizing the iron from $Fe^{2+}$ to $Fe^{3+}$, the cerium is reduced in oxidation state from $Ce^{4+}$ to $Ce^{3+}$.

Redox reactions are exploited by biology (Sec. 6B) but are also a very important part of interpreting the paleo-environment. Examining the oxidation state of elements in the geologic record helps us infer the conditions under which a rock formed. For example, lithified soils ("paleosols") older than 2.4 Ga are generally free of iron, while lithified soils younger than 2.5 Ga are rich in iron, but only iron that is in the $Fe^{3+}$ oxidation state. In the lab, we notice that $Fe^{2+}$ is soluble in water, but $Fe^{3+}$ is not. The standard interpretation of these two facts is that ancient rainwater did not contain dissolved oxygen and, hence, would have leeched soils of soluble $Fe^{2+}$. Oxygenated rainwater (*i.e.*, rainwater containing dissolved $O_2$) would oxidize iron to the insoluble $Fe^{3+}$ oxidation state and trap the iron in the soil. This geochemical interpretation is one of the arguments for a dramatic change in the overall oxidation state of the atmosphere approximately 2.4 Ga (Sec. 2D). Thus, the geochemistry of Fe and other (especially rare earth) elements in well-preserved rocks can constrain the "redox state" of entire reservoirs such as the atmosphere and ocean. In this context, environments are said to be "reducing" or "oxidizing." A reducing environment has an excess of available reductants (usually hydrogen), while an oxidizing environment has an excess of available oxidants (generally oxygen).

### *Geologic Time and the Geologic Record*

Geologic time can be measured on a relative or absolute scale. The relative scale (Fig. 2.2) is derived from the layering of rock sequences and the evolution of life. Index fossils—from widespread life forms that existed for limited periods of time—can be used as a guide to the relative ages of the rocks in which they are preserved. The eons, eras, and periods of the scale reflect specific regimes in the history of life. Ammonites, for example, were common during the Mesozoic Era (251–65 Ma) but went extinct during the K-T extinction event (65 Ma) at the same time as the dinosaurs.

The absolute or radiometric time scale is based on the natural radioactive properties of chemical elements found in some rocks on Earth. Radiometric dating utilizes the instability of certain atoms and isotopes that decay to more stable elements over time. The time it takes for half the amount of a radioactive element to degrade to the more stable form is called its "half-life." Using the half-life, it is possible to calculate the age of a rock based on the proportions of radioactive elements it contains. The geologic time scale of Earth takes into account radiometric age dating of appropriate rocks and yields a chronology for the entire 4.6 Ga of Earth history.

## 2D. Early Earth Environments (SG)

There have been four eons in Earth's history (Fig. 2.2): the Hadean, which spans the time before any of the rocks we see today were formed; the Archean, marked by deposits laid down in the absence of free $O_2$ in the atmosphere; the Proterozoic, during which aerobic and multicellular life developed; and the Phanerozoic, the most recent eon, in which dinosaurs, mammals, and eventually humans inhabited the planet. Here we focus on the Hadean and Archean, which cover the first 2 billion years of Earth's 4.6 Ga history.

### *Formation of the Continents, Atmosphere, and Oceans*

According to radiometric age dating of meteorites and the Moon, Earth reached a large frac-tion of its final size by about 4.5 Ga. The moon formed by ~4.44 Ga via the impact of a Mars-sized object with the proto-Earth. Although the bombardment of Earth continued, the vast majority of Earth's mass had been accreted by this time.

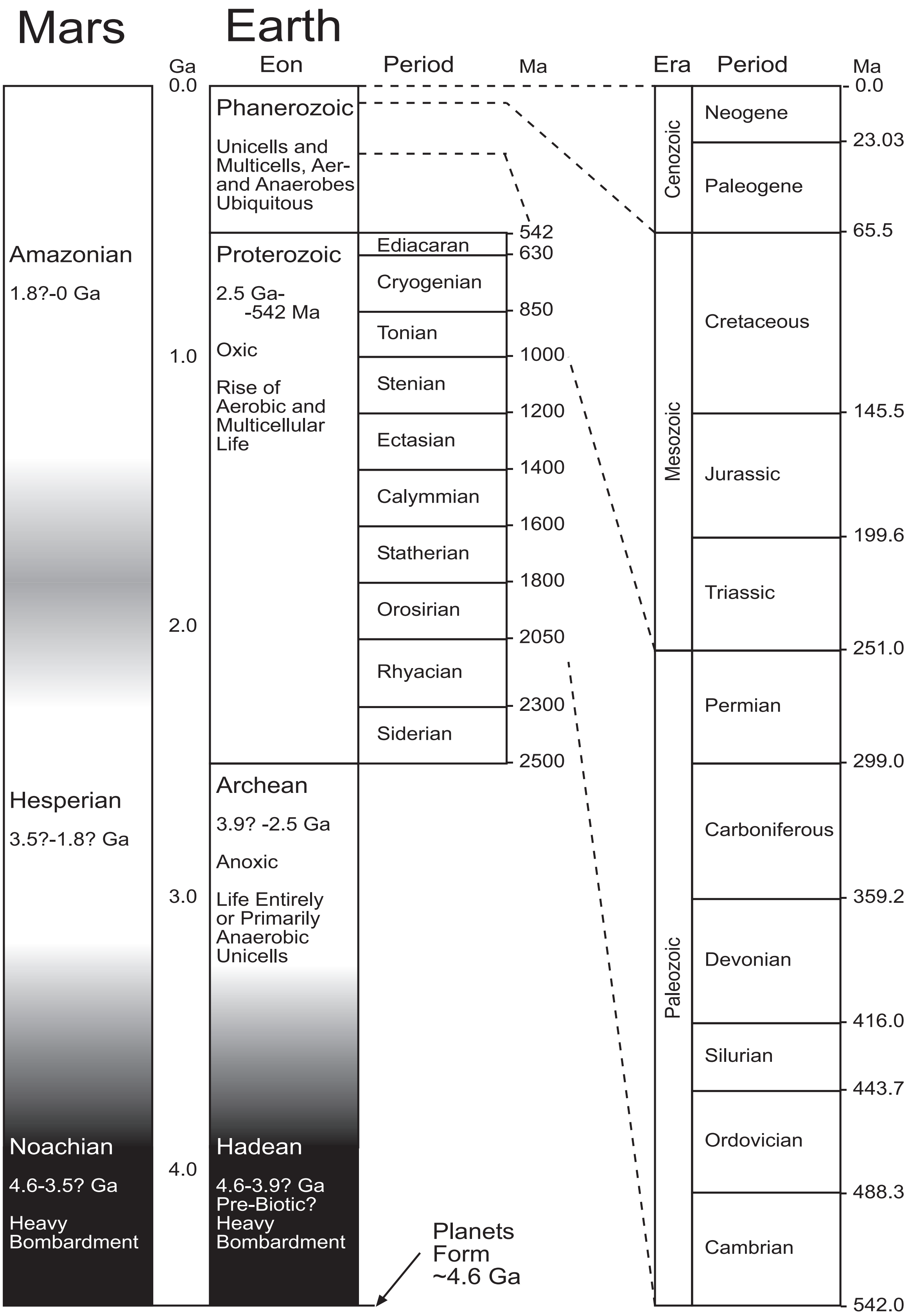

FIG. 2.2. Geologic time scale (Earth and Mars).

While the formation of the Earth-Moon system is fairly well understood, the formation of Earth's crust, oceans, and atmosphere is not nearly as well constrained. Almost all of the materials that composed the original crust have since been destroyed through subduction or giant impacts; however, there is reason to believe that the oceans and some crust were in place by 4.44 Ga. Debates center upon how quickly the crust built up and what type of plate tectonics operated during the early stages of crustal evolution. These unknowns would have had significant effects on the global cycling of many elements important to life (*i.e.*, C). If this recycling of elements is important for habitability, as it seems it was on the early Earth, then we need to understand how Earth's system of plate tectonics came into being. The importance of this can be seen by examining our neighbor planets, Mars and Venus.

Current thought indicates that Mars is geologically inactive. Consequently, recycling of bio-essential elements such as C, N, O, and P is very limited. Venus, on the other hand, has extensive geologic activity, but it is too rapid and violent to be conducive to life as we know it. Venus seems to build up interior pressure until it is all released at once in a giant magma flow that completely covers the planet and forms a new surface. Earth has a steady, continuous release of mantle fluids and gases. The difference results from an abundance of water in Earth's crust, which reduces viscosity and allows for the efficient transfer of heat from the interior by way of plate tectonics. One of the main problems that the next generation of astrobiologists will face will be to bridge the gap between models of planetary accretion and formation and the initiation of geologic activity (or lack thereof) on rocky planets.

The key problem to understanding how Earth formed a hydrosphere and an atmosphere has to do with the way in which volatiles were initially delivered to Earth. As described in Sec. 2A, Earth's "building blocks" were likely completely dry simply because temperatures were too high for water to condense or form hydrated minerals (but see Drake and Righter, 2002, for a different opinion). Thus, Earth's volatiles had to have come from material delivered from colder parts of the Solar System (further away from the Sun). There are two possible sources for this volatile-rich material: the asteroid and cometary regions. Carbonaceous chondrite meteorites sourced from C-type asteroids in the outer asteroid belt (2.5–3.3 AU) have D/H (*i.e.*, deuterium/hydrogen or $^2H/^1H$) ratios similar to Earth's oceans. Models show that Earth could have received its current volatile inventory via incorporation of planetesimals formed in the outer Solar System and gravitationally perturbed inward (Morbidelli *et al.*, 2000; Raymond *et al.*, 2004). Comets have high concentrations of volatiles, which can explain the isotopic composition of the noble gases on the rocky planets but not the D/H ratios.

A thick, steam-laden atmosphere probably resulted from the high flux of volatiles to Earth's surface via impacts and the increased heat associated with those impacts. There is evidence for the rapid escape of H atoms during a time when the atmosphere was extremely $H_2O$ and $H_2$ rich. Loss of volatiles from the atmosphere was likely larger during this time period than at any other time in Earth's history. This is due to the escape of H and "impact weathering," the loss associated with ejection of matter from the planet during large impacts.

As accretion slowed, the surface of Earth cooled. This would have caused atmospheric steam to rain out and form oceans. It should be noted that these oceans may have been vaporized by large impacts but would have continued to form and reform until the impacts died down. Radiometric age dating of cratered surfaces on the Moon suggests that Earth would have continued to receive a significant number of impacts by >100-km-sized objects until at least 3.85 Ga.

### *Earth's Oldest Sedimentary Rocks? The Oldest Evidence of Life?*

Sedimentary rocks can tell us many things about the surface chemistry and biology of Earth at the time of deposition. The earliest known sedimentary rocks have been dated to 3.85 Ga. The lack of sediments prior to this time could be ascribed to a global resurfacing event (such as on Venus), a lack of continental crust prior to that time, or the elimination of earlier sediments through processes such as subduction and metamorphosis.

Earth's oldest putative sedimentary rocks can be found in the Akilia Island Banded Iron Formation (BIF) from Greenland, dated to at least 3.85 Ga (McGregor and Mason, 1977). BIFs are a type of sedimentary deposit that contains alternating Fe-rich and Si-rich layers and were mostly formed prior to 1.8 Ga (the significance of their temporal distribution is discussed below).

The Akilia Island BIF is best known for harboring the oldest reported indicators of biological activity on Earth (Mojzsis *et al.*, 1996): isotopically light carbon consistent with the isotopic fractionation found in photosynthetic organisms. It should be noted, however, that the age and biological fractionation attributed to the Akilia Island formation have been questioned (Fedo and Whitehouse, 2002; van Zuilen *et al.*, 2002).

### *Some Like It Hot . . .*

Three and a half billion years ago, Earth was likely extremely hot. Recent measurements (Knauth and Lowe, 2003) of oxygen isotopes in cherts (very hard sedimentary rocks) are consistent with a surface temperature of ~55–85°C at 3.5–3.2 Ga. This is the temperature range within which heat-loving organisms, "thermophiles," proliferate. This early hot environment may explain why most organisms near the base of the tree of life (see Sec. 6A) are thermophilic; they were the only organisms capable of thriving.

This hot early Earth hypothesis is quite controversial. Because the sun was fainter in the Archean (see Sec. 2E), a massive $CO_2/CH_4$ greenhouse forcing would have been required to keep the Earth above 50°C (Pavlov *et al.*, 2000). The temperature evolution of the Earth proposed by Knauth and Lowe (2003) is also inconsistent with the glacial record described in the next section. Alternate explanations for the evolution of oxygen isotopes have been proposed (Land, 1995) but have their own problems. Future work on interpreting this signal is a crucial area of research in astrobiology, as temperature has a profound effect on a variety of other systems at the Earth's surface.

### *. . . Others Like It Cold*

Twice in its history, Earth was likely extremely cold. There is evidence for tropical glaciers at ~2.4 Ga and between 0.7 and 0.5 Ga (Evans *et al.*, 1997; Hoffman *et al.*, 1998). [In the last ice age, glaciers only advanced as far south as New York City (~40°N latitude)]. The global extent of the glaciation during these time periods has led some to call them "Snowball Earth" episodes, since Earth would have resembled a gigantic snowball (Hoffman *et al.*, 1998).

We have a fairly good understanding of how a Snowball Earth event would have been possible.

If $CO_2$ levels were decreased enough, glaciers would have started to advance from the poles. The growth of these glaciers would have led to further cooling; glaciers have a high albedo (fractional reflectance), and their growth would have reflected more of the Sun's energy back into space. Thus, Earth would have cooled even further in a self-reinforcing trend until glaciers covered the whole planet (Harland, 1965; Kirschvink, 1992).

On modern Earth, $CO_2$ outgassing by volcanoes is roughly balanced by the removal of $CO_2$ through weathering of silicate minerals on Earth's continents (Sec. 2E). During a Snowball Earth event, however, the oceans would have been covered with ice, very little water would have evaporated, and rainfall would have been greatly decreased. Under these conditions, weathering would have diminished severely so that removal of $CO_2$ from the atmosphere would have been almost nonexistent, and $CO_2$ levels would have built up to extremely high levels. Since $CO_2$ is a greenhouse gas, this build-up would eventually have been enough to overcome the cooling caused by the reflectivity of the glaciers, and the glaciers would have melted and left behind a hot, $CO_2$-rich surface environment (Kasting and Caldeira, 1992). (See Sec. 4D for impacts of Snowball Earth on evolution.)

### *Firsts*

One area of ongoing research and almost continuous debate revolves around the earliest known examples of a variety of biologically important phenomena. For example, it is important to know when the continents first developed—the oldest known rocks are ~4.0 billion years old (Bowring and Williams, 1999)—and when liquid water first formed on Earth—the oldest known minerals, zircons dated at ~4.4 Ga (Mojzsis *et al.*, 2001; Wilde *et al.*, 2001), have an isotopic composition consistent with the presence of liquid water at the time they formed. The earliest reported sedimentary deposits were discussed above; these rocks are at least 3.85 Ga (McGregor and Mason, 1977). It is also extremely important to know by what time life had evolved; the earliest chemical signs of life are in rocks dated at ~3.8 Ga (Mojzsis *et al.*, 1996; Rosing, 1999), and most accept evidence that life was around by ~3.5–3.2 Ga (see Secs. 4D and E). Perhaps the greatest debate has been waged over the date of the earliest fossils. The claim that "cyanobacteria-like" fossils have been discovered in Australian rocks that are 3.5 billion years old (Schopf, 1993) is heavily disputed (Brasier *et al.*, 2002) (see Sec. 4E). The last "firsts" in the Archean primarily involve the final deposits before the rise of $O_2$ (and some would assert the rise of oxygen-producing photosynthesis) (see Sec. 2E).

## 2E. Global Climate Evolution (MC)

In this section, we examine how the atmospheres of rocky planets change in response to increases in stellar luminosity and intrinsic changes in planetary biogeochemical cycles. Astrobiologically, we choose to focus on atmospheres, given that the bulk composition of the atmospheres of extraterrean rocky planets may be observable within the upcoming decades. We discuss Earth's possible responses to the "faint young Sun," the apparent stability of Earth's climate, Earth's changing oxygen levels, how atmospheric constituents can escape from rocky planets, and the impact of these topics on our general understanding of planets.

### Earth's Climate over Time

#### *Planetary temperature = effective temperature ($T_{eff}$) + greenhouse warming*

The $T_{eff}$ of a planet at a certain distance from a star is calculated by equating the stellar energy flux received at the surface (minus planetary albedo, energy flux directly reflected back into space) with the thermal radiation of the planet itself (see also Sec. 5B). For a planet at 1 AU from a 4.5-billion year old G2 star (*i.e.*, Earth), whose average albedo is 0.3, the effective temperature is 255 K. Earth's global average surface temperature is currently 288 K, which implies an addition 33 K of warming by greenhouse gases. A "greenhouse" gas is transparent to the dominant stellar wavelengths [near-ultraviolet (UV) and visible for stars of astrobiological interest], but absorbs strongly in the infrared (IR). The UV/visible light passes through the atmosphere and heats the surface of the planet, which then radiates in the thermal IR. Greenhouse gasses (*e.g.*, $CO_2$, $CH_4$, $NH_3$, $H_2O$) absorb these outgoing IR photons and trap heat in the atmosphere. In Earth's modern atmosphere, most greenhouse warming is caused by $H_2O$ and $CO_2$ (approximately two-thirds and one-third of the warming, respectively) with $CH_4$ responsible for approximately 1–2°.

#### *The faint young Sun problem*

As described in Chapter 1, main sequence stars convert hydrogen to helium in their cores. Over long periods of time, the mean molecular weight of a star's core increases, which causes it to contract and increase energy output. Over the ~4.6 billion years that the Sun has been on the main sequence, stellar models predict that its radiative energy per second ("luminosity" in astronomers' jargon) has increased by 30%. Consideration of the effects of stellar evolution on the $T_{eff}$ of Earth draws us to the inescapable conclusion that the composition of Earth's atmosphere must change with time if temperature is to remain constant. When solar luminosity is dropped by 30%, Earth's $T_{eff}$ becomes 233 K. Consequently, the planetary temperature (233 + 33 = 266 K), would be below the freezing point of water (0°C = 273 K) given that our atmosphere had not changed with time. As discussed in Sec. 2D, there is evidence of aqueous process on the surface of the early Earth, with the first evidence for widespread glaciations only appearing in the geologic record around 2.3 Ga. Hence, greenhouse gases must have played a stronger role on the early Earth.

Carbon dioxide ($CO_2$) is a dominant constituent of volcanic outgassing, and therefore a thick atmosphere of $CO_2$ was, for many years, the accepted hypothesis for the early Earth greenhouse warmer. Geologic evidence, however, limits paleo-$CO_2$, and recent research has focused on $CH_4$ as a plausible greenhouse gas present in the atmosphere of early Earth. The multiple lines of evidence that support methane include: (1) methanogenesis is an ancient, methane-producing metabolism whose energy source ($H_2$, $CO_2$, acetate, formate) was widely available during the Archean; (2) the redox state of (and hence gases emanating from) ancient volcanoes and crust was more reduced (Holland, 2002); (3) the lack of oxygen in the Archean atmosphere (see below) allowed methane concentrations to rise to higher levels (Pavlov *et al.*, 2000); and (4) collapse of a $CH_4$ greenhouse could potentially explain "Snowball Earth" events. (The above arguments implicitly assume that Earth's albedo has remained 0.3 for the course of Earth's history—an assumption in almost all climate models, but an unproven hypothesis.) A general theory of cloud cover on rocky planets remains elusive.

### Atmospheric Processes: Feedback and Atmospheric Loss

#### *Climate feedbacks: what stabilizes planetary climates?*

Other than brief intervals of global glaciation, Earth's climate appears to have remained clement over most of Earth's history. We discuss this in terms of "feedback loops" on the climate: a positive feedback occurs when change in a variable reinforces itself and the process accelerates, while a negative feedback occurs when an increase in a variable leads to an eventual decrease in that same variable (or vice versa). Strong negative feedbacks on climate must balance any positive feedbacks in order for climate to remain steady.

The dominant positive feedback process over long periods is the ice-albedo feedback. At low global temperatures, the extent of glaciation is enhanced. Ice is very reflective (albedo of ~0.6), so increased ice coverage increases the planetary albedo. The amount of absorbed energy is, therefore, reduced, cooling the planet and resulting in the formation of more ice.

The ice-albedo feedback is usually offset by the carbonate-silicate cycle, the dominant long-term negative feedback on $CO_2$ levels (and hence temperature via the greenhouse effect) that keeps the planetary climate stable. $CO_2$ dissolves in raindrops to form carbonic acid ($H_2CO_3$), which breaks down silicate rocks on the continents and releases $Ca^{2+}$ ions, bicarbonate ($HCO_3^-$), and silica ($SiO_2$) into the oceans. These products are used to create carbonate rocks, both organically and inorganically. The burial of carbonate rocks removes $CO_2$ from the atmosphere/ocean system. The temperature dependence of weathering makes the process a negative feedback with an approximately 1 million year time scale. Higher temperatures cause more weathering, which leads to more carbonate production and, hence, larger $CO_2$ sink; lower temperatures inhibit weathering and, hence, limit carbonate production, which allows $CO_2$ to build up in the atmosphere. Over geologic time, the $CO_2$ is re-introduced to the ocean/atmosphere system via volcanism and, thus, closes the carbonate-silicate cycle. Under most circumstances, the carbonate-silicate cycle acts to keep climate stable. In brief intervals of geologic time referred to as "snowball earth" episodes (see Sec. 2D and below), the ice-albedo positive feedback may have overcome the negative feedback of the carbonate-silicate cycle.

### *Planetary atmospheres: loss processes*

Atmospheres of rocky planets are so-called "secondary atmospheres." All planets obtained "primary atmospheres" during their formation by the direct gravitational accretion of gas (mostly H and He) from the solar nebula. Massive planets like Jupiter retain their primary atmospheres, but rocky planets do not. Massive impacts, such as the one that created the moon, can hydrodynamically blow the atmosphere off of rocky planets into interplanetary space. New atmospheres develop and are characterized by post-formation processes such as planetary outgassing, biospheric activity, atmospheric photochemistry, and subsequent thermal escape. By number and mass, H is the most important element that escapes from Earth and Venus—less massive rocky planets like Mars may lose heavier elements as well.

## *History of Oxygen in Earth's Atmosphere*

### *Evidence for changing $O_2$ levels in Earth's atmosphere*

Beyond gas bubbles trapped in ice cores, we have no direct evidence of paleo-atmospheric composition. Of the main components of the atmosphere, N and Ar do not react with surface rocks and, hence, leave no trace of their total levels in the bulk rock record. $O_2$ is an extremely reactive molecule, degrades most organic compounds, is highly flammable, and generally leaves evidence of its existence. To most observers, the geologic record preserves clear evidence of progressive oxygenation of the atmosphere with a significant rise in oxygen levels at ~2.4 Ga. A history of the pertinent paleo-constraints on Earth's oxygen levels and potential evolutionary curve is presented in Fig. 2.3 (see Ohmoto, 1996, for an alternative viewpoint).

### *$O_2$ in the modern atmosphere*

Today, free $O_2$ exists in the atmosphere because the net sources are larger than the net sinks. The oxygen we breathe derives from oxygenic photosynthesis. If photosynthetic activity were to stop, free oxygen would be scrubbed from the atmosphere in approximately 2 million years. Another way of saying this is that large quantities of free oxygen in a planetary atmosphere cannot exist in long-term thermodynamic equilibrium. For this reason, the detection of $O_2$ (or its by-product, ozone, $O_3$) is considered a remotely observable biomarker (by future missions like Terrestrial Planet Finder/Darwin; see Sec. 5A) as no large-scale abiotic sources of free oxygen are known. The gross source of free oxygen is oxygenic photosynthesis, with an average "reaction" of $CO_2 + H_2O \rightarrow CH_2O + O_2$. This reaction reverses almost completely by the complementary processes of respiration and decay. The only reason that a net source of free oxygen exists is that some organic carbon escapes re-oxidation by being buried in sediments on the sea floor. This net source is balanced by a combination of oxidation of reduced material on the continents (usually referred to in bulk as "weathering"), and by reaction with reduced gases (such as $H_2$, $H_2S$, $CH_4$, CO) that emerge from the planet via volcanism and metamorphism.

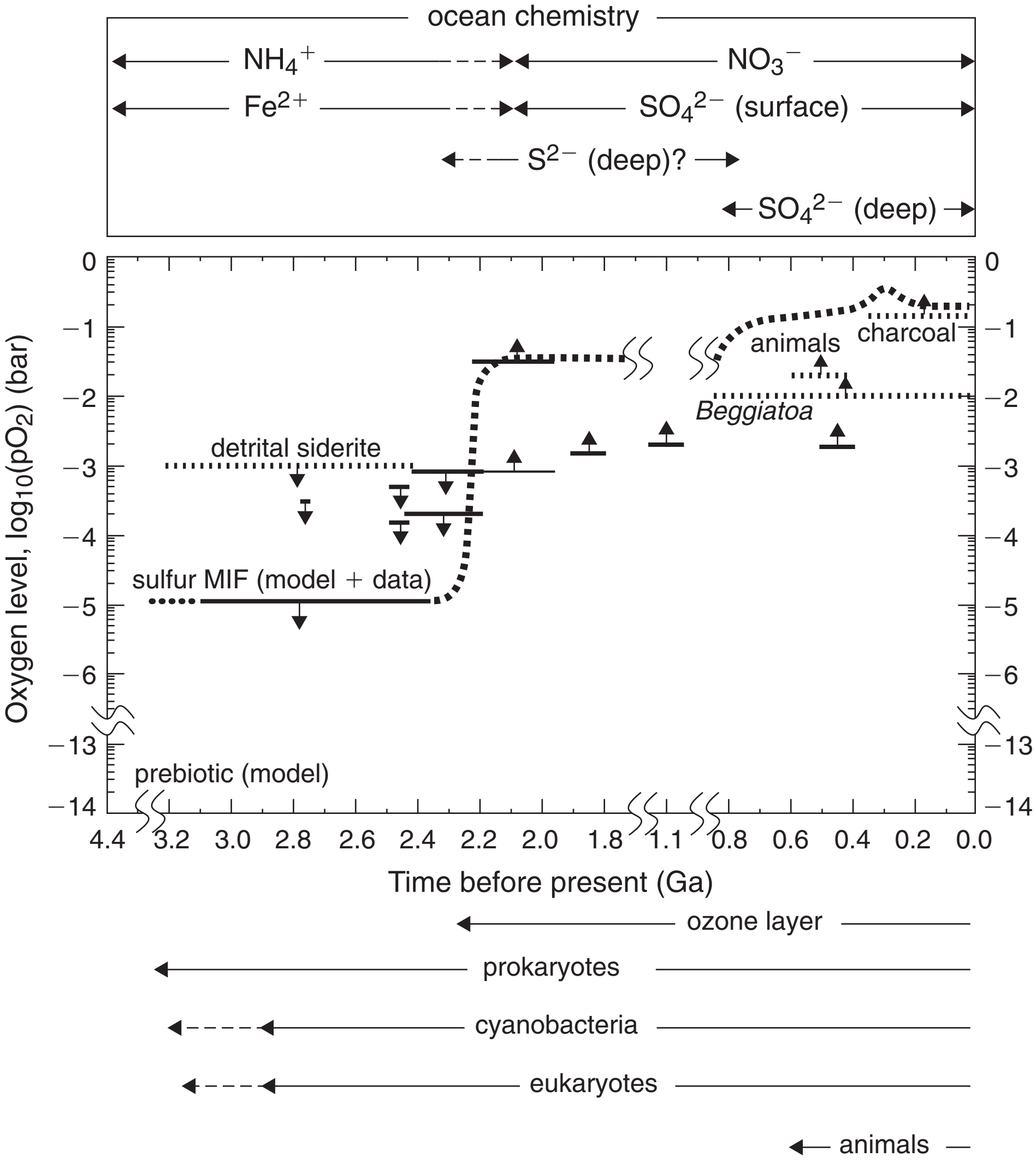

**FIG. 2.3. The history of atmospheric $O_2$.** The thick dashed line shows a possible evolutionary path of atmospheric $O_2$ that satisfies biogeochemical data. Dotted horizontal lines show the duration of biogeochemical constraints. Downward-pointing arrows indicate upper bounds on the partial pressure of oxygen ($pO_2$), whereas upward-pointing arrows indicate lower bounds. Unlabeled solid horizontal lines indicate the occurrence of particular paleosols, with the length of each line showing the uncertainty in the age of each paleosol. See Catling and Claire (2005) for description/references for the constraints. Ga, billions of years ago; MIF, mass-independent fractionation.

### *Theories for the rise of $O_2$*

Oxygen levels were smaller in the past because either the oxygen source was smaller or the oxygen sink was larger. The simplest explanation for a rise in oxygen at 2.4 Ga might be that oxygenic photosynthesis evolved then, but there is biomarker evidence that oxygenic photosynthesis existed by 2.7 Ga (hopanoids) (see Sec. 4C). The direct evidence for oxygenic photosynthesis comes in the form of "molecular fossils" or "biomarkers," remnants of specific organisms that resisted decay in the geologic record. 2-$\alpha$-Methylhopane biomarkers derived from oxygenic photosynthetic cyanobacteria and steranes derived from eukaryotic sterols are present at 2.7 Ga (Brocks *et al*., 1999).

The overall driver for the oxidation of rocky planets is likely to be hydrogen escape. Earth as a whole formed in a chemically reducing state. The only way for any system to become irreversibly oxidized involves a permanent loss of reductants, as occurs when hydrogen (derived from neutral material like $H_2O$) escapes the planet. The progressive oxidation of Earth's mantle and crust via H escape reduces the proportion of reductants in volcanic and metamorphic gases, which reduces the $O_2$ sink over geologic time. In this consensus viewpoint, the early Earth oxygen sink decreases gradually through the Archean until ~2.4 Ga, when the net sources of oxygen to the biosphere were larger than the net sinks. It is unclear what determines the timing of this transition and how the time scale will change for other, Earth-like planets. The details of this interdisciplinary story are being actively investigated by astrobiologists interested in biogeochemical cycling, isotope systematics, field geology, atmospheric modeling, and the search for life on extrasolar planets.

### *Rise of $O_3$ and the Proterozoic*

Concentrations of $O_3$ rise nonlinearly with oxygen levels, with the result that modern levels of $O_3$ are present even if the partial pressure of $O_2$ is 1%, the approximate level predicted after the rise of $O_2$ at 2.4 Ga. Prior to the rise of $O_2$ (and hence $O_3$), most UV radiation from the Sun reached Earth's surface, a condition that many modern organisms would find unpleasant. Oxygen blocks the incoming radiation. So the rise in oxygen, brought about largely by oceanic cyanobacteria, also created conditions more amenable for life to exist on the continents.

### *Comparative Planetology*

The history of Earth is similar, but not identical, to that of neighboring planets. It is unknown whether Venus received the same amount of water as did Earth; planet formation simulations show that multiple planet systems can form with varying water contents (Raymond *et al.*, 2004). If Venus did have the same amount of water as Earth, the higher $T_{eff}$ apparently led to a "runaway greenhouse" condition (Kasting, 1988)—positive temperature feedback from greenhouse warming of water vapor winning over all other feedbacks. The time scale for Venus' global ocean loss (presuming it happened) is unknown, as almost all features of the planet remain underexplored.

Mars suffers from being slightly outside the classical HZ (see Fig. 5.2 and Sec. 5B) and, as a small rocky planet (~1/10$^{th}$ the mass of Earth), appears geologically dead. The root cause of geologic activity is the outward heat flow from radioactive elements in the interior. Smaller planets have less radioactive material and cool more quickly. (This same line of reasoning applies to planetary magnetic fields, which are caused by convective processes in liquid cores.) There is mounting evidence that water flowed on the martian surface during some epochs of that planet's history, but these may have been transient events. The faint young Sun problem is much more severe in the context of early Mars, which necessitates a significant content of greenhouse gasses for the presence of a global ocean. More work must be done to differentiate between the "warm, wet early Mars" scenario and the more conservative "cold with transient liquid water" scenario. (See also Sec. 5C.)

## Abbreviations

AU, astronomical units; BIF, banded iron formation; Ga, billions of years ago; HZ, habitable zone; IR, infrared; KBO, Kuiper Belt object; K-T, Cretaceous-Tertiary; LHB, Late Heavy Bombardment; Ma, millions of years ago; $T_{eff}$, effective temperature; UV, ultraviolet.

## Chapter 3. Astrobiogeochemistry and the Origin of Life (FS)

The fundamental structures and processes that constitute life rely on a relatively small number of molecules. Formulating hypotheses about the origin of life requires a basic understanding of the role these molecules play in modern life, as well as how they may have formed or appeared in the prebiotic conditions of early Earth. This chapter begins with a summary of the chemical building blocks of life and how they interact (Sec. 3A). There follows a discussion of how those interactions may have first started (Sec. 3B). The biochemical complexity of even the simplest organisms makes it difficult to conceive of a less complicated ancestor. Nonetheless, attempts to understand how life originated on a prebiotic earth are extremely important for understanding both the evolution and the definition of life. The end of the chapter discusses current evidence for the development of complexity (both chemical and structural) through the 4.5 billion years of Earth's history.

### 3A. Life's Basic Components (FS, LM)

Life on Earth is distinguished from inanimate matter in part by its chemical complexity. Living organisms consist of a tremendous variety of molecules interacting to maintain physical structure, acquire resources, utilize energy, and maintain a constant environment. Despite this complexity, the molecular constituents of all terrean life share several fundamental properties.

Problems arise when trying to differentiate between the chemical processes that occur only in living systems and those processes that occur at large. The vocabulary of biological chemistry requires important distinctions. "Organic chemistry" includes all chemical processes using two or more covalently bonded carbon atoms (covalent bonds involve the sharing of electrons, while ionic bonds only involve the attraction of positive and negative charges.) Organic molecules usually form a carbon "backbone," which supports any number of other elements—primarily hydrogen, oxygen, nitrogen, phosphorus, and sulfur. At the same time, "biochemical," "biological," and "biotic" all refer to expressly living processes, and thus, both inorganic biochemistry and abiotic organic chemistry occur. The term "biomolecules" refers to the set of compounds common in, but not exclusive to, living organisms.

Most biochemistry involves long-chain molecules (polymers) made up of identical or nearly identical subunits (called monomers). The so-called biogenic elements (C, H, N, O, P, and S), along with trace amounts of other elements (*e.g.*, iron, magnesium), combine to form the three primary "biomonomers" or building blocks of life: sugars, amino acids, and nucleotides. The three monomers in turn form polymers or chains: polysaccharides, proteins, and nucleic acids, respectively. A fourth class of biological molecule, lipids, includes those organic molecules that have both polar and nonpolar groups, thereby allowing reactions with water and organic solvents. (Note that polar groups are parts of molecules with charges that can interact with other charged molecules, like water. Nonpolar groups have no charge and thus can associate with other nonpolar molecules, like oil.)

"Metabolism" refers to the suite of reactions that affect organic molecules in a cell. Metabolism therefore includes the energy-requiring synthesis of biomolecules (anabolism) as well as the energy-yielding degradation of nutrients (catabolism).

#### *Carbohydrates (Sugars)*

Carbohydrates, the most abundant biomolecules on Earth and the primary energy sources for many organisms, are named for their principal components—carbon, hydrogen, and oxygen—which typically conform to the ratio $(CH_2O)_n$. The three major size classes of carbohydrates are monosaccharides, oligosaccharides, and polysaccharides. The monosaccharides, or simple sugars, consist of an asymmetric chain or ring of carbons with accessory oxygen and hydrogen atoms. The most abundant monosaccharide in nature is the six-carbon sugar, D-glucose, used in energy storage for most organisms. Oligosaccharides consist of several simple sugar subunits bound together. The most common oligosaccharides contain only two monomers (*e.g.*, glucose + fructose = sucrose). Polysaccharides, on the other hand, consist of many monomers (often hundreds or thousands) joined in linear or branched chains.

The ability of organisms to form biomolecules from abiotic components defines "primary producers" or "autotrophs." Autotrophs use light energy (photosynthesis) or chemical energy (chemosynthesis) to string together carbon atoms (inorganic) to form monosaccharide subunits (organic). These subunits can then be used to form metabolic and structural components of the cell, including the polysaccharides used as structural support (*e.g.*, cellulose in plants, peptidoglycan in bacteria) or for energy storage (*e.g.*, starch). Organisms that cannot produce their own biomolecules are called "heterotrophs." They break down polysaccharides into glucose and then break down glucose to release energy (respiration). Complex biomolecules are built by stringing together two carbon molecules or larger pieces formed in the breakdown of food. Thus the biosphere runs on captured energy from solar and chemical sources that is stored as sugars and re-released in respiration. (For more on auto- and heterotrophy, see Sec. 6B.)

### *Lipids (Fats)*

Lipids, the primary component of cell membranes, are a chemically and functionally diverse group of biomolecules weakly soluble in water and highly soluble in organic solvents. Most lipids are derivatives of fatty acids: highly reduced carboxylic acids (molecules containing a —COOH group) with long chains of four to 36 carbons. Particularly abundant are the phospholipids, a class of lipids that consist of a sugar molecule (glycerol) linked to two fatty acids and to a polar alcohol molecule via a phosphodiester (—O—$PO_2$—O—) bond. These molecules, therefore, are amphiphilic (*i.e.*, they contain both polar and nonpolar domains). Phospholipids form sheets and bubbles by lining up with nonpolar domains inward and polar domains outward. Indeed, the membrane that surrounds every living cell is essentially such a lipid sheet formed into a bubble. Another common type of lipid, terpene, is essential for energy transfer. Formed from five-carbon segments called isoprene, terpenes have multiple shared electron orbitals, which makes them good for energy capture and redox reactions (see Sec. 6B). Important terpenes in biological systems include all steroids (*e.g.*, cholesterol) and most pigments that absorb or react with light.

### *Amino Acids → Polypeptides (Proteins)*

Amino acids, the building blocks of proteins, consist of a central carbon atom bonded to four different side chains: an amine group (—$NH_2$), a carboxyl group (—COOH), a hydrogen atom, and a highly variable "R" group. Amino acids may function singly (*e.g.*, as the biological signal molecules epinephrine, serotonin, histamine) or link together. Amino acids bond via amine-carboxyl (—$NH_2$—OOC—) connections called peptide linkages to form polymers called polypeptides. Twenty-two standard amino acids form the monomers for almost all polypeptides in terrean organisms. One or more polypeptides folded into a functional three-dimensional structure constitute a protein. In modern cells, proteins are fundamental structural components. Perhaps more importantly, proteins called enzymes catalyze almost all cellular biochemical reactions, including those of energy metabolism, biosynthesis, cell-to-cell signaling, nucleic acid replication, and cell division. (A catalyst is a molecule that speeds up a reaction, but is not consumed by it.)

### *Nucleotides → Nucleic Acids [Deoxyribonucleic Acid (DNA) and Ribonucleic Acid (RNA)]*

Nucleotides, the primary units of biological information systems, are composed of a pentose molecule (five-carbon sugar) connected to a phosphate group and a nitrogen-containing base. Single nucleotides often function as signaling mechanisms in the cell. Adenosine triphosphate (ATP), a nucleotide ring with three phosphate groups, is the principal energy currency within every cell. Energy from sunlight, reduced chemical compounds, or food can be stored in a phosphate bond (—$PO_2$—O—$PO_2$—). Breaking a phosphate bond in ATP releases energy for metabolic reactions. For example, the degradation of glucose by animals (heterotrophs) yields 36 ATP equivalents, while the synthesis of glucose by plants (autotrophs) consumes 66 ATP equivalents.

Nucleotide polymers are called nucleic acids. In modern organisms, DNA and RNA constitute the primary information storage and delivery molecules in the cell. These nucleic acids differ with regard to their pentose molecules—ribose or deoxyribose—and their attached bases. The nitrogenous bases in these molecules belong to one of two types, purines or pyrimidines. In DNA and RNA, the two major purine bases are adenine (A) and guanine (G). The primary distinction between DNA and RNA lies in the nature of their pyrimidine bases. Both utilize cytosine (C), but where DNA has thymine (T), RNA has uracil (U).

Nucleotides are complementary, such that A can form (hydrogen) bonds with U or T while C can form bonds with G. Two nucleic acid strands can thus bond together all along their length and form the characteristic double helix. This complementarity is essential to the storage and transmission of genetic information.

### *The Central Dogma (DNA → RNA → Protein)*

The so-called "central dogma" theory of molecular biology states that information is stored in DNA and transferred to proteins by way of RNA. Almost all information necessary for building and maintaining an organism comes encoded in DNA. Detailed plans for proteins, the active element of biochemistry, can be found as a series of three letter codons. Codons contain three sequential nucleotides (A, C, G, or T/U). Each codon (there are $4^3$, or 64) codes for one of the 22 different amino acids found in contemporary organisms. Because there are more codons than amino acids, the genetic code is referred to as degenerate, with multiple codons coding for the same amino acid. A specific codon (AUG) signals the start of a new polypeptide chain. In contrast, the end of a chain is encoded by one of three possible stop codons (UAA, UAG, or UGA). (Note that this is the most common code; alternate genetic codes, with minor differences, exist for some microorganisms.)

Information transfer occurs in a two-step process. DNA, which is repaired and copied by proteins (*e.g.*, DNA replicase and helicase), forms the primary molecule for information storage. Proteins called RNA polymerases copy codon sequences from DNA to messenger RNA (mRNA) in a process called "transcription." Protein/RNA complexes called ribosomes then direct protein synthesis from RNA by "reading" the codons and inserting the corresponding amino acids into a growing polypeptide chain ("translation").

The direction of information transfer in contemporary organisms, DNA → RNA → protein, applies nearly universally to contemporary organisms, though some information travels in the opposite direction. For example, certain animal viruses called retroviruses store information as RNA and use an enzyme called reverse transcriptase to make DNA from an RNA template. DNA may also store information in regulatory sequences, which do not code for proteins but do affect the timing and rate of transcription.

[A special note on nucleic acid terminology: RNA serves several purposes in the cell. RNA strands used to transfer information from DNA to proteins are called mRNA. Small tangles of RNA that shepherd amino acids to the ribosome for incorporation in proteins are called transfer RNA (tRNA). Larger clusters of RNA essential to the function of ribosomes are called ribosomal RNA (rRNA). Most ribosomes have a large and a small subunit. The small subunit (16S or 18S rRNA) forms the basis of much phylogenetics. See Sec. 4B.]

### *Prebiotic Chemistry*

"Prebiotic chemistry" investigates how synthesis of biomolecules may have occurred prior to the beginning of life as we know it. Abiotic synthesis of amino acids was demonstrated, in 1953, by use of an electrical discharge under reducing conditions. Today, most scientists agree that the atmosphere of the primitive Earth was never as strongly reducing as investigators then suspected, but early experiments showed that common biomolecules could be synthesized.

Several hypotheses have been proposed to account for the abiotic supply of monomers necessary for the first self-replicating biological system to form. It remains difficult to imagine an intermediate system that contains some, but not all, of the elements of modern biochemistry and metabolism. Most researchers suspect that an RNA-based existence or "RNA world" likely preceded the familiar DNA-RNA-protein world, because RNA can function as both an information storage molecule and a catalyst (see Sec. 3B). Thus, one molecule can serve in several roles. Although experimental proof of abiotic amino acid synthesis on early Earth exists, support for abiotic synthesis of nucleotides is lacking. A theory that explains the formation of nucleotides will be a necessary step in any compelling theory for the origin of life.

### *Endogenous Nucleotide Formation*

One nucleotide formation hypothesis suggests that precursors (*e.g.*, ribose, nitrogenous bases) originated on the early Earth (endogenous formation), either in the subsurface in volcanic aquifers or on the surface of metal-sulfide minerals. Still, while several abiotic mechanisms have been proposed, they are difficult to demonstrate. The inability to provide a convincing scenario for the prebiotic synthesis of ribonucleotides remains a major challenge to the RNA world hypothesis.

### *Exogenous Nucleotide Formation*

Alternatively, nucleotide precursors and, potentially, nucleotides may have formed exogenously (elsewhere) and then been transported to Earth. Such material may have arrived via a class of meteorites known as carbonaceous chondrites (see Sec. 2B). Carbonaceous chondrites are over 3% carbon, often in the form of organic material, including amino acids, polycyclic aromatic hydrocarbons, and carboxylic acids. While the ability of this material to withstand passage through the Earth's atmosphere and planetary impact is debated, most scientists agree that meteorites, as well as comets and interstellar dust, must have contributed in part to the buildup of organic matter on the prebiotic Earth (see also Sec. 2A).

## 3B. Evolution of Complexity (FS)

The evolution of modern life from abiotic components must have involved successive increases in biochemical complexity and organization. This section identifies some of the important innovations that occurred during this process and discusses their role in the evolution of complexity. Among these are the inception of information storage and transmission to offspring (heredity), the origin of catalytic activity (enzymes), the formation of cell structure (membranes), and the beginning of energy utilization to make and maintain biomolecules (metabolism). In each theory presented below, one of these four processes is posited as occurring before the evolution of the other three.

### *RNA-First Hypothesis ("RNA World")*

Primitive self-replicating living systems must have been able not only to encode genetic information but also to catalyze biochemical reactions. Neither DNA nor protein can perform these functions simultaneously. It has been hypothesized, however, that primitive forms of life could use RNA as their sole hereditary and catalytic material. "Ribozymes," naturally occurring RNA molecules that exhibit enzymatic activity, suggest the possibility of an RNA world in which organisms have RNA but lack DNA and proteins. Ribozymes have been shown to be capable of synthesizing the sugar-phosphate backbone of RNA (*i.e.*, self-assembly) and catalyzing peptide bonds (*i.e.*, protein formation). While studies to define the catalytic potential of ribozymes are ongoing, it is possible that these molecules were important for both the maintenance of an RNA world and a subsequent transition to protein-based catalysis.

Though abiotic synthesis of ribonucleotide monomers may have occurred on early Earth, successful transition to an RNA world would have required a mechanism for concentration and growth into self-replicating polymers. Mineral surfaces may have provided a regular template upon which organic compounds could adsorb and ultimately polymerize.

### *Proteins-First Hypothesis ("Protein World")*

Alternatively, some support a peptides-first hypothesis, which suggests that proteins were the first catalysts used by life. Only later was the information inherent in protein structure transferred to a nucleic acid (probably RNA) for long-term storage. A peptides-first scenario is supported by the fact that amino acids and peptides form relatively easily in simulated early Earth conditions, whereas the abiotic origin of nucleic acids is much more difficult to explain and demonstrate in an experimental setting. As of yet, however, there is no strong evidence to suggest a primitive polypeptide-to-nucleic acid transition. The hypothesis that information storage and replication began with a self-replicating clay mineral has been put forward, but this scenario lacks experimental verification.

### *Lipids-First Hypothesis ("Lipid World")*

Modern organisms are defined, in part, by their structure. All known organisms (excluding viruses) are composed of cells with membranous boundaries composed of lipid molecules. Lipid membranes play a vital role in compartmentalization, material transport, signal transduction, energy production and storage, and metabolic reactions. Amphiphilic molecules (with both hydrophobic and hydrophilic ends), such as lipids, may have formed on the early Earth. Long-chain fatty acids and fatty alcohols have been shown to form under hydrothermal conditions from gaseous precursors such as CO, $H_2$, and $CO_2$ (*i.e.*, Fisher-Tropsch reactions). Additionally, illumination with ultraviolet light can fuel the oxidation of alkanes (hydrocarbons containing only C—C and C—H bonds) to amphiphilic long-chain acids. Self-aggregation of such compounds could have facilitated the concentration of organics necessary for biochemistry. Alternatively, prebiotic lipids may have been brought to Earth in carbonaceous meteorites. The Murchison meteorite, an extensively studied carbonaceous chondrite (see Sec. 2B), contains amphiphilic molecules that form vesicles in aqueous solution.

The amphiphilicity of lipids and other related compounds (phospholipids, sphingolipids, sterols, pigments) allows these molecules to spontaneously self-assemble in aqueous solutions to form droplets, bilayers, and vesicles. Such structures may have constituted the first biological membranes. These membranes can compartmentalize biochemical functions, which allows for (1) small-scale spatial confinement of important biomolecules and reactions, (2) co-evolution of encapsulated molecules with differing genetic and functional properties (*e.g.*, ribozymes, catalytic proteins), (3) reduction in chemical interference from hazardous molecules in the environment, and (4) lowered molecular exchange with the exterior allowing creation and maintenance of steep chemical gradients. It has even been argued that organic membranes may have hereditary potential; most membranes form by pinching off from other membranes and, therefore, retain part of the biochemical signature of the mother "cell."

### *"Cell" Formation*

At some point, catalytic molecules must have become encapsulated within lipid membranes. One popular theory for cell formation involves clays. Clay mineral particles accelerate the formation of fatty-acid vesicles (membrane-bound sacs) and become encapsulated within the vesicles they create. A template for catalytic reactions (minerals) may have developed in this way. Alternatively, cyclic environmental changes may have been involved. For example, lipids undergoing hydration-dehydration and freeze-thaw cycles have been shown to encapsulate other biomolecules (proteins, nucleic acids) in lipid vesicles.

Selective permeability to nutrient molecules must also have been an essential component of early cells. In the absence of modern membrane-spanning proteins, which facilitate rapid solute transport into cells, early membranes may have been formidable barriers between encapsulated catalytic systems and the external pool of molecules needed for growth. This permeability problem may have been resolved, however, if the amphiphilic molecules in early membranes were significantly shorter than their modern counterparts. Permeability to ionic molecules has been shown to increase by several orders of magnitude when membrane phospholipids are reduced from 18 to 14 carbon atoms in length. Thus, primitive cell membranes may have been considerably thinner than those of modern organisms.

Cells, regardless of how they first formed, played a key role in the development of metabolism. Many key biochemical reactions depend on energy gradients; when high and low concentrations of a molecule are separated, potential energy forms. The most common gradient in biology arises when protons are pumped into or out of a vesicle. As the concentration on one side of the membrane increases, protons will seek to cross to the other side to restore balance. Organisms use this force to power ATP synthase, an enzyme that generates ATP from adenosine diphosphate. Both phototrophy (energy from sunlight) and respiration (energy from breakdown of sugars) first pump protons, then generate ATP with ATP synthase. Thus, the ability to capture energy depends on membrane-bound vesicles. Indeed, it remains difficult to imagine an organism in the absence of lipid membranes.

### *Rise of Metabolism*

Considerable debate surrounds the question of whether metabolism arose before or after the first self-replicating hereditary molecule. Proponents of a metabolism-first hypothesis argue that replication-first scenarios (including RNA first) require sufficient concentrations of helper molecules and, therefore, metabolic reactions. Such reactions, if they occurred, were not analogous to the enzyme-catalyzed metabolic processes of contemporary organisms but, rather, may have been catalyzed by abiotic peptides or ions in the environment.

### *Potential Habitats for the Origin of Life*

Several theories have been put forth to explain where on Earth life originated. High-temperature environments, which were common on a volcanically active Earth during heavy bombardment (the Hadean), have been suggested as potential sites for the origin of life. Deep-sea hydrothermal vents may be contemporary analogs to such environments. Spewing super-heated (up to 400°C) and metal-rich water from mid-ocean spreading centers and deep-sea vents provides a rich-source of reduced chemicals (*e.g.*, $H_2S$, $H_2$) that modern prokaryotes (Archaea and Bacteria) (see Sec. 6A) use to fuel metabolism. Opponents of the theory, however, claim that vent effluent is too hot to allow structural and functional stability of the organic molecules and enzymes that support life. The possibility that life formed in the cooler water at the periphery of vents could be a solution to this problem.

Several other theories exist. One involves life arising in the deep subsurface, perhaps at aquifers in igneous rock. Such habitats could protect cells from ultraviolet light and provide access to rich sources of $H_2$. A more recent hypothesis suggests that life, or at least its molecular precursors, may have first formed on the surface of atmospheric aerosols (liquid droplets suspended in a gas). Experiments have shown that marine aerosols can support an exterior film of amphiphilic molecules, whose synthesis may be driven in part by solar energy. Collapse of this film results in the division (fission) of the system, which suggests a simple mechanism for self-replication. The relevance of this process to the evolution of cellular life remains speculative.

## 3C. Definition of Life (FS)

What biochemical and physical processes are fundamental to life? Determining which parameters are necessary to identify life is vital to origin of life studies as well as astrobiology and the search for life elsewhere. Most scientists agree that no single parameter (*e.g.*, information storage) defines life. Rather, life results from multiple mechanisms acting in concert. As discussed above, all life on earth shares common molecular building blocks, chemistry, and energy generation mechanisms.

### *Evolution*

A substantial fraction of biological work is devoted to the complex process of self-replication. Replication is mediated by a detailed set of chemical instructions encoded within the genetic material of the cell. These instructions are duplicated and transferred to successive cell generations. Significantly, the instructions are susceptible to change (mutation) over time because of imperfect copying as well as external influences. Such changes create a mosaic of cells with differing genetic material (*i.e.*, genotypes). The resulting differences in reproductive efficiency result in dynamic interactions between organism and environment that seem to be characteristic to life. (See Secs. 4A and 4B for a fuller treatment.)

### *Common Properties of Life*

Salient features of known life include: a carbon-based molecular structure, complex chemical interactions, compartmentalization via membranous boundaries, energy and biosynthetic metabolism fueled by external energy and nutrient sources, self-replication, information storage in genetic material, and progressive adaptation via Darwinian evolution. But a definition of life based on contemporary organisms is prohibitive in that it fails to consider forms that preceded modern cellular life (*e.g.*, autocatalytic, self-replicating RNA molecules) or forms that we cannot conceive of presently. Indeed, NASA's broader definition of life as "a self-sustained chemical system capable of undergoing Darwinian evolution" is also somewhat limiting. Luisi (1998) argued that inclusion of the criteria of Darwinian evolution, operating at the level of populations, excludes life in the form of a single individual as well as potential processes of non-Darwinian adaptation. A central challenge in origin of life research is, therefore, to arrive at a definition of life that moves beyond contemporary organisms to include novel forms that likely occur outside our planet.

### *Exceptions to the Rule*

Several contemporary examples of "organisms" that do not meet the conventional criteria for life exist. Viruses, which consist primarily of nucleic acids surrounded by a coat of proteins, do not reproduce on their own. Rather, following invasion of a host cell, they co-opt the machinery of the host for use in their own reproduction. Likewise, autonomous genetic elements such as plasmids replicate within cells but need not benefit their host. Rather, many plasmids (small circular strands of DNA) code for reproduction and transportation to other cells. Similarly, the prion, which consists only of proteins, is an infectious molecule that depends on its host cell for replication. The means by which these molecules replicate, disperse, and evolve remains a mystery. Likewise, their inclusion in the category of "life" is debated.

## Abbreviations

A, adenine; ATP, adenosine triphosphate; C, cytosine; DNA, deoxyribonucleic acid; G, guanine; mRNA, messenger ribonucleic acid; RNA, ribonucleic acid; rRNA, ribosomal ribonucleic acid; T, thymine; tRNA, transfer ribonucleic acid; U, uracil.

# Chapter 4. Evolution of Life Through Time (JR, OZ)

Terrean life reflects billions of years of development. Life has continually reacted to, and ex-tensively restructured, all the environments it has inhabited. Evolutionary biology in general (phylogenetics and population genetics in particular) investigates the process and character of this development. This chapter begins with an overview of evolution as a concept (Sec. 4A). The next section, Evolutionary Dynamics (Sec. 4B), deals with the nuts and bolts of mutation at the genetic level. Sections 4C, 4D, and 4E explain how the history of life on Earth can be reconstructed using genetic, biochemical, and structural evidence. (Editor's Note: Those unfamiliar with biochemistry may wish to read Chapter 3, for an introduction to the nucleic acids, proteins, and metabolism before starting Chapter 4.)

## 4A. Overview (LM)

No subject could be more central to astrobiology than the evolution of organisms through time. The concepts of descent with variation and natural selection appear central to our best definitions of life (though see Sec. 3C). At the same time, it must be noted that modern theories of evolution can be difficult to grasp. Often the mechanisms are obscure and the theories counterintuitive. More than change through time, divergence of species, and adaptation, evolution reflects a probabilistic (technically stochastic) process, in which life explores a seemingly infinite number of possible states. Each organism has descendents, all of which vary slightly in their composition and biochemistry. Complex mechanisms have arisen to encourage variation (such as sex in animals and conjugation in bacteria). Entropy, however, ensures the imperfect transmission of information and guarantees that even the most careful replicating system will result in some errors.

Evolutionary dynamics (Sec. 4B) studies the manner and method of variation as organisms produce offspring. The discovery of deoxyribonucleic acid (DNA) led to an understanding that the basic unit of evolution, the mutation, occurs at the level of nucleotides. As an information storage molecule, DNA is susceptible to change by outside factors (*e.g.*, radiation, chemical mutagens) as well as imperfect transmission of data to offspring. The genotype, or full complement of genetic information contained in an organism, changes. Mutations, which occur at the level of DNA, become expressed in proteins and biochemical regulation of the cell. The phenotype, or full complement of expressed characteristics of an organism, may be unchanged, impaired, or improved.

Phylogenetics (Sec. 4C) involves the reconstruction of historical relationships among organisms on the basis of shared genotypic and phenotypic traits. By looking at large numbers of individuals, it is possible to discover how they are related and trace the process of mutations backward through time.

A few rare events in the history of terrean life (Sec. 4D) appear to have had a disproportionately large effect on modern life. Catastrophic events (such as ice ages and meteor impacts) have caused large-scale extinctions and speciations, and endosymbiosis—a state of being wherein one organism lives inside the cells of another—has been central to the development of modern multicellular life.

While many traits of ancient life can be reconstructed based on modern genetic evidence, physical remains can also be found. Geologists and paleontologists study the chemical and structural remains of past life. Chemical fossils (Sec. 4E) consist of chemical traces that provide evidence that a specific type of organism lived at a given place and time. Several organic compounds have been discovered that, apparently, can only be produced by biological processes. When those processes no longer appear active, these molecules and their degradation products are called "molecular fossils" as they are the chemical remains of life. Structural remains, a more traditional kind of fossil, also give evidence of the nature, distribution, and abundance of past life. Paleontology (Sec. 4F) investigates remnants of organisms that once lived and tries to piece together the history of terrean life.

As with any historical science, the study of terrean evolution is not amenable to well defined variables and clean experiments. Often researchers must make inferences about past events on the basis of limited data. This leads to some contention over the best possible theories of inquiry and reasoning. It is essential to be aware of fundamental assumptions that contribute to popular theories, especially when those assumptions remain controversial.

## 4B. Evolutionary Dynamics (OZ, LM, JR)

At the most basic level, evolutionary changes occur by modification of nucleotide sequences.

The simplest change that can occur to a sequence is a point mutation: the change of a single nucleotide (*e.g.*, adenine → cytosine). Other changes involve deletion or transpositions of nucleotides. (A transposition occurs when a nucleotide or series of nucleotides moves from one place in the genome—the organism's complete set of genes—to another). Mechanisms of mutation (imperfect copying, radiation, chemical mutagens) tend to be more common at certain sites or at certain nucleotides, which biases mutation. If a DNA sequence codes for a protein, some mutations may result in an amino acid change in the resulting protein sequence, though others do not (because of the redundancy of genetic code). Some mutations have no noticeable effect, while others are detrimental to the carrier or (very rarely) provide a benefit. When variation disappears and one sequence becomes the sole genotype for all members of a population, it is said to be "fixed" in that population. A mutation that has reached fixation can be referred to as a substitution.

### *Neutral, Purifying, and Positive Selection*

Most mutations found in a population and most substitutions are selectively neutral or nearly neutral (the ones that are detrimental disappear from the population). The theory of neutral evolution, first proposed by Motoo Kimura in 1968, laid the fundamentals for this point of view. Neutral mutations can be fixed in a population due to random genetic drift. The proponents of the opposing (less popular) school of thought, the selectionists, argue that mutations are fixed because of the selective advantage they provide.

Because of functional constraints on genes, substitutions do not occur in all parts of a gene uniformly (this phenomenon is called among site rate variation). Usually, the sites that are important for the function of a protein accumulate relatively few substitutions; such changes, more likely than not, would decrease the efficiency of the protein. Substitutions also do not occur with the same rate and pattern in all genes. In very important functional genes, almost any mutation appears to be detrimental. As organisms with mutations fail to survive, only the pure copies remain and the gene is said to be under purifying selection. Alternatively, when conditions change or when a gene acquires a new function, a different process may occur. If one type of mutation confers a benefit to the organism, then that organism will be more successful and that mutation will become more common in the population. Then the gene is said to be under positive selection.

One popular way to assess whether a gene is under positive, purifying, or no selection is to calculate the $K_a/K_s$ ratio, that is, the ratio of nonsynonymous substitutions ($K_a$; resulting in an amino acid change) to synonymous substitutions ($K_s$; not resulting in an amino acid change). Selection can be judged as follows:

- $K_a/K_s \approx 1$ implies the gene is under no selection (evolving neutrally)
- $K_a/K_s > 1$ implies the gene is under positive selection
- $K_a/K_s < 1$ implies the gene is under purifying selection

With time, sequences that accumulate substitutions will diverge. By comparison of sequences derived from a common ancestral sequence (homologous sequences), it is possible to infer the relationships among the sequences (see Sec. 4B).

### *Horizontal Gene Transfer*

All members of a population exchange genes and compete for limited resources, which makes it impossible to consider any one organism in isolation. Genes are normally transferred "vertically," passed from parents to offspring, but also travel horizontally (or laterally) between organisms in the same environment. In horizontal transfer, an incoming gene may replace or randomly recombine (hybridize) with an already present gene, or simply be inserted into the genome. Some genes, so-called selfish genetic elements (such as mobile genetic elements or plasmids), even have mechanisms for promoting their own incorporation at the expense of the host.

A number of mechanisms have evolved to facilitate horizontal gene transfer. They fall into, roughly, three categories: transformation, conjugation, and transduction. Transformation involves the uptake of small snippets of DNA from the surrounding environment. Conjugation occurs when one cell constructs a bridge to another cell and passes genetic elements through the bridge (transposons, plasmids). Transduction results from the transfer of DNA or ribonucleic acid (RNA) by a virus that has incorporated sequence data from one host and passed it to a subsequent host. Horizontal gene transfer has become a key mechanism for increasing variation and speeding up evolutionary change.

See literature cited for Sec. 4C.

## 4C. Molecular Phylogenetics (OZ, JR, LM)

### *Making Trees*

During the early 19th Century, ideas on how to visualize relationships among organisms started to crystallize; biologists had long dreamed of the tree-like diagram that would depict the evolutionary history of all living organisms on Earth, but had trouble developing an organizing principle. Initially, morphological (variations in form and shape) characters were used for classification, but with the discovery of the microbial world, it became apparent that this would be insufficient. In the middle of 20th Century, technological advances allowed proteins and, later, genes to be sequenced. Biologists became enthusiastic about using molecular data to extract evolutionary information.

One can infer the history of a group of genes using changes accumulated in DNA (or proteins) of homologous genes from different organisms (see Sec. 4B). Tracing the divergence of genes and gene families backward through time can be modeled by a bifurcating tree of relationships. First, a data set must be carefully assembled from genes descended from a common ancestral sequence (aka homologous genes). As an evolutionary term, homology is a "yes" or "no" decision, *i.e.*, either two sequences are homologous, or they are not. Similarity should never be confused with homology. Phylogeneticists often infer homology on the basis of sequence similarity—similar sequences are probably homologous. The reverse, however, cannot be said—homologous sequences can be so divergent that no sequence similarity is detectable.

Homologous sequences, once identified, need to be aligned; amino acid or nucleotide positions (alignment sites) that correspond to a single ancestral position in the sequence are lined up. The quality and accuracy of the resulting phylogenetic tree crucially depend on the quality of the alignment. To date, no reliable alignment algorithm has been developed, and any alignment remains subjective. While many computer algorithms for creating alignments exist, additional manual adjustments are usually necessary.

### *Choosing Trees*

Phylogenetic reconstruction methods can be grouped into three broad categories: distance, parsimony, and maximum likelihood. Distance methods convert molecular data into a matrix of pairwise distances between all sequences (a "distance matrix"), and then a tree is constructed that best fits pairwise distances to a bifurcating tree. Distance methods have the advantage of speed. Parsimony methods find the tree with the shortest branch lengths (smallest number of substitutions). They are based on the idea of Occam's razor—all things being equal, the simplest hypothesis is the best. In this case the best tree is the one that explains the aligned sequences with the fewest substitution events. Maximum likelihood attempts to find a tree that has the highest probability under the given model for sequence evolution. The best tree is the one for which the data are most likely to have been produced.

Tree space—the set of all possible bifurcating trees relating a set of genes, proteins, or organisms—is vast. For large data sets (more than 20 analyzed sequences), it is impossible to explore all of the possible tree topologies. Most programs attempt to sample areas of tree space intelligently by using heuristic searches. This makes it important to assess the researcher's confidence in any one tree, given the quality of the data, the evolutionary model used, and the phylogenetic methodology. One of the most widely used techniques is bootstrapping. "Pseudosamples" of a data set are created by resampling from the sites in an alignment and analyzing all pseudosamples with the same phylogenetic reconstruction method. After completion of all analyses, the results are compiled, and each possible branch on the tree receives a bootstrap support value equal to the percentage of pseudosamples for which that branch was reconstructed by the algorithm. Bootstrap support values, therefore, provide a measure of how much support the data provide for different parts of the tree when working with a given model of evolutionary dynamics and a fixed methodology for phylogenetic reconstruction.

## *Current Debates*

### *Phylogenetics in the genomic era: credibility of ribosomal RNA (rRNA) as a universal molecular marker and impact of horizontal gene transfer on inferences about organismal evolution*

Since the beginning of molecular phylogenetics, researchers have hoped that gene-based trees could be used to unambiguously recreate the history of terrean life. It was originally believed that all genes would reflect the same tree, given a perfect algorithm. The discovery of horizontal gene transfer, however, showed that this was not the case. In the absence of a single tree for all genes, phylogeneticists began to look for a single gene that was not transferred horizontally and, therefore, could be used to recreate the true or basal tree of life.

Carl Woese suggested rRNA as a universal molecular marker that spans all domains of life (Woese, 1987) (see Sec. 6A). As rRNA is highly conserved, essential to cell function, and strongly tied to almost all biochemical processes, horizontal transfer seemed unlikely (although recently several instances of mosaic rRNAs have been described). Differences between small subunit rRNAs, along with other evidence (including biochemical characteristics and phylogenetics analyses), led to the three domain classification system (Sec. 6A). rRNA has become the gold standard in taxonomy (especially in microbial taxonomy, where morphology cannot be used), and over 100,000 partial and complete rRNA sequences are logged and available through the Ribosomal Database Project.

Often, however, phylogenies based on other genes contradict the rRNA phylogeny. The differences are attributed either to horizontal gene transfer or methodological artifacts. In recent years, large-scale sequencing projects that sought to determine the sequence of nucleotides in every chromosome for a given organism led to the hope that complete genomes would provide an unambiguous Tree of Life (see Figs. 6.1–6.3). Unfortunately, these sequencing projects only reinforced the idea that different genes in one genome have different histories; horizontal gene transfer appears to be quite common. The frequency of horizontal transfer and the extent to which it damages the credibility of phylogenetic trees is still debated.

### *How accurate is the molecular clock: can we reliably date evolutionary events using inferred phylogenetic trees and sparse fossil record?*

In 1965, Zuckerkandl and Pauling introduced the concept of a molecular clock. If genes accumulate changes in a constant, clock-like fashion across all lineages, then the available fossil record can be used to assign dates to some nodes in a phylogenetic tree ("calibration"). From these dates, it should be possible to extrapolate or interpolate when species diverged by assuming a constant rate of genetic change.

More recent data suggest that most genes do not evolve in a clock-like way. Currently, there are various modifications of molecular clock models (*e.g.*, models that allow clocks to "tick" at different speeds in different parts of a phylogenetic tree as well as along different sites of an alignment). The question of how reliably one can place dates on speciation events remains highly debated, especially when extrapolating to early evolutionary events. Integrating or constraining molecular phylogeny with paleontology, and vice versa, presents an exciting synergistic opportunity but requires caution and clearly defined assumptions.

## 4D. Rare Events in Evolution (LM, JR, OZ)

### *Catastrophic Events*

Environmental catastrophes give insight into how the environment can strongly constrain or direct evolution, even if by chance. Perhaps the earliest catastrophe encountered by fledgling life was the so-called heavy bombardment period during early planetary accretion, a period characterized by high rates and sizes of asteroid impacts that may have effectively vaporized and, thereby, sterilized Earth's oceans and shallow crust. Heavy bombardment persisted until roughly 3.9 billions of years ago (Ga), which is remarkable in that it preceded the earliest (albeit controversial) signatures for life by only a few hundred million years. More recent impact events may have been important in driving adaptive change or even extinction, such as the Chicxulub impact that occurred at the time of the Cretaceous-Tertiary mass extinction (see Sec. 2B and Fig. 2.2).

Global-scale glaciations—so-called "Snowball Earth" events—represent periods of near-total freezing over of Earth's oceans, which may have occurred during two distinct periods during the late Proterozoic (see Sec. 2D for details about Snowball Earth theory). Both of these events are closely associated with major diversifications in the complexity of life. The first (Paleoproterozoic) event appears related to the advent of oxygenic photosynthesis and the ensuing development of aerobic respiration and complex multicellular eukaryotes, including metaphytes (plants). The second (Neoproterozoic) event included the evolution of the other major group of multicellular eukaryotes, namely, macroscopic metazoans (animals). The appearance of these multicellular groups preceded the Cambrian explosion of well-skeletonized animal phyla that has dominated the Phanerozoic Eon, up to the present time.

Oxygenic photosynthesis first arose in a primitive cyanobacterium-like organism, probably between 2.2 and 2.7 Ga, though possibly earlier. The ensuing rise of oxygen in the atmosphere undoubtedly precipitated a major extinction event. It also precipitated the evolution of aerobic respiration and made possible a previously untenable range of metabolic capabilities based on $O_2$ as a high-energy electron acceptor (see also Sec. 2E).

### *Endosymbiosis, Mitochondria, and Chloroplasts*

Frequently, associations between different organisms impact the way those organisms live and evolve. The term "symbiosis" literally means living together and refers to any arrangement where two or more organisms maintain a long-term association. Symbiosis sometimes connotes mutual benefit (technically a mutualistic relationship), though this is not necessary. Other symbioses include parasitic relationships—beneficial to one at the expense of the other—and commensalisms—benefit to one without cost or benefit to the other.

Numerous lines of evidence suggest that mitochondria and chloroplasts represent cases of endosymbiosis, wherein one organism has evolved to live within the cells of another. Mitochondria, present in almost all eukaryotes, are bacteria-like entities responsible for respiration. They share similarity in structure, metabolism, and function to alpha-proteobacteria (see Figs. 6.2 and 6.3). Chloroplasts and other plastids, responsible for photosynthesis within eukaryotes, resemble cyanobacteria (see Figs. 6.2 and 6.3). Chloroplasts and mitochondria cannot survive without their host, as some essential genes have been transferred to the host nucleus. At the same time, they are capable of reproduction and cannot be replaced by the host cell if lost.

Although no mitochondria are present in the most basal members of the eukaryote lineage (see Fig. 6.3), those organisms do contain remnants (*e.g.*, hydrogenosomes), and molecular phylogenetics support a common ancestor for all mitochondria. Therefore, one endosymbiotic event probably occurred, which involved the encapsulation of an alpha-proteobacterium by the ancestor of all eukaryotes.

Chloroplasts present a more complex story. One primary endosymbiosis has been hypothesized. A cyanobacterium was engulfed by a eukaryote that would become ancestor to red algae, green algae, and plants. A quick perusal of the tree of life, however, shows that photosynthesis occurs not only in these organisms, but throughout the Eukarya. Secondary endosymbioses must have occurred; algae were engulfed by other organisms, which led to euglenids, diatoms, dinoflagellates, and a host of other photosynthetic protists. Some details of these events have been traced, but the full extent and number of these secondary (and perhaps tertiary) endosymbioses remains unclear.

## 4E. Chemical Fossils (JR)

Biomarkers (aka "chemical fossils") are molecules associated with a particular biochemical process or property. Their utility, which spans multiple disciplines, ranges from measuring physiology or disease states in living organisms to identifying the inhabitants of ancient ecosystems. The last has particular importance for astrobiology. Though complex hydrocarbons extracted from drill cores are the most commonly cited biomarkers, atmospheric gases are often discussed. One such involves molecular oxygen derivatives as potential targets for NASA's planned Terrestrial Planet Finder mission.

In general, biomarker studies focus on:

**Burial and preservation/alteration.** As with any fossil, a biomarker must be adequately preserved before it can be rediscovered. Most organisms have highly tuned enzymatic salvage pathways for recycling even complex molecules. Furthermore, most biomolecules are soluble compounds that quickly degrade and are rapidly lost once the host cell dies. Even under ideal conditions in sedimentary rocks, biomarkers undergo diagenesis (processes that affect sediments at or near the surface at low temperature and pressure; compaction, cementation, etc.). They form a variety of derived compounds and can sometimes migrate into different rock strata. Considerable research, originating in no small part from the petroleum exploration industry, is dedicated to understanding how increasing pressures and temperatures (so-called "thermal maturation") alter and eventually erase the information content of a biomarker. Typically covalent bonds break. Target rocks for biomarker analysis often have high organic matter (kerogen and/or bitumen) content as this indicates a relatively mild degree of thermal alteration.

**Isolation.** Isolation focuses most often on rocks where minimal or no metamorphism has occurred. Because lipids and other hydrocarbon-rich molecules are generally more resistant to alteration and loss from host rocks, these are frequent target biomolecules. Biological as well as chemical contamination is, and should be, of foremost concern in biomarker studies. Such contamination is painstakingly guarded against by, for example, combining sterile techniques (whenever possible) with well-designed replicate and control studies, or by avoiding, *e.g.*, lipid-derived lubricants in drill machinery that could potentially leach into core samples. While this topic is too detailed to be adequately covered here, the question as to what steps were taken to prevent contamination should be foremost on readers' minds when interpreting biomarker studies.

**Characterization.** Field samples contain a rich mixture of molecules that must first be physically separated. First, solvent extraction is utilized for broad-scale separation (*e.g.*, into amorphous/insoluble kerogen versus volatile bitumen components); then one or more chromatographic steps are used for fine-scale separation. Most often, mass spectrometry is then used to identify biomolecules against known libraries based on mass-to-charge ratios, fragmentation patterns, and stable isotope analysis. Intensive interrogative techniques, such as nuclear magnetic resonance and a gamut of spectroscopic methods, are also used commonly in analyzing biomarker structure.

**Identification.** Using a biomarker as an indicator for ancient organisms or environments proves most useful when its biosynthesis, abundance, and distribution in modern organisms are well constrained; there must be a library of information against which to compare a characterized biomarker. Furthermore, the processes that alter and, thereby, erase the information contained within a biomarker's structure must be taken into account. While progressive breakdown of biomarkers, even those preserved in the most mild of conditions, is invariable, some classes of biomarkers—lipid derivatives notable among them—are more recalcitrant to degradation. Correlating biomarker distribution with organismal taxonomy is ongoing, with the goal of understanding the specificity of certain biomarker modifications and the presence or absence of biomarker classes across the tree of life. Among the most informative ancient biomarkers are late Archean (~2.7 Ga) derivatives of hopanes and steranes—cholesterol-like molecules with chemical modifications. They appear to be highly specific to several taxonomic groups, including eukaryotes and cyanobacteria. Furthermore, the enzymatic reactions that comprise an organism's metabolism often leave isotopic fingerprints on biomarker compounds. These can be used in a diagnostic fashion. For example, specific types of autotrophic carbon fixation leave biomarkers. Such research can offer an additional level of information for potentially constraining biodiversity or redox conditions in ancient environments (see Sec. 2C).

### *Brief Survey of Biomarkers*

*Alkanes/alkenes*

Alkanes/alkenes and their modifications are abundant biomarkers because they are essential in membrane synthesis in most organisms. Based on the number of carbon atoms and branching patterns, some alkane biomarkers are broadly, if not specifically, diagnostic (*e.g.*, algaenan as a biomarker for algae).

*Isoprenoids*

Isoprenoids are five-carbon compounds used ubiquitously in the biosynthesis of larger biomolecules (*e.g.*, carotenoids and phytol). Isoprenoid polymers of varying length and branching pattern are commonly retrieved biomarkers, with some—such as archaeol, thought to be specific for the domain Archaea—providing taxonomic information despite their abundance in organisms. Isoprene derivatives can also be highly specific to some organisms. For example, the carotenoid isorenieratane is found only in a single family of photosynthetic (green sulfur) bacteria. It has been used recently to support anoxic sulfidic ocean conditions that accompanied the Permian-Triassic mass extinction (Grice *et al.*, 2005).

*Sterols*

Sterols are an incredibly diverse group of biomolecules (*e.g.*, cholesterol, lanosterol, cycloartenol), synthesized ultimately from isoprenoids and found almost exclusively in eukaryotes (but also in a few bacteria). They function in cell-cell signaling and as controls of membrane fluidity. Though sterol biosynthesis shares the same basic biosynthetic steps with a variety of isoprene derivatives, $O_2$ is required for sterol synthesis in all known organisms. Thus the presence of sterols, which extends back perhaps 2.7 billion years, may indicate not only the rise of eukaryotes but also the appearance of $O_2$ in the early atmosphere (Brocks *et al.*, 1999).

*Hopanoids*

Hopanoids can be thought of as "prokaryotic sterols." They are thought to serve an analogous role to sterols in regulating membrane fluidity (but not requiring $O_2$ in their biosynthesis). Some modifications, in particular the position and geometry of methyl groups, are specific to certain taxonomic groups, including some cyanobacteria and proteobacteria (Summons *et al.*, 1999).

### *Precambrian Biomarkers*

While the gamut of biomarkers, especially from the Phanerozoic, is too extensive to be detailed here, substantially fewer informative Precambrian biomarkers are available. They are, nonetheless, of particular importance to understanding early, prokaryote-dominated environments on Earth (and possibly elsewhere). A few of these are summarized below. Brocks and Summons (2003) and Simoneit (2002) give a much more detailed cross section of these and other examples.

**2.7–2.5 Ga.** Both hopanoid and sterol derivatives have been recovered from Archean-aged rocks that have been subject to only mild thermal alteration (Brocks *et al.*, 1999; Summons *et al.*, 1999). Hopanoids are found with the $3\alpha$-methyl and $3\beta$-methyl substitutions, consistent with biomarkers found in modern cyanobacteria and methane-oxidizing bacteria. Sterol derivatives have side chain modifications only found in modern eukaryotes and explicitly require $O_2$ in their biosynthesis. Taken together, this evidence suggests that at least low levels of oxygen were available from cyanobacterial oxygenic photosynthesis.

**2.2–2.0 Ga.** Branched alkanes have been recovered that may correlate with prokaryotic sulfide oxidation. While that is present across the tree of life, it may be a proxy for the use of oxygen in biological electron transfer chains (Kenig *et al.*, 2003). This particular biomarker class is also found extensively in more recent rocks.

**1.7–1.5 Ga.** Derivatives of carotenoid pigments consistent with purple sulfur and green sulfur bacteria have been recovered. These groups include many species of anoxygenic phototrophs (Brocks and Summons, 2003).

**FIG. 4.1. Important events in the Precambrian.** Ga, billions of years ago.

**Hadean Eon (4.6-3.9? Ga)**

4.6 accretion of Earth

*Heavy Bombardment period*

**Archean Eon (3.9?-2.5 Ga)**

3.8 possible earliest isotopic evidence for biological carbon chemistry

3.5 isotopic evidence for biological sulfur reduction, first stromatolites

3.2 filamentous fossils in deep-water hydrothermal setting

2.8 isotopic evidence for biological methane chemistry (archaea and bacteria)

2.77 biomarker evidence for cyanobacteria, eukarya, and heterotrophic bacteria

2.7-2.6 carbon isotope evidence possibly indicating land based life

**Proterozoic Eon (2.5-0.542 Ga)**

Paleoproterozoic (2.5-1.6 Ga)

2.45-2.22 geological evidence for oxygenation of the atmosphere and oceans

2.1 oldest fossils with modern morphology (cyanobacteria)

1.87 oldest eukaryotic macrofossil (*Grypania*)

1.80-1.45 oldest certain eukaryotic microfossils and biomarker evidence

Mesoproterozoic (1.6-1.0 Ga)

1.2 oldest fossil multicellular photosynthetic eukaryote with modern morphology (red algae)

Neoproterozoic (1.0-0.542 Ga)

0.8-0.7 eukaryotic biomineralization, fossil amoebae, probably fossil fungi, multicellular algae

0.6 animal radiation, geological evidence for glaciations and rising oxygen levels

**[Phanerozoic Eon (<0.542 Ga)]**

**1.6 Ga.** The presence of well-preserved steranes with eukaryote-specific modifications, temporally consistent with eukaryotic microfossils (Summons *et al.*, 1999), has been found.

### *Talking Points*

*Are ancient biomarkers diagnostic for specific organisms?*

Given the small number of modern organisms that have been sampled and characterized, and the vast number of ancient, extinct organisms that never will be sampled, how can we best appraise the diagnostic value of an ancient biomarker?

*Are ancient biomarkers diagnostic for specific processes?*

Furthermore, could the appearance of biomarkers, *e.g.*, for cyanobacteria, correlate with the evolution of an unrelated process, such as oxygenic photosynthesis?

## 4F. Paleontology (EJ)

The search for past or present life beyond Earth requires a solid understanding of life's origin and evolution on the only planet on which life is known—Earth. Life had a profound effect on the landscapes and atmospheres of Earth and has left biosignatures (objects, molecules, or patterns that can be explained only by biology), which allow for the detection of past life-related activities (see Fig. 4.1).

Many biological characters used to differentiate members of the three domains of life—Archaea, Bacteria, and Eukarya (Sec. 6A)—rarely survive fossilization and so are not generally available to the palaeontologist. Moreover, geological processes such as metamorphism (high temperature and pressure alteration of rocks in Earth's crust), diagenesis, and tectonics (processes that cause major structural features of Earth's crust) erased most of the early record of life. Nevertheless, in a few areas of the globe, sedimentary rocks are better preserved and contain traces of past life. They indicate that microorganisms inhabited Earth during the Archean (before 2.5 Ga) and had developed metabolisms similar to many living microbes. Combined approaches, including geology, geochemistry, and paleontology, as well as insights from molecular phylogenetics and microbiology, are necessary to better understand the early evolution of terrean life.

### *Isotopic Record*

Each chemical element is defined by how many protons can be found in the nucleus. An element may have differing numbers of neutrons, however, and thus various isotopes. The name of an isotope reflects the total number of protons and neutrons so that carbon, with six protons, has three isotopes, $^{12}C$, $^{13}C$, and $^{14}C$ with six, seven, or eight neutrons, respectively.

Isotopic signals in preserved organic matter (*e.g.*, bioprecipitated minerals) suggest an early microbial biosphere. Living organisms use preferentially lighter isotopes in their metabolism. Photosynthesis, for example, prefers to incorporate $^{12}C$ over $^{13}C$, a process called "fractionation," which causes a different ratio of the two isotopes in organic matter than that found in abiological systems. Carbon, sulfur, and iron isotopes have been particularly important to astrobiology.

Carbon isotopes from the oldest known sedimentary rocks (3.8–3.6 Ga) in the Isua Greenstone succession, Greenland, have values fitting the range of fractionation consistent with biological activity (autotrophic carbon fixation), though some debate remains over what this represents. Most of the Greenland record has been reassessed, and the biological interpretation seems to hold for only one area. Caution is required as abiological chemical reactions in hydrothermal fluids may produce abiological organic matter with similar carbon isotope fractionation patterns.

The isotopic record also gives information on the composition of the early atmosphere and oceans. Several independent lines of geologic evidence suggest that significant levels of oxygen first accumulated in the atmosphere in the late Archean (see Sec. 2E and Fig. 2.3 in particular). Recently, a study of sulfur isotope fractionation dated this oxygenation event at 2.32 Ga (for a different view, see Ohmoto, 1996).

### *The Earliest Fossils*

The earliest fossil record is controversial. The famous 3.5-Ga Apex chert microfossils, first interpreted as fossilized filamentous cyanobacteria or bacteria, have been recently reassessed. Several authors have presented evidence that they are abiological organic matter that was produced by hydrothermal processes around mineral casts. Careful reconstruction of environmental deposition of sediments or precipitation of minerals is crucial in the interpretation of biosignatures.

Stromatolites (layered carbonate formations caused by microbial communities) occur throughout the rock record, starting in the early Archean (3.4 Ga). They are generally interpreted to be built or occupied by bacteria, although abiological processes can produce similar structures and the biological nature of the oldest examples has been questioned. Other sedimentary structures in 2.9-Ga siliciclastic (containing mostly silicate minerals) rocks also preserve signatures of bacterial mats.

### *Fossil Record of Bacteria and Archaea*

Dating the earliest structural evidence for unicellular organisms presents multiple difficulties. Abundant filamentous (thread-shaped) and coccoidal (rod-shaped) carbonaceous structures have been described from the last half of the Archean Eon; however, their biological origin and pattern of deposition have been debated. Abiotic processes can synthesize products that resemble the morphologies of some Archean microfossils. This suggests that simple morphology is not sufficient to prove biogenicity (biological origin). Problems in determining biogenicity include artifacts of sample preparation, contamination by younger microorganisms, and isotopic fractionation by nonbiological organic chemistry in hydrothermal settings. Biogenicity of microstructures from well-dated rocks within well-constrained paleoenvironments of deposition can be clearly established only when morphology and organic chemistry are considered along with degradational criteria, such as distinctive cell division pattern, plasmolysed cell content (dark blob inside fossil cells corresponds to shrunk cytoplasm), pigmentation gradients in colonies, spatial distribution, and population variation. Development of reliable criteria of biogenicity and instrumentation are crucial for assessing claims of ancient or extraterrean life.

Specific morphological evidence for the domain Archaea is not known from any part of the fossil record. Inference from phylogenies and carbon isotope signals suggest their presence by at least 2.8 Ga.

The oldest fossils well enough preserved to be recognized as members of an extant (still living) group appear in the early Proterozoic. Endolithic cyanobacteria occur in cherts from the 2.15 Ga Belcher Group, Canada. Akinetes (reproductive spores) of cyanobacteria provide evidence of cyanobacteria with cell differentiation in 2.1-Ga cherts from Gabon.

### *Fossil Record of Eukaryotes*

The oldest remains of possible eukaryotes are carbonaceous impressions of coiled ribbon called *Grypania*. Reported in 1.87-Ga iron deposits, they have been interpreted as green algae (based on their large macroscopic size and regular coiled shape), though their biological nature has been questioned. Younger, better-preserved *Grypania* with transversal striations occur in 1.4-Ga rocks.

A moderate diversity of protists (unicellular eukaryotes) represented by acritarchs (organic-walled microfossils with unknown biological affinities) occurs in mid-Proterozoic shales (siliclastic rocks) and demonstrate the evolution of cells with a nucleus and a cytoskeleton. These protists may be plain spheres or have walls ornamented with concentric ridges, irregularly distributed protrusions, or polygonal plates.

Evidence for photosynthetic eukaryotes abounds in the late Proterozoic. Exquisite preservation of 1.2-Ga red algae (affiliated with bangiophytes) documents the evolution of multicellularity, sexual reproduction, and eukaryotic photosynthesis (involving a primary endosymbiosis). Xanthophyte vaucherian algae from the 1-Ga Lakhanda Formation, Siberia, and 0.7–0.8-Ga shales in Spitsbergen indicate the appearance of stramenopiles (a group including diatoms, xanthophytes, and brown algae) and a secondary endosymbiosis (see endosymbiosis in Sec. 4D).

Fossil evidence for fungi has also been recently suggested to date from at least the Neoproterozoic [~750 millions of years ago (Ma)] and possibly the Mesoproterozoic (~1,500 Ma), though the Mesoproterozoic record remains to be confirmed. Fossil testate amoebae (amoebae producing an organic or mineral test or "shell") in 750-Ma rocks document early eukaryotic biomineralization, predation (suggested by the presence of hemispherical holes), and heterotrophy (amoebae eat other organisms). Fungi and testate amoebae provide a firm calibration point for the great clade that includes animals, fungi, and the amoebozoans.

The Phanerozoic is marked by a remarkable diversification of animal life. Through the Proterozoic, there is also a rich record of macroscopic carbonaceous compressions interpreted as multicellular algae, or parts of microbial mats. Following the Neoproterozoic glaciations, the early Ediacarian acritarchs (~630–550 Ma) increase in diversity with the evolution of forms bearing regularly arranged processes. Macroscopic compressions record a high diversity of forms, which include metazoans and multicellular algae. The ~600-Ma Doushantuo Formation of China hosts multicellular green, red, and possibly brown algae, as well as animal embryos, possible stem group cnidarians (*e.g.*, corals), and putative sponges. Ornamented acritarchs disappear just before the Phanerozoic and only large simple sphaeromorphs (sphere shapes) are found. The Phanerozoic starts with a burst in diversity and abundance of protists (including skeletal forms) and marine animals.

## Abbreviations

DNA, deoxyribonucleic acid; Ga, billions of years ago; $K_a$, nonsynonymous substitutions; $K_s$, synonymous substitutions; Ma, millions of years ago; RNA, ribonucleic acid; rRNA, ribosomal RNA.

# Chapter 5. Planet Detection and Characterization (JA)

## 5A. Methods of Planet Detection (CL)

The observation of extrasolar planets can be extremely difficult. The distances involved, as well as the faintness of any planet relative to its parent star, make it nearly impossible to make direct observations. Recent advances in technology and elegant new methodologies make it possible to infer, and sometimes even directly see, these distant objects (Fig. 5.1).

### *Radial Velocity (RV) (aka "Doppler Searches")*

Extrasolar planets have been observed indirectly by tracking the RV of stars. Every star with planets is affected by their mass. As planets orbit their parent star, they periodically pull it back and forth. Movement toward and away from Earth changes the nature of the light that we receive from the star in question. These "Doppler shifts" contract and expand the wavelength of light waves, which makes their light bluer or redder when observed than it was when emitted. (Wavelength contraction corresponds to increased frequency and bluer light. Wavelength expansion corresponds to decreased frequency and redder light.) RV search programs look for very small, periodic wavelength shifts in the spectra of the parent star (<50 m/s to as low as 1 m/s with current instrumentation). Because the planet's gravitational pull must be large enough to move its parent star significantly, the potential for detection of planets with this method increases with planet mass and decreases with orbital radius. Additionally, the method is sensitive to the inclination of the planet's orbit relative to the observer on Earth. Because we are looking at the effects of the star moving closer and farther, we would prefer to be close to the plane of orbit. Only the radial component of the star's velocity is measured, and the mass of the companion object is known only to within a factor of *sin i*, where *i* is the inclination angle of the planet's orbit as seen from Earth. (If the orbital inclination is not measurable by other means, a lower limit may be assigned.)

RV searches provide estimates of the planet's mass, eccentricity, and average distance from its host star. They have been, by far, the most successful of the search methodologies and have discovered the overwhelming majority of the 200+ currently known extrasolar planets. However, no current or planned RV search program is capable of detecting Earth-mass planets in the habitable zone (see Sec. 5B) around solar-type stars; at 1 astronomical unit (AU) from a 1 solar-mass star, no planet smaller than roughly a dozen Earth masses can currently be detected through the Doppler shift it induces on its parent star.

### *Astrometry*

Astrometry can infer the existence of a planet from periodic motions by the host star. The motion observed in astrometry is the movement of the star in the celestial sphere. Because of the vast distances, these motions are very small—perhaps less than a few microseconds of arc ($1/3{,}600{,}000{,}000^{th}$ of a degree viewed from Earth). Such precision necessitates either space-based observations or the use of adaptive optic systems.

Additionally, the minimum mass of a detectable planet increases directly with stellar distance, meaning that only the nearest few thousand stars can be studied by existing or planned astrometric programs. However, unlike the RV method, the minimum mass of a detectable planet increases for larger planetary orbits, so that with sufficiently large time periods for observations, this technique is potentially sensitive to 1 Earth mass planets orbiting any of several hundred stars nearest to the Sun.

Astrometric methods provide estimates of the planet's mass, eccentricity, and average distance from its host star. When combined with RV data, the inclination angle of the planetary system can be highly constrained, which greatly reduces the uncertainty in the planet's estimated mass.

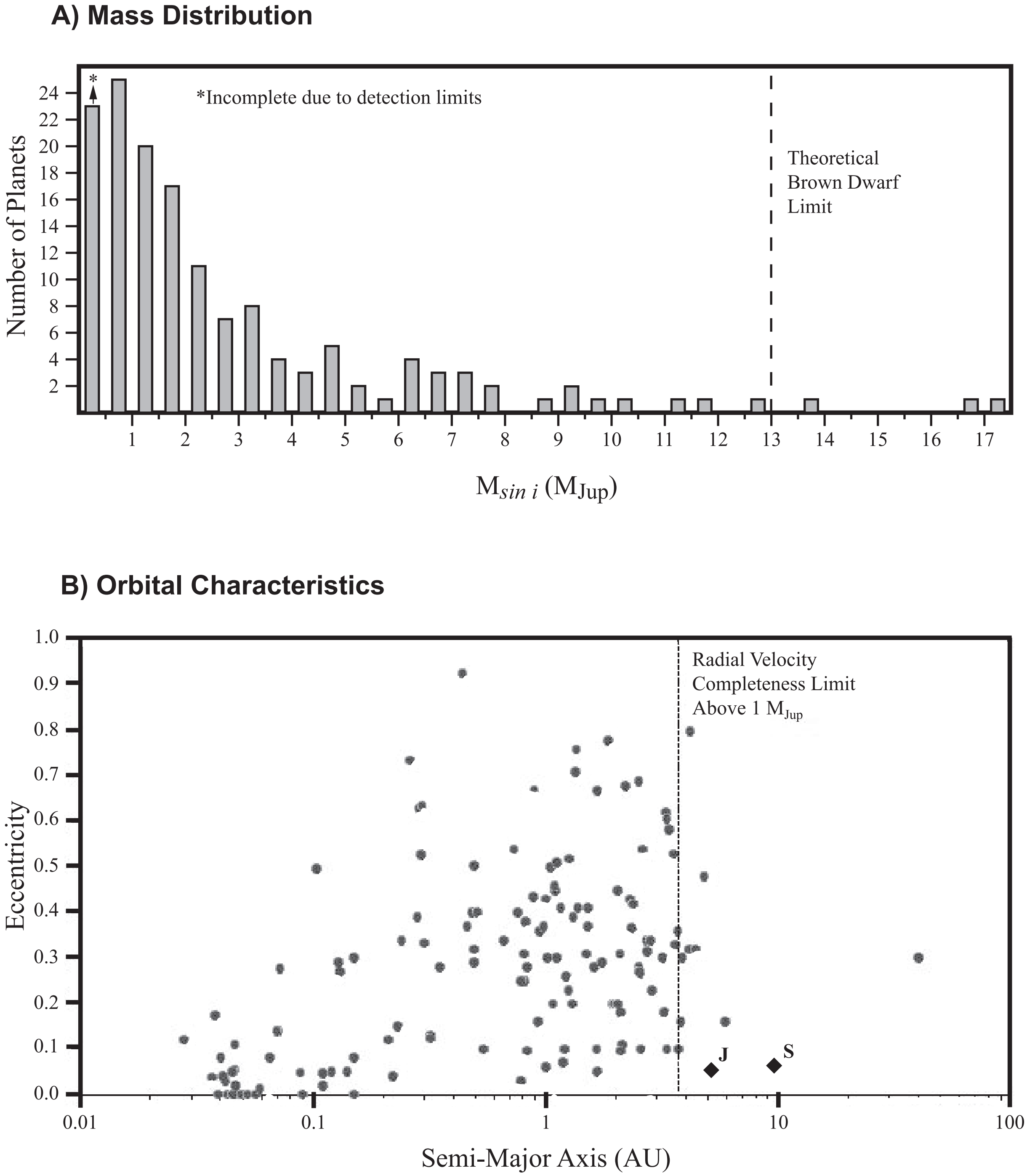

**FIG. 5.1. Known extrasolar planets as of February 2005: (A) mass distribution and (B) orbital characteristics.**

### *Transit photometry*

In planetary systems aligned so that we view them "edge-on," we can detect planets as they transit (or "eclipse") their parent star. Astronomers look for periodic decreases in star brightness as its planet (or planets) temporarily blocks the starlight from reaching Earth. This dimming can be as much as a percent or two of the total light from the star and is readily detectable from even small, ground-based observatories. However, the method is limited to stars with planetary systems "edge-on" to Earth (the orbital plane less than roughly 10° off the line of observation), with planets typically less than ~3 AU away from their parent star.

While these difficulties greatly limit its utility, current precision limits for the photometric method are sufficient to detect low-mass planets in habitable orbits, which makes it unique among planet-search techniques (see the Kepler mission, noted below). Further, details of the transit provide estimates of the planet's radius, which can be combined with RV studies and the inclination angle to provide a highly precise estimate of the planet's mass and density. Follow-up spectroscopy of the host star taken both during and out of a planetary transit can also potentially provide information on the composition of the planet's atmosphere.

### *Direct Imaging*

Direct imaging methods seek to isolate and observe light coming from the planet itself, rather than inferring its existence from effects on its host star. This method offers direct constraints on size and orbital characteristics, and also provides information on the planet's atmospheric composition. Unfortunately, contrast issues between planets and their host stars make this method by far the most technically challenging—even Jupiter-mass planets are a billion times fainter in visible light than their host stars. Strategies to mitigate this have focused on combinations of (a) blocking the light from the host star and (b) conducting observations in the infrared (IR), where the contrast between planet and star approaches a minimum value of $10^6$. These approaches have seen some measure of success; in September 2004 researchers using a near-IR camera on the Hubble space telescope reported the detection of a 3–7 Jupiter-mass planet in a 55 AU orbit around a nearby brown dwarf. This system, however, with a large planet in a distant orbit around a very faint substellar object represents a best-case scenario for detection—direct imaging of a rocky planet in a habitable zone orbit around a solar-type star remains beyond the capacity of any current astronomical instruments.

### *Current Debates*

How biased are our current observations of extrasolar planets? What is the distribution of extrasolar planets as a function of their stellar properties, such as mass, metallicity, age, and galactic environment?

### *Some Important Research Groups*

There are 59 groups listed as carrying out Extrasolar Planet Search programs on the Extrasolar Planets Encyclopedia website (www.obspm.fr/encycl/encycl.html).

Additionally, there are 18 space-based missions with at least a partial charge to search for and characterize extrasolar planets. Notable among these are the following:

- *Space Interferometry Mission (SIM).* SIM, designed for astrometric searches, will have the capacity to detect Earth-like planets around the nearest several hundred stars. It is planned for launch in 2015.
- *Kepler*. This photometric survey for transiting planets will examine over 100,000 nearby stars continuously (in a region of the sky approximately 100 square degrees) in an attempt to detect transiting planets. This survey is notable for its capacity to detect Earth-mass planets within the habitable zone. It is scheduled for launch in 2008.
- *Darwin and Terrestrial Planet Finder (TPF)*. These two missions are being studied by ESA (Darwin) and NASA (TPF). Both involve a space-based interferometer/coronograph telescope. They are intended to provide direct imaging and spectroscopy sufficient to detect Earth-mass planets within the habitable zone as well as characterize their atmospheres. Their mission represents the most probable observational tools to search for biomarkers such as methane, water, and ozone on extrasolar rocky planets; however, they are both still in the design stage, with no target launch date set.

## 5B. Planet Habitability (AvM)

A habitable planet has traditionally been described as one that can sustain terrean-like life on its surface or subsurface for a significant period of time. However, this definition is constantly being redefined as we hypothesize or discover new environments in which life can sustain itself.

The main requirements are the presence and stability of liquid water over long time periods and the availability of the basic organic building blocks of life (C, H, N, O, P, and S; building blocks; and nutrients; see Sec. 3A). Water is the most fundamental constituent for the assemblage and functionality of life on Earth.

### *Basic Habitability Criterion: Temperature*

The most fundamental requirement for liquid water is a clement mean environmental temperature. Temperatures between 0°C (273 K) and 100°C (373 K) are necessary for pure water to form a liquid at standard temperature and pressure. Pressure and solutes, however, can have a large impact, and life has been observed in liquids between −20°C (253 K) and 121°C (394 K) (see Fig. 6.6 and Sec. 6C). On rocky planets or moons, this environment can be on or below the surface. The temperature of a planetary environment is a function of the balance between heating and radiation into space. In this section, the word "planet" will be used to signify a large rocky body that orbits the central star (a traditional "rocky planet") or another large body in the system (traditionally called a "satellite" or "moon").

#### *Surface environments*

For surface environments, heat comes primarily from the radiation absorbed from the central star (or stars), and cooling is mainly limited by the ability of the atmosphere to retain heat. Stellar heat input will be sensitive to the size of the central star (see Sec. 1A and Fig. 1.2) and the orbital distance. We can estimate an "habitable zone" or range of liquid water for a given star type, planet size, and orbital distance (Fig. 5.2). For a G-type star such as our Sun, the HZ lies between 0.95 AU and 1.37 AU.

The surface temperature can be increased if the incident radiation can be retained by the atmosphere. The "greenhouse effect" occurs when the atmosphere is transparent to light at shorter wavelengths [such as ultraviolet (UV) or visible light] but absorbs light at longer wavelengths (such as IR heat). The shorter wavelength light is converted to heat when it hits the surface and is then retained by the atmosphere, which increases the temperature near the surface. The extent of heating depends on the composition of the atmosphere, and for thick atmospheres of molecules such as carbon dioxide and methane, the surface temperature may be raised significantly. The greenhouse effect may, therefore, extend the habitable zone out to 2.4 AU for G-type stars.

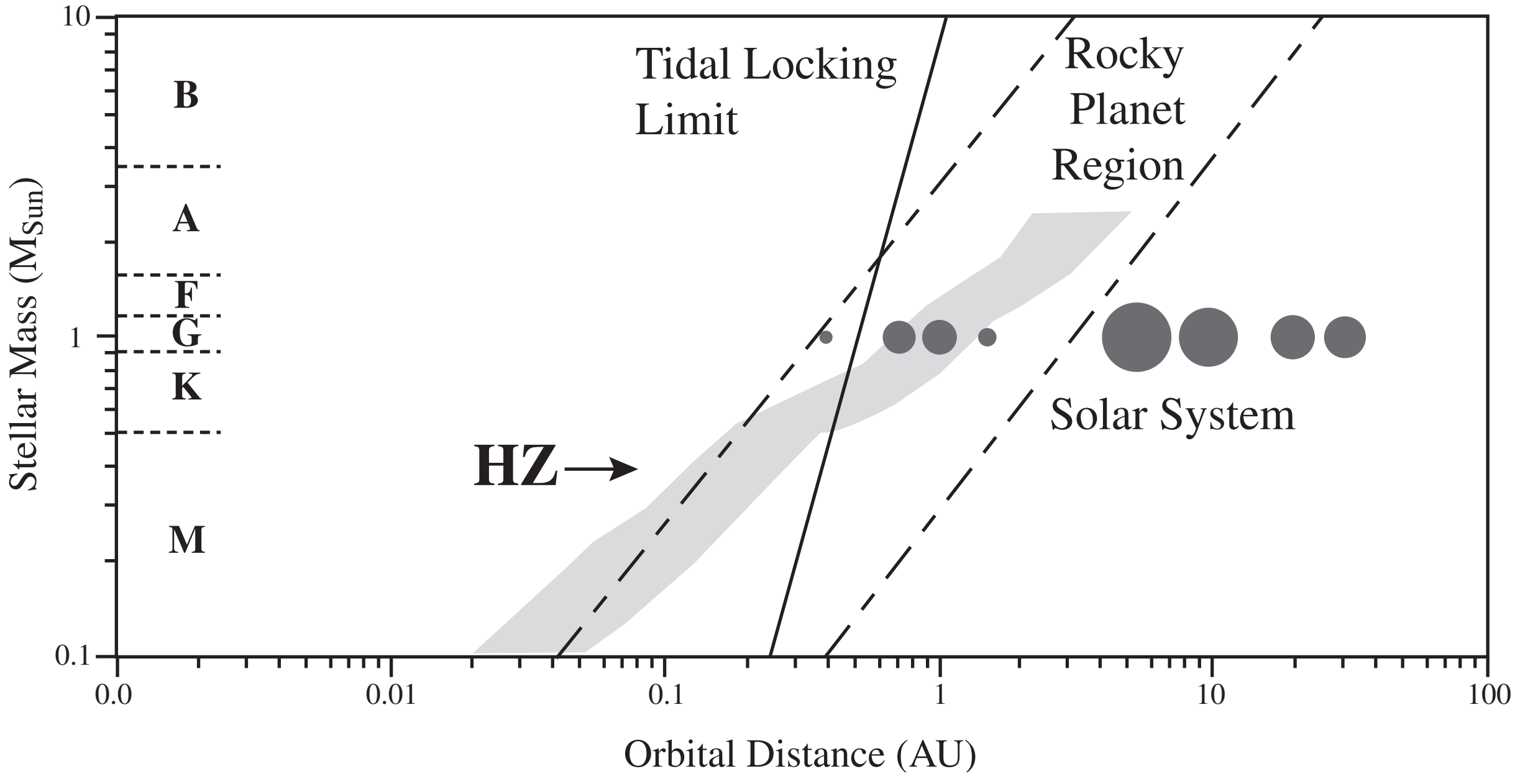

**FIG. 5.2. The habitable zone (HZ) around main sequence stars.**

To complicate matters, changes in stellar radiation output (*e.g.*, luminosity, color), planetary orbit (*e.g.*, inclination, eccentricity), and planetary history (*e.g.*, ice ages, carbon dioxide sinks) may result in global temperature changes over time. It remains unclear to what extent life might tolerate lapses in "habitability," but the term "continuously habitable zone" (CHZ) is often used to denote the region around a star within which planets with stable orbits could remain habitable long enough for complex life to evolve. It is usually defined with respect to some time interval (the CHZ for the Sun over 4.6 billion years is approximately 0.95–1.15 AU), but since the definitions of both "habitability" and "complex life" remain unclear, this terminology can become problematic.

*Subsurface environments*

In subsurface environments, heat comes from within the planet due to internal processes (as on Earth) or an external force such as tidal compression (as on Io). Cooling is limited by insulation from surface layers. Internal heating is initially generated by conduction of latent heat created during the formation of the planet, but additional heating comes from the decay of radioactive isotopes. Radioactive decay makes only a small contribution but may be enough to maintain liquid water in a subsurface layer (a possibility on Mars).

Internal heat can also be generated by tidal forces in orbiting bodies with eccentric orbits (a moon orbiting a planet or a planet orbiting a star). The tidal force (differential gravity, a stretching force) increases and decreases with the body's orbital distance, which causes the internal structure of the orbiting body to compress and expand periodically. The process creates frictional heating, which can be conducted throughout the body and may be enough to maintain a large subsurface ocean (hypothesized for Enceladus—orbiting Saturn—and possibly for Callisto, Europa, and Ganymede—orbiting Jupiter).

The central object will tend to circularize the orbits of closer orbiting bodies (in regions where the tidal field is strong). Thus, the orbiting body needs some additional gravitational effect to maintain an eccentric orbit. For Jupiter's system, this comes in the form of orbital resonances (the orbital periods of Ganymede, Europa, and Io form the ratio 4:2:1) (see Sec. 2A).

*Additional potential habitable environments*

Theoretically, life is possible anywhere liquid water and nutrients exist. Specifically, the surfaces of icy grains at the water sublimation level in the atmospheres of giant planets and the interiors of hydrated or icy small bodies such as asteroids or comets should be considered. The potential for the *in situ* origin of life in these environments, however, is low.

### *Factors That Limit Habitability*

Many features of our Solar System and their effects on habitability have been investigated. It remains unclear, however, whether these details are specific to Earth and terrean life or they can be generalized to all habitable planets.

#### Stellar mass

More massive stars have shorter lifetimes than less massive stars. F-type stars, for instance, have lifetimes of less than 1 billion years, which may limit the probability of life arising on one of their satellites. More massive stars also emit a larger fraction of their light in UV and x-ray wavelengths, which would have a detrimental impact on organic processes not protected by a thick atmosphere. Stars less massive than the Sun give off most of their light at longer wavelengths, which would potentially inhibit biological processes, such as photosynthesis, that provide energy for the biosphere. (See Chapter 1, specifically Fig. 1.2, for properties of stars.)

*Planetary dynamics*

Orbital dynamics and interactions with other planets in the system can affect the habitability of a planet. To maintain a consistent temperature, an Earth-like planet must have a nearly circular orbit, or it will undergo extreme temperature changes. Eccentric orbits also increase the possibility of collisions between planets; an eccentric giant planet can strongly inhibit stable orbits of other planets. All the planets in our system have relatively circular orbits; however, many extrasolar systems contain planets with highly eccentric orbits. Smaller bodies such as asteroids and comets can also affect the formation and survival of life. The composition of these bodies makes them ideal for delivering water (Sec. 2A) and organics (Sec. 2B) during planet formation. On the other hand, bombardment of a planet may kill developing biospheres, a process known as "impact frustration" (Sec. 4D). The possibility of habitable zones around binary and trinary stars remains unclear.

*Presence of satellites*

Earth is unusual with respect to the other rocky planets in our Solar System in that it has one large moon in a circular orbit. This could play a significant role in stabilizing Earth's rotation; the tilt of Earth toward the Sun—its "obliquity"—is relatively stable over very long time periods. Without the Moon, Earth's obliquity could potentially vary drastically over million-year time scales and cause major surface temperature variations.

*Planetary properties*

The internal structure of a planet will have a direct impact on its habitability. On Earth, internal heating mechanisms result in plate tectonics and volcanism, both of which play an integral role in the exchange of materials in the atmosphere and oceans. On Mars, the cessation of plate tectonics may have been critical in the loss of a thick atmosphere and surface water. Additionally, the differentiation of interior layers of a planet can affect energy and material transport. This differentiation causes Europa and Ganymede to have the potential for liquid water layers. Callisto is undifferentiated and does not. (See also Fig. 2.1B for planetary properties.)

*Biological feedback*

Once life forms, its impact on the environment will have profound effects on habitability. It is hypothesized that biological production of methane on early Earth may have been critical for maintaining a high surface temperature and exposed liquid water (see Sec. 2E). Similarly, the rise of oxygen probably triggered global extinctions of anaerobic organisms; however, the presence of oxygen also produced an ozone layer, which provides a barrier against harmful UV radiation (see also Fig. 2.3 and Secs. 2E and 4D). Thus, an oxygenic atmosphere opened the way for aerobic and multicellular life as well as land-based life, which does not have water to protect it from solar radiation.

## 5C. Exploration and Characterization of Mars (JA)

### *Current Conditions and Exploration History*

Mars, the fourth planet from the Sun, is roughly one-half the size of Earth and 1/10th the mass. Its surface area is about equal to that of the continents on Earth. A martian day (called a sol) lasts 24 h 40 min, and a martian year lasts 687 Earth days (667 sols). The tilt of the martian rotational axis (25°) is nearly equal to Earth's; thus, Mars experiences similar seasonal variations.

Mars has retained a thin atmosphere, roughly 6 millibars (6/1000ths as dense as on Earth). The atmospheric composition is 95% $CO_2$, 2.7% $N_2$, and 1.6% Ar, with trace amounts of $O_2$, CO, $H_2O$, and $CH_4$. The thin atmosphere and distance from the Sun (1.52 AU) cause the surface temperature to vary from the $CO_2$ frost point of $\sim$140 K ($-$133°C) at the poles to nearly 310 K (37°C) for brief periods in the southern hemisphere. These radical temperature swings cause the $CO_2$ atmosphere to condense in the polar regions in winter and sublimate in summer; this seasonal

"breathing" causes a nearly 25% fluctuation in atmospheric pressure every year.

Stable liquid water on the surface is unlikely under current Mars conditions. Metastable liquid water (protected against boiling but not evaporation), however, may be possible under a range of present-day conditions.

### Evidence for Water Ice on Mars

Observations of the martian polar regions reveal spectroscopic evidence for water ice outcrops in the southern polar regions within predominantly carbon dioxide ice. Observations of receding ice pits in the southern polar deposits appear to require a water ice foundation; the steep slopes would be unstable if supported by carbon dioxide ice.

Evidence also exists for water ice within the martian regolith. Water ice can be detected by a reduction in the observed number of neutrons of a specific energy emitted. These observations indicate 10–50% water ice by weight in the regolith (loose particles over bedrock). It is unevenly distributed—with higher concentrations near the poles and lower concentrations near the equator. Surface features—such as glacial-like deposits, debris mantles, and polygonal structures—also suggest subsurface water ice.

### Climate Change and Early Mars

Unlike Earth, Mars experiences extreme variations in solar flux at the surface. Aside from changes in the solar output, three factors influence this variation: the planetary tilt (obliquity), the noncircularity of the orbit (eccentricity), and the timing of the planet's closest approach to the Sun (perihelion). Of these three effects, changes in the obliquity are the main contributor to changes in climate. Numerical calculations show that the martian obliquity over the short term can vary from 0° to 60°, and over longer time scales can reach values as high as 80°. Oscillations in the obliquity of up to 15° occur on 150,000-year time scales, and the mean value of the obliquity can shift by 25° or more over 10 Ma. The short-term oscillations may be responsible for recent periods of climate change recorded as variations in polar layered deposits, and longer-term variations may be responsible for more extreme climate variations, including changes in the location of ice on the surface and in the amount of atmospheric $CO_2$.

Mars shows evidence of transient liquid water today on very small scales in the form of gullies. Currently, the main debate in Mars science is whether Mars supported a warmer, wetter climate in the past. Several lines of evidence indicate that Mars has experienced enough warming to produce episodic and prolonged periods of liquid water on the surface. Old features, such as the valley networks that cover the ancient cratered lands of the south, appear to require more extended periods of liquid water flowing at or near the surface. Younger, catastrophic outflow channels near the boundaries of the cratered southlands and smoother lands to the north probably represent episodic floods that may not have required a large change in climate.

Three lines of evidence suggest the presence of liquid water near the surface: cross-bedded, fine-scale layered minerals; mineral sequences that resemble evaporation patterns; and iron-rich mineral inclusions. This evidence has been found at the landing sites for both Spirit and Opportunity missions, which indicates a global phenomenon (see Sec. 7B).

The difficulty arises in producing an environment warm enough to support the conditions that formed these features. Models designed to produce an early Mars warm enough to have the mean temperature above 273 K (the melting point of water) require unrealistic assumptions or very large greenhouse atmospheres. However, proposals for a "cool, wet" Mars allow for warm enough temperatures to get episodic formation of the observed features without requiring the mean temperature to rise above 273 K.

### *Potential Habitats*

Aside from the highly controversial evidence for microfossils and biomarkers in the Mars meteorite ALH84001, there is no conclusive evidence that life exists or ever existed on Mars. Low surface temperatures, extreme radiation, and lack of liquid water on the martian surface are assumed to be too hostile to support life as we know it. The martian subsurface, however, may provide a number of potential habitats. A permafrost layer may be as close as a few meters below the surface, and a water ice/liquid water interface could be as shallow as a few kilometers. Hydrothermal systems could make communication possible between the deep and near subsurface habitats, and provide a potential environment for life.

While Mars may have supported life in the past, one of the biggest arguments against the existence of current life has been the lack of measurable, biologically produced gases in the atmosphere. Recent observation of methane in the atmosphere, however, may provide such support. It should be noted that several other explanations for this—such as hydrothermal outgassing or release from clathrate deposits—are being explored.

### *Exploration History, Strategies, and Plans*

Viking orbiters and Landers 1 and 2 gave the first comprehensive view of the martian surface, which indicated a geologically dynamic place (see Sec. 7B). The landers measured surface temperature, pressure, and wind speeds for nearly 2 martian years at two distant places. They also sampled the atmosphere and measured bulk elemental composition at the surface.

Each lander carried a suite of three experiments, which were carefully designed to detect a specific kind of biochemistry. They looked for the formation and destruction of molecules incorporating $^{14}C$, a radioactive isotope of carbon. The results were consistent with the results expected if life were present, though abiological peroxide chemistry is thought to produce similar results. The ambiguity of these results demonstrates the difficulty of looking for alien life in an alien environment.

Another experiment, using a gas chromatograph-mass spectrometer, failed to detect any evidence for organic compounds in the martian soil. This suggests that terrean-like life has not been recently present on the surface of Mars and the radioactive carbon experiments detected abiological chemistry.

The next generation of Mars exploration began in the 1990s; its focuses are the search for liquid water and exploring strategies for finding life. Current mission goals are orbital reconnaissance (Mars Global Surveyor, Mars Odyssey, Mars Reconnaissance Orbiter, and Mars Express) and exploring the surface using flexible, mobile instrument platforms (Mars Exploration Rovers). Future plans include more targeted, on-site experiments and sample return.

http://www.marsprogram.jpl.nasa.gov/missions

### *Major Mars Missions (Ongoing and Planned)*

- Mars orbital resources:
  - Mars Global Surveyor—high-resolution imagery, altimetry, thermal spectroscopy
  - Mars Odyssey—imaging IR spectroscopy, gamma-ray spectrometer
  - Mars Express—stereoscopic imagery, radar sounding
  - Mars Reconnaissance Orbiter—high-resolution imagery, meteorology/climate mapping, shallow radar
  - Mars Telecommunications Orbiter—planned launch 2009
- Mars surface resources
  - Mars Exploration Rovers—suite of surface exploration tools: stereoscopic imagery, thermal emission spectrometer, alpha/x-ray spectrometer, microimager
  - Mars Phoenix—planned launch 2007
  - Mars Science Laboratory—planned launch 2009
  - Beyond 2009: Mars Sample Return, Deep Drilling, Astrobiology Science Lab, Scout Missions—novel surface exploration strategies.

## 5D. Exploration and Characterization of Europa (SV)

### *History of Europan Exploration*

In 1610, Galileo Galilei and Simon Marius observed Europa and Jupiter's three other large inner moons—the first objects known to orbit a body other than the Sun (or Earth). Subsequent observations have revealed a bevy of smaller moons that also orbit the planet.

Modern space exploration has afforded a closer look. Voyager 1 and 2 (1979) provided detailed pictures of Europa's surface ice and clues of a possible liquid water ocean underneath. From 1995 to 2003, Galileo orbited Jupiter, equipped with a 1,500-mm telescopic camera, a photometric polarimeter for looking at surface texture, near-IR mapping capabilities, UV spectrometers for surface and gas composition, and a magnetometer. Galileo made 84 orbits of Jupiter, with 19 passes near Europa, coming closer than 1,500 km to the surface. (See Sec. 7B for mission details.)

http://www.galileo.jpl.nasa.gov/ (NASA's Galileo Legacy site)

### *Bulk Composition: Thickness of the $H_2O$ Layer*

Europa's radius is 1,565 km ($\pm$ 8 km), and the average surface temperature is 103 K. Gravity and inertial moment at Europa were inferred by tracking the Doppler shift of Galileo's radio signal (due to gravitational acceleration from Europa) during each pass. Europa appears to have an inner rock/metal shell of density greater than 3,800 $kg/m^3$ covered by a layer of partially or entirely frozen $H_2O$ ($\rho \sim 1{,}000\ kg/m^3$) between 80 and 170 km thick. Inertia measurements from four close passes by the Galileo satellite constrain the size of Europa's metal core, if it has one, to no more than half the moon's radius.

Galileo's magnetometer detected a field near Europa's surface, induced by Jupiter's intrinsi-cally generated magnetic field. Europa's magnetic field signature has a strong component near the surface, which suggests an ionic fluid (*i.e.*, salty water) as the likeliest inductive material. A europan ocean, 10 km deep with salinity equivalent to Earth's (~3.75%), would have conductivity sufficient to support the induced field.

### *Surface Spectra: An Indicator of Ocean Composition?*

Non-ice material is visible on Europa where liquid water may have reached the surface, most prominently along linear cracks and chaos (or melt-through) regions. Thus, the non-ice component most likely comes from within the moon; its composition may be representative of the composition of an underlying europan ocean. Near-IR spectra of the non-ice component on Europa's surface appear to match spectra of hydrated sulfate and carbonate salts ($MgSO_4 \cdot H_2O$, $Na_2SO_4 \cdot H_2O$, $CaCO_3$). The match is not unique (other materials fit reasonably well), but sulfate salts are the prime candidates because they are predicted to be the dominant component of an ocean on a moon formed from C1 chondrites, the oldest known meteorites thought to represent the non-hydrogen bulk composition of the Solar System (see meteorites in Sec. 2B).

### *Age and History of Europa's Surface*

Though Europa has not been mapped at high resolution, Galileo achieved pole-to-pole swaths of imaging for most of the moon's surface. Analysis of composite images reveals many features at the 20 km and greater scale. Among these are the previously mentioned cracking features, the wider ($d \sim 100$ km) described as bands, and narrower ($d \sim 20$ km) as ridges. Domes and pits are regions that appear to have been lifted or dropped, respectively. Chaos regions appear on ~20% of Europa's surface. All of the features mentioned above are believed to have hosted liquid water at some time, and non-ice deposits are concentrated at them.

Craters are few in comparison with those on the surface of Ganymede and Callisto, which indicates that Europa has been resurfaced within 30–70 Ma. Stratigraphic relationships (temporal sequences inferred from analysis of overlying features) indicate that cratered regions and the broadest bands are the oldest features on Europa's surface. Chaos regions are the youngest. Apparently, melt-through features are a recent mode of alteration, whereas cracking features have formed continuously.

The dominance of extensional (as opposed to compressional) features on Europa's surface is consistent with a cooling and, thereby, expanding and thickening shell that overlies a liquid water ocean. The discovery of compressional features in future geologic surveys, however, would call into question the inference of a cooling and thickening shell. Changes in the thickness of Europa's ice shell may be expected due to orbital evolution of the moon.

### *Can Seafloor Heat Cause Surface Melt-Through Features?*

Is the ocean fully convecting? Does material from the seafloor reach the ice ceiling? Can a heated hydrothermal plume—a parcel of water that is less dense than the surrounding ocean—transport enough energy to the ice ceiling to cause melt-through? If so, there should be regular exchange of materials between the surface of Europa and the underlying ocean.

If the ocean is sufficiently dilute and the ice shell thinner than ~20 km, the peculiar thermal expansion properties of water would prevent convection; near the ice surface ceiling, the water would be warmer than the water below. The presence of a stagnant layer would limit physical interaction between ocean and ice lid. It is also possible that a fully convecting ocean would have little variation in temperature and the whole ocean would be close to freezing.

### *Ocean Composition*

A model of Europa's formation by the accretion of primordial meteors predicts that the resulting ocean would be initially warm and draw salt from Europa's rocky component. Subsequent cooling would drive the ocean to a saturated state, with perhaps kilometers of salt precipitate settling to the seafloor.

Because of escape of light molecules ($H_2$, $CH_4$, and CO) on formation, the mantle is expected to be oxidized at the present (see redox in Sec. 2C). Sulfate ($SO_4^{2-}$) and perhaps sulfite ($SO_3^{2-}$), but not sulfide ($S^{2-}$), are believed to be the dominant salts in the ocean.

### *Life in Europa's Ocean?*

Hydrothermal systems may persist in Europa's ocean and represent a potential source of energy for life, but compositional or redox (chemical energy) conditions may limit the extent. Acidity or low temperature may create an insurmountable barrier to chemical reactions necessary for metabolism. On the other hand, bombardment of Europa's surface by solar wind particles may instill Europa's ice surface with chemical energy, which, filtered to the ocean, could power life.

## Abbreviations

AU, astronomical unit; CHZ, continuously habitable zone; IR, infrared; RV, radial velocity; SIM, Space Interferometry Mission; TPF, Terrestrial Planet Finder; UV, ultraviolet.

# Chapter 6. Diversity of Life (AnM)

This chapter summarizes the current understanding of life on Earth and highlights research that is challenging the assumptions of life's limitations; namely, the search for life in the harshest environments on earth where extreme physical and chemical conditions suggest the existence of life is unlikely. The chapter begins with a presentation of biodiversity (Sec. 6A), and then moves from biological to metabolic diversity and the variety of mechanisms used to harness energy (Sec. 6B). Finally, Sec. 6C forms an overview of the limits of life on Earth with a brief introduction to the organisms that live at or near the theoretical boundaries of life.

## 6A. Biodiversity (OJ)

Recent advances in molecular biology have unveiled a remarkably diverse biological community, which nonetheless shares common ancestry. Modern molecular methods that compare conserved deoxyribonucleic acid (DNA) sequences common to all life indicate that Earth's entire biota can be represented by a single family tree (see Sec. 4C and Figs. 6.1–3). Current approximations place the number of species inhabiting the Earth at 3–30 million, though only 2 million have been identified. As each new species is identified and its characteristics are defined, astrobiologists gain another data point to use in the search for life beyond Earth.

Everything we know about life in the universe comes from our understanding of Earth. It is fair to say that each species lost to extinction limits our ability to understand the universe and the potential for extraterrean life. Earth conservation is truly conservation of our knowledge of life in the universe.

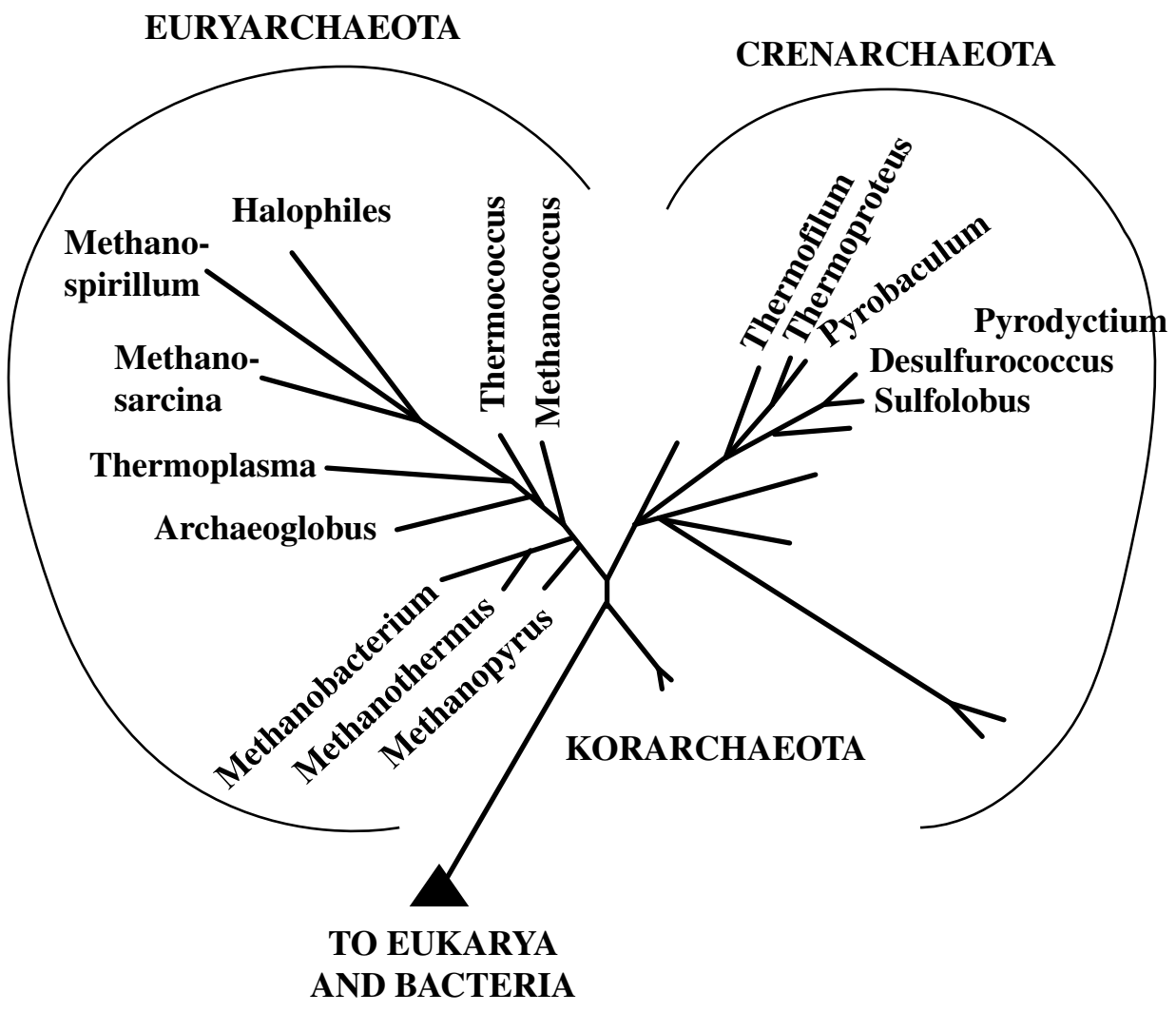

**FIG. 6.1. Tree of Life: Archaea.**

### *Diversity Studies—History and Techniques*

Current classification schemes and models of diversity differ radically from early approaches, which were based on physical traits (phenotypes). Carolus Linnaeus (1707–1778) developed the first systematic approach to identifying organisms using "binomial nomenclature." In this system, every organism is given a two-part name based on categories of similarity, placing organisms into a genus and species. Debate ensued over the best system of naming, however, and with Charles Darwin's discovery of evolution by natural selection, taxonomists came to believe that classification should reflect the history of life. The classification system was adapted into family trees or phylogenies (see Sec. 4C).

In 1959, Robert Whittaker advanced a classification system that arranges organisms into a hierarchical taxonomic scheme, where the largest categories are called kingdoms. The five kingdoms were called Plantae (plants), Animalia (animals), Fungi, Protoctista (one celled eukaryotes and their immediate multicellular descendants), and Monera (all modern prokaryotes). Beneath each kingdom, life was further subdivided into Phyla, Class, Order, Family, Genus, and Species.

More recent advances in molecular phylogenetics (see Sec. 4C) have again reshaped taxanomic classification. The result is a tree of life that contains three primary lines of descent referred to as "domains" by Carl Woese: Bacteria, Archaea, and Eukarya. Four of the five previously recognized kingdoms belong in the domain Eukarya. Organisms with nuclei (an internal, membrane-bound organelle containing the cell's genetic material), which include all multicellular organisms, fall within the Eukarya (literally meaning "possessing a good or true nucleus"). The genetic material of the two other domains, called prokaryotes (literally meaning "prior to nuclei"), is centrally located within the cell but not bound by an internal membrane.

By far the most significant result of the new tree of life is recognition of the great diversity of microbial life. Single-celled organisms have occupied Earth for far longer (3.4–3.9 Ga) than their macroscopic cousins, the complex multicellular eukaryotes (0.6 Ga, see Fig. 2.2), and are far more abundant and varied. Within prokaryotes, smaller categories (called kingdoms, phyla, or divisions depending on the school of thought) often represent greater molecular and metabolic diversity than that found in plants or animals. Archaea and Bacteria are now known to contain the overwhelming majority of metabolic diversity on Earth. Members of the two groups are capa-ble of utilizing virtually every available source of chemical energy to fuel their biological needs (Sec. 6B).

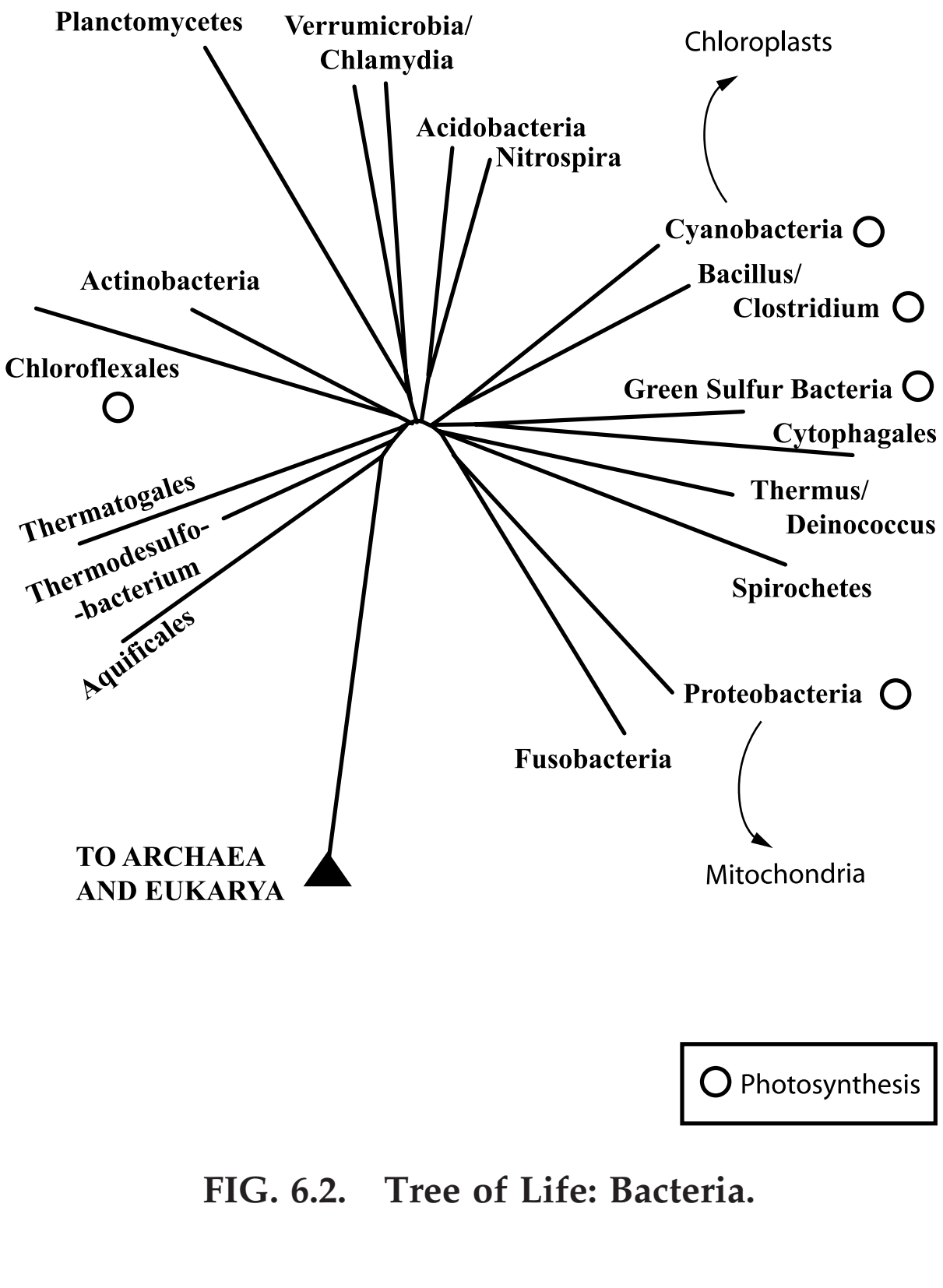

**FIG. 6.2. Tree of Life: Bacteria.**

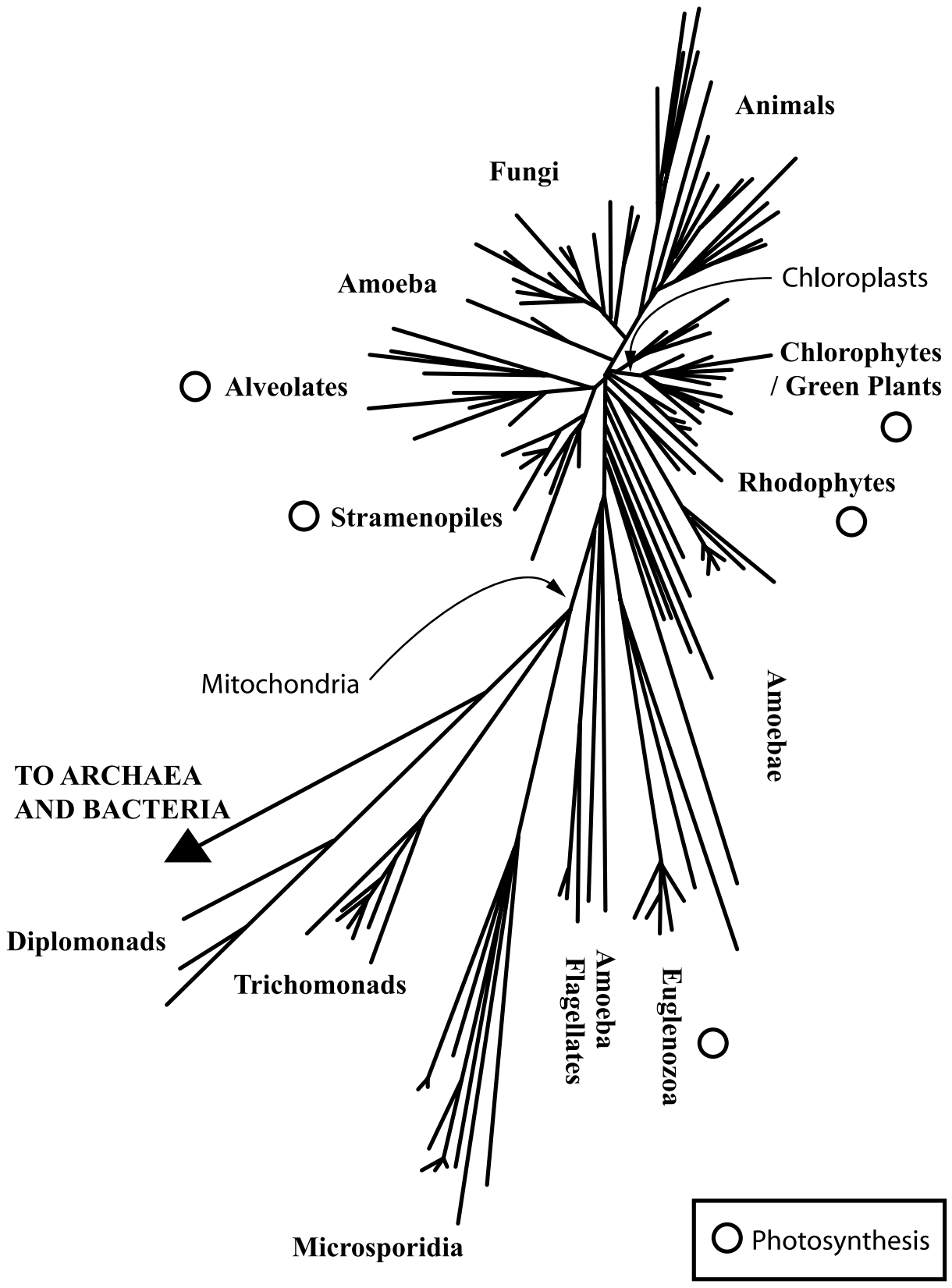

FIG. 6.3. Tree of Life: Eukarya.

The microscopic size and limited morphology of prokaryotes and many eukaryotes requires the use of molecular and chemical data to define groups at all levels. The domains are officially defined using ribosomal RNA sequences. A number of additional properties are characteristic, if not always definitive, of each group (Fig. 6.4). In addition, a large majority of prokaryotes are not yet cultivated, which makes the recognition of their diversity totally dependent on molecular methods.

### *Diversity of Extant Life*

Sequence-based diversity studies indicate that the bulk of Earth's biodiversity belongs to Bacteria and Archaea. Microbial life encompasses over 90% of genetic diversity, which accounts for essentially all metabolic diversity and an overwhelming majority of the total biomass on Earth. There are currently over 40 phylum level divisions within Bacteria, over 20 within Archaea, and more than a dozen that describe single-celled eukaryotes. Within the new framework, Animalia, Plantae, and Fungi now appear as a relatively young, yet highly branched twig on an ancient tree. It remains true, however, that eukaryotes are much more diverse than prokaryotes at the morphological and behavioral scale.

### *Modern Techniques and Perspectives*

The field of genomics aims to understand the evolutionary history and the metabolic and physiological potential of organisms from their DNA. Modern advances in DNA sequencing technology have made it possible to sequence genomes (an organism's entire set of genetic information) in one day (for some prokaryotes) or one week (for some eukaryotes). The resulting sequence data can then be sorted into a map of the organism's genome that indicates the location of genes, marker sequences, viral DNA, transposons, and other features. An enormous amount of "putative" information about the function of genes can be inferred from their similarity to known sequences in well-studied organisms.

In spite of all of the advances made possible by modern computing power, much remains to be discovered. Even in the simplest self-replicating organism known, *Mycoplasma genitalium*, at least 100 of 512 genes still have unknown function, and in many relatively well-studied prokaryotes, 50% or more of the genome remains labeled as "hypothetical" or unknown.

**FIG. 6.4. Characteristics of the three domains.** DNA, deoxyribonucleic acid; RNA, ribonucleic acid.

| PROPERTY | TAXONOMIC DOMAIN | | |
|---|---|---|---|
| | *Eukarya* | *Bacteria* | *Archaea* |
| *RNA-polymerase* | complex-multisubunit structure; shares greatest homology with Archaea | relatively simple structure containing few subunits that is highly diagnostic of bacteria | complex-multisubunit structure; shares greatest homology with Eukarya |
| *Amino acid initiating protein synthesis* | methionine | n-formyl methionine | methionine |
| *DNA Structure* | multiple linear chromosomes | circular | circular |
| *Nucleosomes/histones* | present | none | present |
| *Operons* | none | present | present |
| *Transcription factors required* | TATA box | none | TATA box |
| *Cell membrane* | ester-linked phospholipid bilayer | ester-linked phospholipid bilayer | ether-linked phospholipid bilayer or monolayer with branched fatty acids |
| *Ribosome size* | 80S | 70S | 70S |
| *Cell wall* | none (animals), polysaccharide (plants), chitin (fungi) | peptidoglycan (an amino acid-carbohydrate polymer containing muramic acid) in most bacteria | composed of amino acid carbohydrate polymers not containing peptidoglycan |
| *Cell division* | mitosis and/or meiosis | binary fission | binary fission |

Whole community sequencing efforts, known as "metagenomics," represent powerful attempts to understand the community structure and metabolic potential of ecosystems. Investigators treat the genetic component of all organisms in the community as a whole. Rather than sequencing the genomes of particular individuals, they take random samples from the environment and infer properties of the whole community. As yet, these methods provide only a survey of ecosystem diversity in all but the most species-limited environments. Even there, the high number of putative gene functions remains a formidable obstacle.

## *Current Debates*

### *On what basis do we define "species"?*

Differences between binomial taxonomy and new molecular phylogenetics have generated intense debate over the classification of organisms.

Some authors have suggested grouping sequences into "species" by 97% similarity of 16S (or 18S) ribosomal RNA genes. While this generates a useful organizing principle, there are many examples in which such "species" show little congruity with biologically interesting phenotypic and metabolic traits, and others in which phenotypically similar organisms are quite different as judged by molecular methods. The definition also excludes organisms for which there are no molecular data. Lastly, the extent and significance of horizontal gene transfer (see Sec. 4B)—the transfer of genetic material between otherwise unrelated groups—is highly significant in biodiversity studies and contributes as much as 44% of an individual's protein coding genes. Such data make molecular phylogenetics and, in particular, species concepts wide open for discussion.

## 6B. Redox Chemistry and Metabolic Diversity (LM, GD, FS)

Organisms on Earth have evolved to harness energy from a wide range of thermodynamically favorable (energy-yielding) chemical reactions, and the vast majority of this metabolic diversity lies within microorganisms. Despite a variety of metabolic strategies, all terrean life shares common biochemical mechanisms for translating chemical reactions into usable energy. Specifically, all known organisms generate an electrochemical gradient across a cellular membrane (see Sec. 3B). The gradient is used to drive the formation of adenosine triphosphate (ATP), the principal currency for energy storage in all cells (see nucleotides in Sec. 3A).

### *Redox (Oxidation-Reduction) Chemistry*

Chemical reactions that involve electron transfer are central to metabolic processes. Oxidation refers to the loss of electrons (a molecule that has lost electrons is oxidized), and reduction refers to the gain of electrons (a molecule that has gained electrons is reduced). Oxidation must be coupled to reduction—such reactions are also known as redox reactions. (See Sec. 2C for an introduction to inorganic redox chemistry.)

By coupling the oxidation of a fuel (a reduced electron donor) to the reduction of an oxidant (an electron acceptor), energy can be generated for life. The tendency of a chemical species to gain or lose electrons is referred to as the redox potential. Some chemical species have a great tendency to donate electrons and are, therefore, great fuels (*e.g.*, $H_2$). Others tend to gain electrons and so are great oxidants (*e.g.*, $O_2$). Most eukaryotes can only use organic compounds as fuel and $O_2$ as an oxidant, whereas bacteria and archaea are able to utilize a number of alternative fuels and oxidants. [Photosynthetic eukaryotes rely on internal plastids for photosynthesis. Plastids, including chloroplasts, appear to be descended from cyanobacteria engulfed by a eukaryote billions of years ago (see Sec. 4D).] These pseudo-bacteria provide nutrients for the host cell. Theoretically, any fuel can be coupled to any oxidant so long as the redox potential of the fuel is more negative than that of the oxidant (Fig. 6.5).

### *Metabolic Nomenclature*

Every organism can be described on the basis of its metabolism, how it processes energy, and how it makes basic biomolecules (see Sec. 3A). Terrean organisms get the energy necessary to live and reproduce either by using solar radiation (sunlight) or reduced chemical compounds (*e.g.*, reduced sulfur species such as sulfide) available in their environments. While it remains theoretically possible that other forms of energy—such as heat or kinetic energy—could be used, examples of such organisms have not been observed.

An organism is often described by its overall energy metabolism, including energy source, electron source, and carbon source (Table 6.1). Energy can be derived from light or from reduced chemical compounds. Electrons can be taken from organic compounds (containing at least one C—C bond) or from inorganic compounds. Organisms obtain carbon via two major processes: conversion of $CO_2$ (or other single carbon molecules) into organic carbon (autotrophy) or direct uptake of organic carbon (heterotrophy). When an atom such as C or N is incorporated into an organic compound, it is said to be "fixed." "Primary producer" is another word for autotroph and describes any organism that fixes carbon. All other organisms (heterotrophs) build structures using organic molecules synthesized by the primary producers.

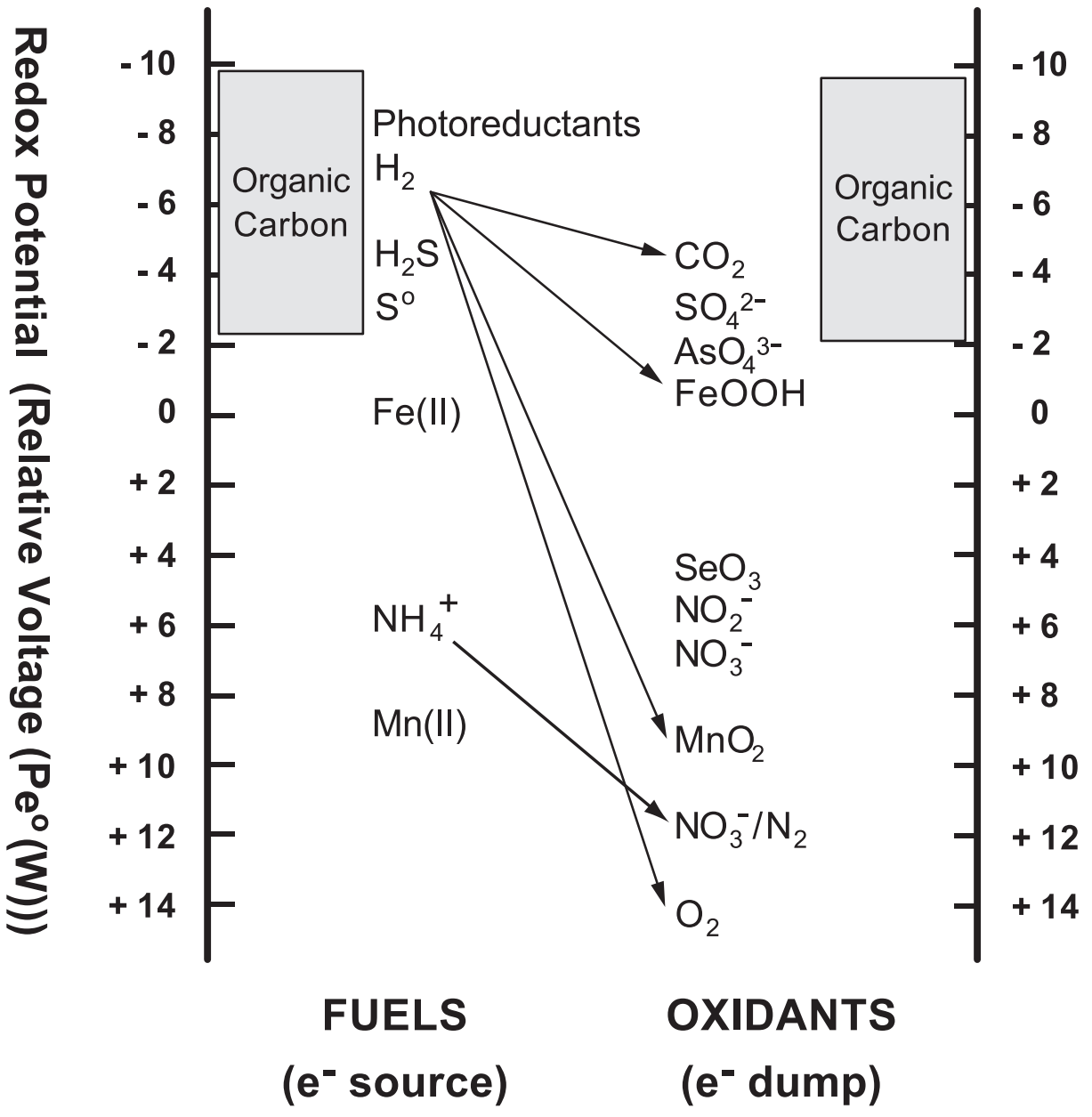

**FIG. 6.5. Redox potentials and life.**

The names can be combined, and the best descriptions involve all three categories. For example, plants and cyanobacteria are photolithoautotrophs, animals are chemoorganoheterotrophs, and iron-oxidizing bacteria are chemolithoautotrophs.

## *Energy from Light*

The ability of organisms to derive energy from light is called phototrophy. (The term "photosynthesis" properly refers only to photoautotrophs that synthesize organic molecules using light energy). More than 99% of the energy budget for terrean life comes from the Sun through phototrophs. Phototrophy involves a complex protein apparatus, which uses light to run a proton pump as well as to store energy in charged molecules (NADH and NADPH). The proton pump creates a reservoir of protons on one side of a membrane. Protein gates allow protons back through the membrane (driven by osmotic pressure), which creates ATP in the process. The charged molecules, including ATP, can later be used to reduce other molecules and power reactions.

Phototrophy has been observed in five groups of bacteria (Fig. 6.2). Oxygenic phototrophs (*e.g.*, cyanobacteria) absorb light at high enough energies to split water, which produces oxygen as a by-product. The remaining phototrophs are anoxygenic. The "endosymbiont hypothesis" explains how phototrophic eukaryotes acquired the ability to photosynthesize by incorporating cyanobacteria (or a close relative) into their cells at some point early in evolution. Over time, these cyanobacteria evolved into modern chloroplasts, eukaryotic organelles in which photosynthesis occurs. Eukaryotic phototrophy, consequently, is oxygenic photosynthesis (see Sec. 4D). No photosynthetic archaeans have been observed.

Overall, the diversity of metabolic pathways for harvesting light energy has only recently been fully appreciated. Some organisms in all three domains of life possess light-mediated proton pumps, which do not directly store energy in charged molecules (bacteriorhodopsin and proteorhodopsin). It remains unclear whether, and to what extent, organisms can function using this as their sole source of energy.

TABLE 6.1. ORGANISM DESCRIPTION BY OVERALL ENERGY METABOLISM

| *Energy source* | *Electron source* | *Carbon source* |
|---|---|---|
| Light: phototroph | Inorganic: lithotroph | Inorganic: autotroph |
| Chemical: chemotroph | Organic: organotroph | Organic: heterotroph |

### *Energy from Chemicals*

Those organisms that do not derive energy from sunlight use the energy locked in chemical bonds (or available electrons) to fuel metabolism and biosynthesis. Such "chemotrophs" obtain energy from the oxidation of organic compounds (chemoorganotrophy) or inorganic compounds (chemolithotrophy).

Chemoorganoheterotrophic metabolism occurs in each of the three domains of life and is employed by all animals, including humans. Heterotrophs acquire the organic molecules (*e.g.*, glucose) necessary for fuel by predation (killing other organisms), scavenging (eating decaying organisms), or by using the metabolic byproducts released by living organisms (*e.g.*, photosynthate from algae). In aerobic heterotrophs, organic compounds ("food") provide a rich source of reducing power used to pump protons and create energy storage molecules such as NADH. As in photosynthesis, ATP formation occurs by the movement of protons ($H^+$) through a membrane. In respiration, electrons pass from the food, through a series of intermediate molecules, to the terminal acceptor, oxygen ($O_2$). Alternatively, some organisms use nitrate ($NO_3^{2-}$), sulfate ($SO_4^{2-}$), or reduced metal species (*e.g.*, iron and manganese oxides) as an electron sink in anaerobic conditions.

Many single-celled, anaerobic heterotrophs, including yeasts and many species of bacteria, use fermentation with organic molecules acting as both electron donor and electron acceptor. In every major ecosystem on the planet, metabolism by both aerobic and anaerobic heterotrophs plays a substantial role in the cycling of organic matter. Even synthetic molecules only produced by humans in the last century represent a source of food for some species (*e.g.*, *Dehalococcoides ethenogenes*, which removes halides from ethene).

In contrast to chemoheterotrophs, chemolithotrophs, which are restricted to the domains Bacteria and Archaea, obtain energy by oxidizing inorganic compounds such as hydrogen ($H_2$), sulfide ($HS^-$), or ferrous iron ($Fe^{2+}$). As in chemoorganoheterotrophy, ATP formation is coupled to the transfer of electrons from donor compounds to a terminal oxidant. Typically, this is oxygen, though some bacteria can use nitrate, sulfate, and ferric iron ($Fe^{3+}$). Most chemolithotrophs use captured energy to fix $CO_2$ into biomass. These organisms are called chemolithoautotrophic or "chemosynthetic."

Chemosynthetic bacteria and archaea inhabit some of the most inhospitable environments on Earth, including the acidic waters leaching from mines, the deep (>1,000 m) subsurface, and the hot, metal-rich fluids emanating from deep-sea hydrothermal environments. In most instances, chemosynthetic microorganisms are free-living in the water column or attached to surfaces. But in some habitats, such as hydrocarbon seeps and hydrothermal vents, chemosynthetic bacteria also live directly within the cells of marine invertebrates; one example is the giant tubeworm *Riftia pachyptila* that colonizes vents along mid-ocean ridges in the Pacific. In these symbiotic associations, the host invertebrate facilitates access to the substrates (sulfide, oxygen, and $CO_2$) that fuel chemosynthesis by the bacteria. In return, the bacteria fix carbon used by both bacteria and host. The occurrence of chemosynthetic microbes (in both symbiotic and free-living forms) in extreme environments has recently attracted much interest from astrobiologists owing to the potential for chemically analogous habitats to occur and support life on other planets.

## 6C. Life in Extreme Environments (AnM)

### *Life on Earth*

Scientists are reassessing long-held assumptions that life cannot exist in certain harsh environments. The boiling waters of Yellowstone hot springs, ice systems in Antarctica, and the highly alkaline waters of Mono Lake are examples of environments previously considered uninhabitable but are now known to yield life. The study of microbes found in these and other severe environments has deepened the understanding of microbial tolerance for extreme conditions and broadened our understanding of life's origin. It is hypothesized that microbes found in high-temperature environments such as hydrothermal vents were the springboard from which all life has evolved. It is these microbes that form the root of the family tree discussed earlier in this chapter.

Speculation of extraterrean life is driven by the knowledge of life on Earth. That is, life in extreme environments analogous to those on other planets informs all hypotheses for extraterrean life. As scientists explore the limits of life on Earth, the possibilities for life on other planets is redefined.

### *Defining the Environmental Limits of Life on Earth*

Terrean life, as we currently understand it, requires at least three fundamentals to survive: a liquid medium (*e.g.*, water or brine), an energy source for metabolism, and a source of nutrients to build and maintain cellular structures (see also Sec. 3C). Of course, where these fundamentals are found one doesn't always find life. Other critical physical and chemical forces impact the ability of life to survive.

Scientists are increasingly seeking life in harsh environments where the subsistence of life is thought questionable or even impossible. Exploration in these extreme environments is pushing the preconceived boundaries of life. And wherever the definition of life's limits is being redefined, astrobiologists are keenly looking on.

The latest definition of limits for biological activity is outlined in Fig. 6.6. This definition accounts for discoveries of microbes in harsh environments—those of high and low pH, extreme temperature, or low oxygen for example—where previous understandings of the limits of biological activity suggested subsistence was not possible. Scientists have learned that life exists at an incredible range of physical-chemical extremes, including pH values ranging from near 0 to greater than 11 and pressures more than 1,000 times greater than the average atmospheric pressure at sea level. (It is important to note that the limits defined in the table are based on activity and not mere survival. For instance, prokaryotes can survive in temperatures much colder than −20°C but they are not necessarily metabolically active at those temperatures.)

Defining just how extreme an environment can be and still sustain life is critical to astrobiology. Identifying limits for growth and organism survival helps guide the search for life, both on Earth and elsewhere in the universe. For instance, the survival of biota in the Earth's atmosphere and their tolerance for desiccation and UV radiation is an example of extremophiles informing the search for life on other planets. Some scientists now believe that life could have arisen in aerosols and past or present life may be found in the atmosphere of Venus.

Microbial diversity defines the ultimate habitability of an environment. In simple terms, the more microbial diversity, the more habitable the environment. Molecular, phylogenetic, and culturing surveys in extreme environments depict the diversity—or lack thereof—of biota that the environment supports. Examples of organisms found in a broad spectrum of physical and chemical conditions are summarized in Fig. 6.7. Also described in the table are specific terms used to classify extremophiles according to their tolerance for conditions such as high salinity (halophile) and high pressure (piezophile).

**FIG. 6.6. The limits of known life on Earth.** UV, ultraviolet. Table adapted from Marion *et al.* (2003) and references therein.

| *Factor* | *Environment/Source* | *Limits* |
|---|---|---|
| High Temperature | Submarine Hydrothermal Vents | 110 to 121°C |
| Low Temperature | Ice | -17 to -20°C |
| Alkaline Systems | Soda Lakes | pH > 11 |
| Acidic Systems | Volcanic Springs, Acid Mine Drainage | pH -0.06 to 1.0 |
| Ionizing Radiation | Cosmic Rays, X-rays, Radioactive Decay | 1,500 to 6,000 Gy |
| UV Radiation | Sunlight | 5,000 $J/m^2$ |
| High Pressure | Mariana Trench | 1,100 bars |
| Salinity | Low Temperature Systems | $a_{H2O} \sim 0.6$ |
| Desiccation | Atacama Desert (Chile), McMurdo Dry Valleys (Antarctica) | ~60% relative humidity |

### *Size Limits and Life*

The controversial report of possible microfossils in the Martian meteorite ALH84001 highlights the importance of understanding cellular size limits. Typical molecular and biochemical analyses are useless on fossilized molecules like those found in ALH84001. Without the ability to examine the viability or biological framework of these "cells," the physical construct has to be assessed. The size of the fossil, therefore, becomes a critical tool in defining the character of life.

Ranging in length from 10 to 200 nm, the diameter of the microfossils is smaller than most, or perhaps all, modern prokaryotes and may be too small to support a viable cell. Current theoretical estimates of the minimal diameter of a viable cell (bacterial or archaeal) range from 250 to 300 nm, with a minimum of approximately 250–450 essential genes. A number of factors impose significant constraints on minimal cell size, including the size of the genome required to encode essential macromolecules, the number of ribosomes necessary for expression of the genome, the number of proteins needed for essential functions, and physical constraints (such as membrane width).

An increasing number of studies report evidence for nanobacteria and nanoarchaea—organisms falling at or below the theoretical lower size limits. Nanobacteria have been linked to serious health problems, including kidney stones, aneurysms, and ovarian cancer. A thermophilic nanoarchaea, *Nanoarchaeum equitans*, was isolated from hydrothermal vents in 2002. This symbiont has a diameter of 400 nm, and its genome is only 490,885 nucleotides long.

**FIG. 6.7. Extremophiles.** UV, ultraviolet. Table adapted from Cavicchioli (2002), Rothschild and Mancielli (2001), and references therein.

| *Factor* | *Class* | *Defining Growth Condition* | *Example Organisms* |
|---|---|---|---|
| High Temperature | Hyperthermophile | > 80°C | *Pyrolobus fumarii* |
| | Thermophile | 60 to 80°C | *Synechococcus lividis* |
| Low Temperature | Psychrophile | < 15°C | *Psychrobacter*, *Vibrio*, *Anthrobacter* |
| High pH | Alkaliphile | pH > 9 | *Natronobacterium*, *Bacillus*, *Clostridium paradoxum* |
| Low pH | Acidophile | pH < 5 (typically much less) | *Picrophilus oshimae/torridus*, *Cyanidium caldarium*, *Ferroplasma* |
| Radiation | - | High ionizing and UV radiation | *Deinococcus radiodurans*, *Rubrobacter*, *Pyrococcus furiosus* |
| High Pressure | Barophile | High weight | Unknown |
| | Piezophile | High pressure | *Pyrococcus* sp. |
| Salinity | Halophile | 2 to 5 M NaCl | *Halobacteriaceae*, *Dunaliella salina* |
| Low Nutrients | Oligotroph | *e.g.*, < 1 mg $L^{-1}$ dissolved organic Carbon | *Sphingomonas alaskensis*, *Caulobacter* spp. |
| Oxygen Tension | Anaerobe | Cannot tolerate $O_2$ | *Methanococcus jannaschii* |
| | Microaerophile | Tolerates some $O_2$ | *Clostridium* |
| Chemical Extremes | - | Tolerates high concentrations of metal (*e.g.*, Cu, As, Cd, Zn) | *Ferroplasma acidarmanus*, *Ralstonia* sp. CH34 |

Studies of nanobacteria and nanoarchaea have proven contentious, and many argue that nanosized particles cannot harbor the components necessary to sustain life. However, it is well accepted that the early forms of life were far simpler than the simplest free-living organisms presently known. It is hypothesized that Earth's primitive life forms used RNA as their sole hereditary and catalytic material (see Sec. 3B) and could have been as small as 50 nm in diameter. Thus, the minimum size of cells on present-day Earth may not be applicable in setting limits for biological detection in extraterrean environments such as Mars and Europa.

### Adaptations to Extreme Conditions

Microorganisms inhabit some of the most severe environments on Earth. These organisms have evolved mechanisms that enable survival within their respective environments—mechanisms that resist freezing, desiccation, starvation, high-levels of radiation, and many other environmental challenges. These mechanisms include:

- Keeping extreme conditions out of the cell (*e.g.*, spore formation)
- Quickly addressing threats to the cell (*e.g.*, pumps to reduce intracellular metal concentrations)
- Altering internal chemistry (*e.g.*, adjusting cellular concentrations of solutes to minimize desiccation at high salinities)
- Protein or membrane alterations (*e.g.*, high proportions of fatty acids in thermophiles making them more stable in hot conditions), and
- Enhancing repair capabilities (*e.g.*, accelerated DNA damage repair under high levels of radiation).

### Extreme Environments on Earth as Analogues to Possible Extraterrean Habitats

Astrobiologists study harsh earthly environments that possess features similar to those thought to exist on other planets and moons. Study of these environments informs their hypotheses of extraterrean habitats and refines the strategies applied in exploration of the Solar System. For instance, surface temperatures on nearby planets vary from more than 400°C on Mercury and Venus to below −200°C in the Kuiper Belt (see Fig. 2.1B and Secs. 2A and B). Only the harshest environments on Earth compare to these extreme temperatures.

### Europa

Characteristics of ice-covered Lake Vida and Lake Vostok in Antarctica are thought to be similar to the thick (1–200 km) ice layer of Europa's saline ocean (see Sec. 5D). A 4-km-thick ice sheet covers Lake Vostok while a thinner, less than 19-m sheet, covers Lake Vida. Microorganisms have been found in the deep ice of both lakes, which illuminates the potential for life in Europa's ice cover.

Additionally, vents similar to hydrothermal vent systems on Earth may occur on the ocean floor of Europa (see Sec. 5D). On Earth, dynamic temperature and chemical gradients form as subsurface water mixes with seawater in the vent chimneys. These vents support an ecosystem in which chemolithoautotrophs (see Sec. 6B) convert reduced chemicals and inorganic carbon into energy and biomass. The vents sustain dense communities of clams, worms, shrimp, crabs, lobsters, fish, and other animals.

The probability of photosynthetic life on the surface of Europa is particularly low, given the extreme levels of radiation. The discovery of life capable of processing energy chemosynthetically, however, opens the possibility that life could be found in the subsurface.

### *Mars*

The structure of the martian subsurface is believed to possess similarities to Earth's subsurface. The discovery of chemosynthetic organisms deep within Earth's crust implies the possibility that organisms will be found below Mars' surface. Permissive subsurface temperatures and the possibility of liquid in rock pores suggest that Mars could support lithotrophic organisms like those found on Earth.

The subsurface of Mars is also likely to contain significant amounts of water ice (see Sec. 5C). Documentation of life in permafrost on Earth and permafrost's ability to preserve molecular signals of ancient life (millions of years old) suggest that the search for past or present life on Mars should include exploration of permafrost. Organisms isolated from Siberian permafrost show growth at −10°C and metabolism at −20°C, temperatures within the range seen at high latitudes on Mars at obliquities greater than ~40°.

## *Current Debates*

### *Is the word "extremophile" a misnomer?*

Extremophiles inhabit environments considered too harsh for humans. Likewise, conditions suitable for humans are too harsh for survival of many extremophiles. Should the terminology be modified to be less anthropocentric?

### *Are extremophiles relevant to the origin of life?*

The last common ancestor at the base of the phylogenetic tree of life is thought to be a thermophile (see *Some like it hot . . .* in Sec. 2D). If that is the case, life on Earth may have originated in a hot environment, such as hydrothermal vents. Alternatively, life may have originated in a more neutral environment but only thermophiles survived the warm temperatures that resulted during a major asteroid collision with Earth.

*Must an organism survive* continual *exposure to an extreme condition to be called an extremophile?*

For instance, should an organism that lives within ice in spore-form be called an extremophile if it is only withstanding the harsh environment and not necessarily thriving within it?

## Abbreviations

ATP, adenosine triphosphate; DNA, deoxyribonucleic acid; NADH, nicotinamide adenine dinucleotide (reduced); NADPH, nicotinamide adenine dinucleotide phosphate (reduced).

# Chapter 7. Science in Space (LM)

## 7A. Space Biology (JoR)

While it is unknown whether life exists elsewhere in the universe, a tiny fraction of Earth's organisms have been exposed to the harsh conditions of spaceflight. Many specimens that have experienced spaceflight were noticeably changed by the event. Space biology has emerged as a critical component of successful human space exploration, fundamental biology research, and our understanding of the limits of life.

### *Space Biology*

Space biology can be broadly defined as the study of life in the spaceflight environment. Space biologists use a combination of ground studies, specialized hardware, culture systems, and flight experiments to characterize the responses of living systems to lower gravity and increased radiation. By studying the effects of the spaceflight environment on life from Earth, space biologists are able to gain insight into fundamental biological adaptations to gravity and space radiation. This allows astrobiologists to hypothesize about life on other worlds from a perspective that reaches beyond the confines of Earth. In addition, space biology experiments provide critical insight into the engineering requirements for habitats and life support systems. Today, major research platforms designed to accommodate space biology research include the NASA Space Shuttle fleet and the International Space Station (ISS).

### *Environmental Conditions in Space Vehicles*

The conditions inside the Space Shuttle and ISS can be considered extreme. The ISS is a laboratory in free-fall and microgravity dominates (approximately $10^{-3}$ to $10^{-6}$ *g*, as compared to 1 *g* at Earth's surface). This allows researchers (with the use of a centrifuge) to simulate a variety of gravitational forces like those on the Moon and Mars. Although the atmosphere and pressure aboard the Space Shuttle and ISS are similar to ambient conditions on Earth (pp $N_2$ = 78.1 kPa, pp $O_2$ = 20.9 kPa, pp $H_2O$ = 1.28 kPa, pp Ar = 0.97 kPa, pp $CO_2$ = 0.05 kPa, total pressure = 101.3 kPa, T = 76°F; where pp stands for partial pressure), variations in temperature and atmospheric composition may occur due to the lack of normal gravity-induced convection. For instance, warm air rises very slowly in microgravity requiring forced-air convection systems.

In addition to microgravity, organisms are also exposed to high levels of radiation. The intensity of ionizing radiation (including particles trapped in Earth's magnetic field, particles from solar flares, and galactic cosmic radiation) varies with spacecraft position. Over the course of a 6-month mission, ISS astronauts receive about 160 milliSeiverts of ionizing radiation during solar minimum, the period when the sun has the least number of sun spots. For comparison, organisms on Earth receive approximately 2 milliSeiverts/year from background radiation.

### *The Effects of the Space Environment on Life*

Cells appear to possess a tension-dependent infrastructure sensitive to mechanical stimuli. Microtubules and filaments of a cell's cytoskeleton can be altered by vibration and gravity, or the lack thereof. Alteration of the cytoskeleton may cause a cell to respond by producing biochemical signals affecting cell structure and function.

Gravity has been shown to affect metabolism, growth patterns, and relations between organisms. For instance, *Escherichia coli* strains flown aboard seven different space shuttle missions responded to the spaceflight environment with a shortened growth phase, extended reproduction, and a final cell population density approximately double that of ground controls. Movement of nutrients and cellular waste are important elements in biological processes, and it has been suggested that if gravity-dependent convection and sedimentation are absent, metabolites may build up in and around cells in culture. As a result, chemical environmental changes within a cell may result in a new physiological state of sustained equilibrium, *i.e.*, "spaceflight homeostasis."

Biological changes in spaceflight suggest that planetary-scale forces, such as gravity, may play a significant role in the development and evolution of living systems. Microgravity can cause changes in cell shape, alter cell-to-cell contacts, and affect communication. Several different studies reveal possible changes in the growth of multicellular organisms. Even the rate at which deoxyribonucleic acid (DNA) is transcribed to messenger ribonucleic acid (RNA) (see Sec. 3A) can change.

If microgravity has such extensive effects on gene expression, it could ultimately alter all large-scale characteristics of an organism. High gravity during fruit-fly development alters the gravity-sensing system of the adult. On the other hand, some plants seem unaffected by spaceflight. Seeds harvested from *Arabidopsis thaliana* grown in microgravity have been successfully grown on Earth without observable abnormalities.

### *What Happens to the Human Body in Space?*

Short-duration exposure (up to one year) to microgravity is not life-threatening to humans, though the spaceflight environment causes a cascade of adaptive processes that have wide-ranging effects on human biology. The most immediate effect of spaceflight is space adaptation syndrome (SAS), in which the neurovestibular organs that aid in posture, balance, guidance, and movement send conflicting visual and balance information to the brain, which causes symptoms of disorientation, nausea, and vomiting. After an individual adapts to spaceflight for a few days, SAS symptoms disappear, but other changes continue to occur. Body fluids shift toward the head due to the lack of gravity, causing the body to respond as if is there were an overall increase in the volume of fluids. Reduced heart size, decreased red blood cell counts, and reduced numbers of immune cells result in immunosuppression (reduced function of the natural immune response). Weight-bearing muscles and bones that normally function to oppose gravity begin to atrophy. Countermeasures such as exercise mitigate these effects but do not stop the changes entirely. In addition to these microgravity-related problems, space radiation presents risks of cancer, immune system problems, and possible neurological effects. The biological effects associated with spaceflights lasting longer than one year, and subsequent issues of Earth re-adaptation, remain unknown.

### *Biological Experiments in Space*

In general, biological experiments in space assess organisms and tissue cultures for changes in morphology, genetics, physiology, biochemistry, protein composition, and behavior. An experiment performed in space will generally have inflight controls and be repeated on Earth as a "ground control." Organisms are grown in microgravity habitats and in centrifuges (either in space, on the ground, or both) to characterize the biological response to a range of gravitational forces and radiation levels in space and on Earth. The unique environment of spaceflight allows space biologists to vary gravity, a force that has been essentially constant over the span of biological evolution.

### *Major Challenges*

Because access to space is difficult, expensive, and infrequent, our understanding of the biological response to spaceflight progresses very slowly. Experiment operations in space are also greatly limited due to a lack of available crew time, physical space, safety and engineering constraints, and a general lack of spaceflight-rated laboratory tools similar to those found in a typical biology lab. Moreover, very few long-duration experiments have been completed, and opportunities to repeat them are rare.

### *The Future of Space Biology*

The NASA Bioastronautics Critical Path Roadmap has been designed to address major unanswered questions associated with long duration human spaceflight on ISS and possible future lunar and Mars missions. For example, if an explorer on the Moon or Mars suffers a broken leg, would specialized pharmaceuticals be required? Do antibiotics have the same effect in partial gravity as in 1 *g*? Will wound-healing occur at the same rate? Furthermore, can long-term missions alter normally benign microbes in the human (*e.g.*, *Candida albicans*) or other potentially pathogenic organisms known to be present in spacecraft (*e.g.*, *Salmonella typhimurium* or *Pseudomonas aeruginosa*) in a way that affects their ability to cause disease? These health concerns may be especially problematic given the altered human immune system during spaceflight. A great deal of biological research needs to be accomplished to characterize and assess the genetic and structural changes induced by spaceflight.

Gravity has profound effects on life. Spaceflight environments cause biological change, though the underlying mechanisms remain unclear. An understanding of the biological adaptations that occur will help astrobiologists hypothesize about life on other worlds, under different gravitational constraints. It will also aid in the development of the tools, life support systems, and countermeasures required for human exploration of extraterrean environments.

### *Current Debates*

#### *Is gravity a requirement for life?*

Gravity itself may be a critical force involved in the development and evolution of life on this planet and, possibly, others. In some cases, microgravity-like environments are simulated on Earth with a rotating wall vessel (RWV). These culture systems slowly rotate to produce a reduced fluid-shear environment that is somewhat similar to microgravity. The validity of the RWV and other ground-based experiments as useful models for true microgravity is, however, an area of active debate.

Cells may be hard-wired to respond to an external mechanical stress like gravity. Whether this is a consequence of gravity or aids in sensing microgravity is unknown.

## 7B. Planetary Missions (LM)

Between 1962 and 2005, over 100 interplanetary missions have been launched. Despite great technological challenges, 59 have returned images and other data from planets, moons, asteroids, and comets in the solar system. NASA maintains several websites with detailed descriptions and useful tables detailing this information, but a brief overview is provided here as well. The best resource for missions and acquired data can be found at the NASA Space Science Data Center, which maintains links to official mission websites for NASA and available data for both U.S. and foreign missions.

Missions listed in parentheses below section headers indicate all those spacecraft that have returned data on the relevant object. Launch dates are indicated within the text in parentheses, unless otherwise noted. Most interplanetary craft and landers use solar power, though smaller landers and probes use batteries. Outer solar system missions utilize nuclear fission of plutonium, as did the Viking landers.

nssdc.gsfc.nasa.gov/planetary/
nssdc.gsfc.nasa.gov/planetary/timeline.html (complete list in chronological order)
www.solarviews.com/eng/craft2.htm (another detailed chronology)

### *Space Agencies*

NASA, National Aeronautics and Space Administration, United States (1962+)

http://www.jpl.nasa.gov/solar_system/planets/ (missions)
http://history.nasa.gov (history)

RKA, Russian Aviation and Space Agency, Union of Soviet Socialist Republics (1967–1991)
Roscosmos or FSA, Russian Federal Space Agency, Russia (1991+)

http://www.roscosmos.ru/index.asp?Lang=ENG

ESRO, European Space Research Organization, intergovernmental agency (1964–1974)
ESA, European Space Agency, intergovernmental agency (currently Austria, Belgium, Denmark, Finland, France, Germany, Greece, Ireland, Italy, Luxembourg, The Netherlands, Norway, Portugal, Spain, Sweden, Switzerland, and the United Kingdom) (1974+)

http://www.esa.int
http://www.rssd.esa.int/index.php?project=PSA

NASDA, the National Aerospace Development Agency, (1968–2003)
JAXA, Japan Aerospace Exploration Agency, Japan (incorporates NAL, NASDA, ISAS) (2003+)

http://www.jaxa.jp/index_e.html

### *Mercury*

*(Mariner 10, Messenger)*

To date, only one spacecraft has visited Mercury. Mariner 10 (NASA, 1973) swung by the planet three times but was only able to photograph one side of the planet. Messenger (NASA, 2004) is expected to enter Mercury orbit in 2011. ESA and JAXA have slated the BepiColombo orbiters and lander for launch in 2009 and 2012.

### *Venus*

*(Mariner 2, 5, 10, Venera 4–16, Pioneer Venus 1, 2, Vega 1, 2, Magellan, Galileo)*

Venus was the first target for interplanetary missions in 1962, and 22 spacecraft have returned data from the planet. In the 1960s and 1970s, both NASA and the Soviet space agency had extensive programs to Venus. The NASA Mariner program included three Venus flybys—2, 5, and 10 (NASA, 1962, 1967, 1973)—and was an opportunity to gather science data while mastering orbital dynamics. Meanwhile, the Soviet space agency was launching approximately one mission a year and attempting probes, flybys, and landings on the yellow planet. The successes—labeled Venera 4–16 (1967–1983)—constitute the most extensive planetary investigation prior to the current Mars initiative. Eight Venera landers successfully returned data, though survival in the harsh environment (465°C, 94 atm) was limited to a maximum of 127 min. Venera 15 and 16 performed the radar mapping of Venus with 1–2 km resolution.

NASA and the Soviet space agency continued with additional exploratory programs. Pioneer Venus 1 and 2 were launched by NASA in 1978. Venus 1 orbited the planet, while Venus 2 included a series of atmospheric probes. In 1984, the Soviets returned with Vega 1 and 2, both of which dropped balloons into the atmosphere. The balloons survived for 2 days and surveyed weather conditions. NASA returned to Venus in 1989 with the Magellan orbiter, which stayed in orbit for four years and mapped 98% of the sur-face. NASA also acquired data with the Galileo mission (1989), which performed a flyby on the way to Jupiter. Focus has shifted to Mars in recent years, however, in part because of the fact that extreme conditions on Venus make the search for life a somewhat futile endeavor. It should be noted, though, that ESA (Venus Express) and JAXA (Planet-C) plan to revisit Venus in the not-too-distant future.

### Mars

*(Mariner 3, 6, 7, and 9, Mars 2–6, Viking 1 and 2, Global Surveyor, Pathfinder, Odyssey, Spirit, Opportunity, Mars Express)*

Investigation of Mars has involved almost as many missions and far more data-gathering than the Venus programs. NASA and the Soviet space agency flew parallel programs to Mars in the 1960s and 1970s. The NASA Mariner program included three Mars flybys—3, 6, and 7 (1964, 1969, 1969)—and one orbiter—9 (1971). After an initial flyby—Mars 1 (1962, contact lost)—the Soviets did not achieve successful transit to Mars until the early 1970s. Five successful Soviet missions were labeled Mars 2–6 (1971, 1971, 1973, 1973, 1973), including the first lander to return data (Mars 3). In 1975, NASA landed two large research platforms, Viking 1 and 2, at Chryse Planitia (22.697°N, 48.222°W) and Utopia Planitia (48.269°N, 225.990°W). (See also Exploration History in Sec. 5C). After a hiatus of 13 years, the Soviet Phobos 2 spacecraft successfully returned data from Mars orbit, but the attempted landing on the moon Phobos failed.

A new Mars initiative was started by NASA in the 1990s. Launched in 1996, the Mars Global Surveyor orbiter supported Pathfinder, a new generation of lander with a rover. Having proved the concept with Pathfinder, NASA launched Mars Odyssey orbiter (2001) to support two larger rovers, Spirit and Opportunity (2003). Pathfinder set down in Ares Vallis (19.33°N, 33.55°W), Spirit in Gusev Crater (14.82°S, 184.85°W), and Opportunity in Terra Meridiani (2.07°S, 6.08°W). Spirit and Opportunity continue to return data. NASA is also collecting data from orbit by way of the Mars Express (joint project with ESA) and Mars Reconnaissance Orbiter missions.

NASA plans to continue investigation of Mars with the eventual goal of sample return. These missions include Phoenix, Mars Scientific Laboratory, and Mars 2011.

http://marsrovers.jpl.nasa.gov/home/index.html

### Asteroids

*(Galileo, NEAR, Deep Space 1, Hayabusa)*

NASA launched several missions to investigate asteroids starting in the late eighties. The Galileo spacecraft (1989) flew by Gaspra and Ida on the way to Jupiter. In 2001, the NEAR (Near Earth Asteroid Rendezvous, launched 1996) spacecraft achieved orbit of the asteroid Eros and landed a probe on the surface. Deep Space 1 (1998) performed a flyby of 9969 Braille before going on to investigate comets.

The JAXA spacecraft Hayabusa (2003) was the first mission to collect samples on an asteroid surface. It arrived at asteroid 25143 Itokawa late in 2005 and took samples, but suffered a fuel leak shortly thereafter. JAXA hopes to be able to return the spacecraft to Earth in 2010.

The ESA Rosetta (2004) mission plans to pass by asteroids Steins and Lutetia in the coming years, before reaching its target, the comet 67P. Mission managers hope to achieve sample return in 2007. Also, a private company in the US has plans to launch NEAP (Near Earth Asteroid Prospector) before the end of the decade to investigate the asteroid Nereus.

### Comets

*(ISEE 3/ICE, Vega 1, 2, Sakigake, Suisei, Giotto, Deep Space 1, Stardust, Deep Impact)*

The return of comet Halley in 1986 inspired a number of missions to investigate comets. The US spacecraft ISEE 3 (International Sun-Earth Explorer, 1978) flew by the comet Giacobini-Zinner and, later, the comet Halley. Renamed ICE (International Cometary Explorer) in 1991, the spacecraft continued to make observations until 1997. Soviet Vega 1 and 2 (1984), Japanese Sakigake and Suisei (1985), and European Giotto (1985) missions all flew by comet Halley and collected data. Giotto went on to flyby comet Grigg-Skjellerup.

The United States has launched three other cometary missions. Deep Space 1 (1998) failed to rendezvous with comet Wilson-Harrington because of a failed star tracker, but managed a flyby of comet Borelly in 2001, passing through the comet's coma. Deep Impact (2005) has already encountered comet Tempel 1 and recorded the effects of firing a projectile into the comet. Stardust (1999) was designed as a sample collection mission. In 2000, it gathered a sample of interstellar dust; in 2002, it collected dust near comet 5535 Annefrank; and in 2003, it entered the coma of comet P/Wild 2 to take samples. In January of 2006, Stardust return samples landed in Utah. Analysis of the dust particles suggests that comets may be more complex in composition and history than previously expected. One example is the presence of olivine, which is formed at high temperatures and thought not to be present in the outer solar system.

The ESA Rosetta mission (2004) was created to provide an orbiter and a lander for the comet 67P. Rendezvous is expected in 2014. Additionally, NASA is developing the Dawn mission for rendezvous with asteroids Vesta and Ceres (launch in 2006).

### *Outer Solar System*

*(Pioneer 10 and 11, Voyager 1, 2, Galileo, Ulysses, Cassini/Huygens)*

Seven spacecraft have been launched to investigate planets in the outer solar system, starting with the NASA Pioneer missions. Because of the huge distances involved and the distance from the Sun, solar power was judged insufficient to power spacecraft. All use power from redundant radioisotope thermonuclear generators.

http://www.saturn.jpl.nasa.gov/spacecraft/safety/power.pdf (radioisotope thermonuclear generator information for Cassini)

NASA launched Pioneer 10 (1972) and 11 (1973) to investigate Jupiter. Both spacecraft flew by Jupiter and continued onward. Pioneer 11 observed Saturn in a flyby. After passing the Kuiper Belt, both eventually ran too low on power to return signals—11 in 1995 and 10 in 2003. Voyager 1 and 2 (NASA, 1975) flew twice as much mass as the Pioneer missions and began a more comprehensive tour of the outer planets. Voyager 1 visited Saturn and Jupiter, while Voyager 2 visited all four gas giants. Both are passing outside the influence of solar wind and continue to return data. After passing Venus and two asteroids, Galileo (1989) proceeded to Jupiter orbit, dropped a probe into the atmosphere, and flew by Jupiter's four largest moons. In 2003, the orbiter finally succumbed to gravity and dropped into Jupiter's atmosphere. (See also Sec. 5D, for information on Jupiter's moon Europa.)

ESA entered the arena with Ulysses (1990) mission, which performed a flyby of Jupiter before transfer to a heliospheric orbit for investigation of the Sun. ESA also participated in the Cassini/Huygens (1997) mission. Cassini (NASA) briefly passed by Jupiter before arriving in Saturn orbit in late 2004. In addition to observations of Saturn and its moons, it delivered an ESA probe (Huygens) to Titan. Cassini continues to return data.

To date, no missions have visited Pluto/Charon or the Kuiper belt, though the NASA New Horizons mission (2006) is scheduled to reach Pluto in 2015.

## 7C. Planetary Protection (PP) (MR, LB, JRm)

### *Introduction*

PP refers to the practice of preventing human-caused biological cross contamination between Earth and other Solar System bodies during space exploration. PP aims to avoid the transport of microbes that could cause irreversible changes in the environments of celestial bodies (planets, moons, asteroids, comets) or jeopardize scientific investigations of possible extraterrean life forms, precursors, and remnants. In practical terms, two concerns predominate: avoiding (1) *forward contamination*, the transport of terrean microbes on outbound spacecraft, and (2) *back contamination*, the introduction of extraterrean life to Earth by returning spacecraft.

Planetary protection has been a serious concern since the start of the Space Age. International quarantine standards for solar system exploration were drafted shortly after the launch of Sputnik. The U.N. Outer Space Treaty of 1967 articulated the policy that guides spacefaring nations even today. The international Committee on Space Research (COSPAR) is the focal point of PP activities. It maintains and revises policies to reflect advances in scientific understanding and technological capabilities. In the United States, NASA is responsible for implementation and compliance with appropriate planetary protection policy for all space missions. NASA's Planetary Protection

Officer establishes detailed implementation requirements for specific mission types in accordance with COSPAR policy and NASA directives.

### Mission Requirements

COSPAR policy currently specifies PP controls for five categories of target-body/mission types, dependent on degree of scientific interest and the potential for contamination. The most stringent category applies to round-trip missions with sample return. One-way robotic missions have varying and lesser levels of controls depending on their target body (*e.g.*, moon, planet, asteroid, comet, etc.) and mission type (orbiter, flyby, lander, rover, penetrator, etc.). Over the years, NASA has often sought scientific advice on PP from the Space Studies Board of the National Research Council, which has issued numerous reports and recommendations related to both robotic and human missions.

PP contributes to astrobiological research and solar system exploration. Both pursuits seek to determine whether extraterrean bodies host indigenous life and whether such life would be related to Earth life. Answering these questions requires the avoidance of forward contamination—both microbes and organic constituents that could interfere with *in situ* science observations (*e.g.*, living or dead organisms or parts, as well as life related materials such as proteins, lipids, DNA, RNA and organic molecules of biogenic origin; see Sec. 3A). PP requirements aim to minimize contamination by reducing the microbiological burden and maintaining organic cleanliness throughout spacecraft assembly, testing, and prelaunch.

For each mission, NASA develops and implements a Planetary Protection Plan. The process involves the scheduling and implementation of required measures during design, construction, assembly, and prelaunch phases as well as documentation and certification throughout.

### One-Way Missions

Implementation of a robotic mission's PP Plan involves a number of activities. Depending on the particular mission, these may include the analysis of orbital lifetimes and accidental impacts, use of microbiological assays to monitor cleanness during assembly, sterilization and cleaning of the spacecraft and/or hardware (*e.g.*, dry heat), protection of the spacecraft from recontamination prior to launch, and documentation of compliance through all mission phases.

### Sample Return Missions

Round-trip missions—particularly those returning from places with the potential for harboring life—must include measures to avoid both forward contamination of the target body and back contamination of Earth. For example, for sample return missions from Mars, all materials that have been exposed to the martian surface will be considered biohazardous until proven otherwise by rigorous testing in an appropriate containment facility. Conceptual requirements for a test protocol were developed through a series of international workshops.

### Human Missions

PP requirements for human missions have not yet been developed, but appropriate controls will undoubtedly involve ongoing research and technology development based on data from precursor robotic missions about the presence of possible extraterrean life and the prospect for special regions with the potential for liquid water. Although human missions to Mars are decades away, recent workshops have undertaken preliminary discussions on technical, scientific, operational, and policy issues that must be ad-dressed to enable safe human exploration missions in the future.

### Abbreviations

COSPAR, Committee on Space Research; DNA, deoxyribonucleic acid; ESA, European Space Agency; ICE, International Cometary Explorer; ISEE, International Sun–Earth Explorer; ISS, International Space Station; JAXA, Japan Aerospace Exploration Agency; NASA, National Aeronautics and Space Administration; pp, partial pressure; PP, Planetary Protection; RNA, ribonucleic acid; RWV, rotating wall vessel; SAS, space adaptation syndrome.

## Notes on Figures

### FIG. 1.1. Local Numbers (LM) (p. 741)

### FIG. 1.2. Color-Magnitude Diagram (KvB) (p. 743)

Data are discussed in von Braun *et al.* (2002).

### FIG. 1.3. Stellar Properties (KvB, AvM) (p. 744)

Intrinsic properties of the main sequence spectral types, post-main sequence stellar states, and Jupiter and Earth are given for comparison. Luminosity is over all wavelengths (except for pulsars), average temperature is the blackbody temperature, and main sequence lifetime assumes that the Sun has a main sequence lifetime of 10 billion years. Appearance was determined with the help of: http://www.mi.infm.it/manini/dida/BlackBody.html?textBox=5000

### FIG. 2.1. Solar System Bodies (AvM) (p. 749)

#### *A. Orbital Parameters*

Orbital parameters for the Solar System planets, the asteroid/comet parent distributions, and several important planetary satellites.

#### *B. Planetary Properties*

Planetary properties for Solar System planets, the asteroid/comet parent distributions, and several important planetary satellites. Values with question marks are uncertain because of observational limitations. Compositional components are listed in order of concentration by mass.

### FIG. 2.2. Geologic Time Scale (LM) (p. 757)

This representation of the geologic time scale of Earth was generated by using the International Commission on Stratigraphy website. The martian time scale was generated from the primary literature. Thanks to Andrew Knoll for help in interpretation and presentation.

http://www.stratigraphy.org

### FIG. 2.3. The History of Atmospheric $O_2$ (MC) (p. 763)

This diagram also appears in Catling and Claire (2005) and is here by permission of the authors.

### FIG. 4.1. Important Events in the Precambrian (EJ) (p. 779)

Geological time scale with important geological and biological events in the Precambrian incorporates some information from Brocks *et al.* (2003).

## FIG. 5.1. Extrasolar Planets (AvM) (p. 784)

Properties of the 149 extrasolar planets discovered by February 2005. Extrasolar planets are defined as orbiting a parent star and, therefore, do not include free-floating brown dwarfs. Mass and distance limits are defined by the completeness limit for radial velocity surveys.

### *A. Extrasolar Planet Mass Distribution*

### *B. Extrasolar Planet Orbital Characteristics*

http://www.obspm.fr/encycl/encycl.html (maintained by Jean Schneider)

## FIG. 5.2. The Habitable Zone around Main Sequence Stars (AvM) (p. 786)

Location of the habitable zone for different stellar types. Modified from Kasting *et al.* (1993).

## FIGS. 6.1–3. The Tree of Life (LM) (pp. 794–795)

All three "Tree of Life" diagrams come from simplifications of phylogenetic analyses on small subunit ribosomal RNA (SSU rRNA) using the method known as neighbor joining. (See Sec. 3A for the molecule and Sec. 4B for the methodology.) It should be noted that, while most conceptions of the tree of life are based on SSU rRNA analyses, there is currently debate as to whether they are truly representative of historical relationships. The trees have been modified from their web-based versions, and all mistakes should be attributed to the editor.

## FIGS. 6.1 (Archaea) and 6.2 (Bacteria)

These trees were generated using data published by Norman Pace. The trees have been significantly modified.

http://www.pacelab.colorado.edu/PI/norm.html

## FIG. 6.3 (Eukarya)

Copied and modified by permission from Mitchell Sogin's website.

http://www.mbl.edu/labs/Sogin/Pages/rg.html

Mitochondria fall within the alpha-proteobacteria, while chloroplasts fall at the base of the cyanobacteria on the basis of SSU rRNA phylogenies (see previous diagram). It should be noted that research on endosymbiosis—encapsulation of one organism by another as in the case of chloroplasts and mitochondria—is ongoing. Mitochondrial remnants have been found in the organisms at the base of the tree, which suggests that mitochondria entered Eukarya earlier than the arrow suggests and were lost in the oldest branches. Similarly, there is growing evidence that a single endosymbiotic event is responsible for the incorporation of cyanobacteria. Both chlorophytes and rhodophytes have primary endosymbioses—they picked up cyanobacteria—and this may represent a single event preceding the divergence of the two groups. Other eukaryotes acquired phototrophy by picking up chlorophytes or rhodophytes (secondary endosymbiosis), or another eukaryote (tertiary endosymbiosis).

Phylogenetic information is available from a number of publicly available sites. Most notable are the taxonomy browser run by the National Center for Biotechnology Information (taxonomic nomenclature), the Tree of Life Web Project (large-scale phylogenies), and Ribosomal Database Project out of Michigan State University (publicly available RNA sequences). See also *Endosymbiosis* in Sec. 4D.

http://www.ncbi.nlm.nih.gov/Taxonomy/taxonomy-home.html
http://www.tolweb.org/tree/phylogeny.html
http://www.rdp.cme.msu.edu/index.jsp

## FIG. 6.4. Characteristics of the Three Domains (OJ) (p. 796)

## FIG. 6.5. Redox Potentials and Life (GD) (p. 798)

Copied by permission and formatted from a presentation by Kenneth Nealson.

## FIG. 6.6. The Limits of Known Life on Earth (AnM) (p. 800)

Adapted from Marion *et al.* (2003).

**FIG. 6.7. Extremophiles (AnM) (p. 801)**

This list is not comprehensive but does include some examples of extremophiles. Adapted from Rothschild and Mancinelli (2001) and Cavicchioli (2002).

Address reprint requests to:
*Lucas John Mix*
*8700 196th Ave NE*
*Redmond, WA 98053*

*E-mail:* lucas@flirble.org